\documentclass{article}

\usepackage{lscape}
\usepackage{simplewick}

\usepackage{amssymb,amsfonts,amsmath}
\usepackage{cite,enumerate,float,indentfirst}
\usepackage{color}

\def\be{\begin{eqnarray}}
\def\ee{\end{eqnarray}}
\def\nn{\nonumber}

\def\p{\partial}
\def\tr{{\rm tr}\,}
\def\Tr{{\rm Tr}\,}

\def\l[{\phantom.[}

\definecolor{red}{rgb}{1,0,0}
\definecolor{orange}{rgb}{1,0.5,0}
\definecolor{violet}{rgb}{0.7,0,1}

\def\cre{\color{red}}
\def\co{\color{orange}}
\def\cy{\color{yellow}}
\def\cg{\color{green}}
\def\cb{\color{blue}}
\def\cv{\color{violet}}

\hoffset= - 1.0in         

\begin{document}

\title{\vspace{.1cm}{\LARGE {\bf Ward identities and combinatorics of rainbow tensor models
}\vspace{.5cm}}
\author{{\bf H. Itoyama$^{a,b}$},
{\bf A. Mironov$^{c,d,e}$},
\ {\bf A. Morozov$^{d,e}$}
}
\date{ }
}

\maketitle

\vspace{-6.2cm}

\begin{center}
\hfill FIAN/TD-07/17\\
\hfill IITP/TH-06/17\\
\hfill ITEP/TH-12/17\\
\hfill OCU-PHYS 464
\end{center}

\vspace{4.cm}

\begin{center}
$^a$ {\small {\it Department of Mathematics and Physics, Graduate School of Science,
Osaka City University, 3-3-138, Sugimoto, Sumiyoshi-ku, Osaka, 558-8585, Japan}}\\
$^b$ {\small {\it Osaka City University Advanced Mathematical Institute (OCAMI), 3-3-138, Sugimoto, Sumiyoshi-ku, Osaka, 558-8585, Japan}}\\
$^c$ {\small {\it I.E.Tamm Theory Department, Lebedev Physics Institute, Leninsky prospect, 53, Moscow 119991, Russia}}\\
$^d$ {\small {\it ITEP, B. Cheremushkinskaya, 25, Moscow, 117259, Russia }}\\
$^e$ {\small {\it Institute for Information Transmission Problems,  Bolshoy Karetny per. 19, build.1, Moscow 127051 Russia}}
\end{center}

\vspace{.5cm}

\begin{abstract}
We discuss the notion of renormalization group (RG) completion
of non-Gaussian Lagrangians
and its treatment within the framework of Bogoliubov-Zimmermann theory
in application to the matrix and tensor models.
With the example of the simplest non-trivial RGB tensor theory (Aristotelian rainbow),
we introduce a few methods, which allow one
to connect calculations in the tensor models to those in the matrix models.
As a byproduct, we obtain some new factorization formulas and sum rules
for the Gaussian correlators
in the Hermitian and complex matrix theories, square and rectangular.
These sum rules describe correlators as
solutions to finite linear systems, which are much simpler than
the bilinear Hirota equations and the infinite Virasoro recursion.
Search for such relations can be a way to solving the tensor models,
where an explicit integrability is still obscure.
\end{abstract}

\bigskip

\bigskip

\section{Introduction}

The key problem in study of interactions in quantum field theory (QFT)
is an identification of RG-complete actions which contain
all local operators that can be generated by the action of
renormalization group.
Such theories possess rich Ward identities associated with change
of integration variables in the functional integral, which can be
alternatively associated with the diffeomorphisms in the space of
couplings.
In the conventional QFT, this technique is known as the Bogoliubov-Zimmermann (BZ)
renormalization theory \cite{BZren}. In matrix models, it reduces to the theory of
generalized Virasoro/W-constraints \cite{virrel}
and to the formalism of check operators \cite{AMM2,GMM}.
In the both cases, the story consists of two steps:

(1) The original interaction (operator), which we will call the {\it keystone} operator,
is complemented by its {\it tree} descendants
(see s.\ref{BZrem} below for an explanation of this central notion).
These are often not all independent
and, in the theories possessing the space-time, where one often distinguishes
between the UV and IR RG-completeness, many of them can be non-local
and neglected in the study of, say, the UV renormalization group.
Inclusion of {\it the tree descendants} makes the theory {\it quasiclassically} complete
and reduces the Ward relation (diffeomorphism) symmetry to representation
theory of the tree composition algebra in graph theory.
The old-fashioned renormalizability relevant for identification of the RG stable
low-energy theories like the Standard Model implies that only a finite
number of tree operators is generated, but, in generic string/M-theory
operative at the Planck scales, this restriction is not necessarily imposed.
Of course, it is never imposed in the theory of matrix and tensor models
along the lines of \cite{UFN23},
where the space-time degrees of freedom are not present at all, and there is no difference
between local and non-local operators.

(2) Unfortunately, besides the {\it tree} operators, there are also
{\it loop} operators, and in the study of Ward identities
{\it a la} \cite{virrel} they emerge from the Jacobians of change of the
integration variables.
Taming of loop operators is the main problem in the search for the RG-complete theories.
In fact, some of loop operators reduce to tree operators, e.g. all the loop operators
made from the planar or melonic diagrams in matrix and tensor models respectively.
Sometimes, all the other loop operators are algebraically expressed through the
tree operators. The famous example is the Hermitian matrix model,
where the tree and loop operators are respectively the single- and multi-trace operators,
the latter being just the products of single-trace ones.

In one-matrix models, we are accustomed to a very simple description
of all possible "gauge-invariant" operators: they are just products of traces,
$\prod_i\tr M^{k_i}$.
However, already for the two-matrix models the situation changes drastically:
even the single-trace operators $\Tr \left(\prod_i A^{k_i}B^{l_i}\right)$
are labeled by {\it words} and are difficult to enumerate in any efficient way.
In result, while, in the first case, it is easy to define generating
functions: they are products of, say, the resolvents $\Tr\frac{1}{z-M}$,
i.e. the functions of a single variable $z$,
in the second case, the counterpart of the resolvent is far more involved
and depends on infinitely many variables $x_i$ and $y_i$:
$\Tr \prod_i \frac{1}{x_i-A}\frac{1}{y_i-B}$.
In the multi-matrix and tensor models, the set of operators allowed by symmetry
becomes more and more involved.
However, the point is that most of these enumerable operators are in fact a kind
of alien to the original one.
If the keystone operator was, say $\Tr (ABAB)$, then neither $\Tr (AB^2AB)$
nor $\Tr (AB^2A^2B)$ would ever arise as (RG) time goes, if they were not present
from the very beginning.
This does not mean that they have vanishing correlators, this means
that they have vanishing {\it averages} at all times provided they
were not present in the initial state.
In other words, such operators can be excluded from consideration
by a {\it superselection rule}
(like superpositions of neutron and proton,
it can also deserve reminding that the gauge invariance is also present
in Yang-Mills theories in the sense of the same rule:
if the initial state was gauge invariant, then such are all the states
in the course of its evolution).

In fact, the RG-completion is a slightly weaker statement: it admits
not only the operators directly present in the keystone evolution
operator, but also their further descendants.
Allowed are all operators emerging in the course of {\it multi-evolutions}
(many time variables) associated with all the descendants of the
keystone operator. Still, the operators like $\Tr (AB^2A^2B)$ and their
more sophisticated counterparts would never emerge.
In variance with $\Tr (AB^2AB)$, they are not forbidden by any explicit
symmetry of the model.
What forbids them is a {\it hidden} symmetry.
In this case, it is the possibility of generalization \cite{recMM} from the square
to rectangular matrices $A$ and $B$, where such operators simply can not exist,
and, {\it therefore}, do not appear among the descendants; in result
they can not appear even if one makes the matrices square again,
only the generically allowed $\Tr (AB)^k$ can emerge.
Revealing and exploiting such symmetries is the main idea
of studying the {\it rainbow} tensor models, their enhanced symmetries
make properties of the RG-completeness much simpler and explicit.
In \cite{rainbow}, we already mentioned one example of this kind:
in the rainbow models {\it melonic}  are the only {\it existing} among the planar
diagrams, and this is what {\it explains} this their {\it dominance} at large $N$
in {\it all} tensor models.
However, the examples are not at all exhausted by this one, and we exploit
the power of the rainbow models more in the present text.

Thus, in a {\it particular} model, not {\it all} the operators permitted by symmetry
necessarily arise in the course of evolution and the ones which arise can form a smaller set,
much easier enumerated than one could expect.
Moreover, when considered from the right perspective, these sets come
with their own hierarchy, which also provides a useful approach
to the RG-evolution problem.
As already mentioned, only the tree descendants of the
keystone operators matter quasiclassically and already this reduces the set,
when the keystone is carefully selected, especially in the theories
with high symmetry like the rainbow tensor models of \cite{rainbow}.
The question is what happens at the loop operator level.
There are at least three options to consider:

First: all loop operators are algebraic functions of
the tree ones like it happens in the one-matrix model;

Second: some loop operators are independent of the tree ones, but they
also form a comprehensible subset, which can be just added to the
action, while all the rest of emerging operators are expressed
through them;

Third: this happens at the level of averages in certain limits,
as an analogy of the factorization of multi-trace operators
at large $N$, only this time this can be used to formulate a
model that is RG-complete in the limit (the ordinary matrix model
is RG-complete irrespective of any limits and factorizations).

The long-awaited surge in attention \cite{physfirst}-\!\!\cite{physlast} to the tensor models \cite{tensor}, \cite{GrFieldTh,Akhm}, \cite{BGRfirst}-\!\!\cite{BGRlast}
allows one
to begin a systematic investigation of these problems.
They did not receive enough attention within the matrix model context,
because the multi-matrix models \cite{MuMaMo} were long considered as rather exotic objects,
but, in the tensor case, the issue arises already in the indisputably beautiful examples.
The questions are what are the extended partition functions of these models,
where the full sets of symmetry-allowed operators are enormous
and practically innumerable?
Can we restrict our consideration to some nicer subsets?
How do we distinguish between allowed subsets, and
what makes them {\it closed} and the model {\it consistent}?

As already reminded at the beginning of this introduction,
in the conventional renormalizable quantum field theory, the requirements come
from the unitarity of the regularized evolution operator and are guaranteed
by the application of the Bogoliubov-Zimmermann procedure.
The question is what substitutes it (or how it looks) in generic string theory,
i.e. within the context of generic tensor categories and,
to begin with, of generic matrix and tensor models.
This question was addressed in \cite{GMS} in association with the work by A. Connes and D. Kreimer \cite{CK}
(which describes the Bogoliubov-Zimmermann formulas in terms of the Hopf algebra
of Feynman diagrams).
The true motivation was, however, somewhat broader and
included also the search for the QFT reformulation of
the problems of non-linear algebra \cite{NLA}.
In the present paper, we discuss further steps towards constructing the
renormalization group (RG) complete models and the RG-closed sets of operators.
We adopt a simplest option for
the definition of the {\it complete} models:
to request that they possess {\it a sufficiently rich}
set of the Ward identities, which can make them potentially integrable
(in a sense which still needs to be defined).
In other words, one can begin from the search of the tensor models, which are as close
in their solvability to the Hermitian matrix model as possible.

Investigating this problem, we actually discovered a previously unknown
feature of matrix models:
they possess additional, linear and finite, relations between
Gaussian averages, which allow one to find them {\it explicitly} and provide
a tremendously simple character expansion for the extended partition function,
with coefficients made from the dimensions of representations of $GL(N)$.
This is the long-awaited property explaining what lies at the intersection
of KP integrability and Virasoro constraints and what is so peculiar for
{\it the matrix-model $\tau$-functions}.
More important in the present context is that this is a {\it simple}
property, for which one can straightforwardly look in the tensor models,
once one manages to perform explicit calculations,
and at the very end of this paper we provide some initial evidence in favor
of its existence.


We begin in s.\ref{mamorem} by reminding the basics of matrix model theory from
\cite{UFN23} and some of more recent papers.
We also report the discovery of new relations and explicit formulas
for arbitrary Gaussian averages and extended (coupling/time-dependent) partition functions.
Then, in s.\ref{BZrem} we remind the basics of BZ theory
in the formalism of \cite{GMS}, best suited for applications to the
matrix and tensor models.
In the remaining part of the paper, we discuss two simple examples
of the rainbow-type tensor models.
The "red" model in s.\ref{redmod} trivially reduces to a
rectangular complex matrix model,
but another, "red-green" model in s.\ref{redgreenmod}
(which actually has {\it three} colorings and
can be naturally called RGB or Aristotelian)
exhibits interesting deviations from it,
which are already peculiar for tensor models.
Further generalization to the most interesting case with the tetrahedron-like interaction
remains as a next natural step to make.

\section{Combinatorics of matrix models: old results and new claims
\label{mamorem}}

The most interesting tensor models are the far-going generalizations
of the eigenvalue matrix models,
where everything needs to be re-analyzed:
expressions for the averages, recurrent relations between them,
their solutions provided by the $W$-representations,
the genus expansions, the spectral curves and the AMM/EO-topological recursions,
and their interpretations in terms of integrable systems,
the KP/Toda and Hurwitz $\tau$-functions.
Still, there are artificially designed tensor models,
which deviate from the matrix case in a minimal way,
with different directions of deviation, while preserving one or another
of the matrix model properties. Thus, their study is useful
not only for the initial steps in the tensor model theory,
but also for clarifying the origins and universality of particular structures,
revealed in the matrix model studies.
This section
provides a basis for such an analysis,
which is attempted in the remaining part of the paper.

The first question to address in any model is evaluation of the correlators
(averages of various operators).
This can be done either by direct calculation or by using the Ward identities.
We mostly concentrate on the interplay between these two,
with the $W$-representations and integrability mentioned only in passing.
Instead, we suggest to define the Gaussian correlators from very simple and {\it finite}
sets of {\it linear} equations,
which efficiently substitute both the Virasoro constraints and integrability.

\subsection{Hermitian matrix model}

\subsubsection{Partition function and Ward identities}

The model is associated with the integral over the $N\times N$ matrix $M$
\be
Z_H = \int dM \exp\left(-\frac{\mu}{2}\,\Tr M^2\right)
\ee
where the measure is induced by the norm $||\delta M||^2 = \Tr (\delta M)^2$
and the Ward identities are the usual Virasoro constraints \cite{virrel}
for the {\it extended} partition function
\be\label{HMM}
{\cal Z}_H\{t\} = \int dM \exp\left(-\frac{\mu}{2}\,\Tr M^2 + \sum_{k} t_k\Tr M^k\right)
\ee
that is,
\be
\hat L_n^H {\cal Z}_H =
\left(-\mu\frac{\partial}{\partial t_{n+2}}
+\sum kt_k \frac{\partial}{\partial t_{k+n}}
+ \sum_{a=1}^{n-1} \frac{\partial^2}{\partial t_a\partial t_{n-a}}
+2N\frac{\partial}{\partial t_n}
+ N^2\delta_{n,0}
\right)  {\cal Z}_H,
\ \ \ \ n\geq -1
\label{VirH}
\ee
(one often simplifies the formula by introducing the time $t_0$
with the additional constraint
$
\frac{\partial {\cal Z}_H}{\partial t_0} = N {\cal Z}_H
$
but a similar counterpart of this trick is not known for the rectangular and tensor models).

\subsubsection{The simplest averages from Virasoro recursion}

The correlators
\be
{\cal O}_{\Lambda} = \left< \prod_{i=1}^{l_\Lambda} \Tr M^{\lambda_i} \right>
= \left.\frac{1}{{\cal Z}}
\left( \prod_i \frac{\partial }{\partial t_{\lambda_i}}\right) {\cal Z}\right|_{t=0}
\ee
are naturally labeled by the Young diagrams $\Lambda$ with $l_\Lambda$ rows,
\be
\Lambda = \Big\{\lambda_1\geq\lambda_2\geq \ldots \geq \lambda_{l_\Lambda} > 0 \Big\}
\ee
They can be recursively restored by solving the Virasoro constraints and their $t$-derivatives.

The first steps of the recursion are:
\be
\begin{array}{ccc}
\hat L_0^H {\cal Z}_H = 0 & \Longrightarrow
& \mu\,\Big<\Tr M^2\Big> = N^2  \\
\frac{\partial}{\partial t_1}\hat L_{-1}^H {\cal Z}_H = 0 & \Longrightarrow
& \mu\,\Big<\Tr M \ \Tr M\Big> = N \\
\hat L_2^H {\cal Z}_H = 0 & \Longrightarrow
& \mu \,\Big<\Tr M^4\Big> = 2N\Big<\Tr M^2\Big> + \Big<\Tr M \ \Tr M\Big> = \frac{2N^3+N}{\mu}
\\
\ldots
\end{array}
\ee
see \cite{AMM1} for continuation of the list.
The parameter $\mu$ is kept in these formulas to identify the "direction"
of the recursion: every recursion step adds an extra power of $\mu$ in the denominator.

\subsubsection{Pictorial representation\label{mmp}}

One rarely uses pictures in discussing general features of the matrix models: an
analytical language is developed well enough for writing easily readable formulas.
Things are still very different in the tensor models, where at this stage we need
to express many ideas pictorially.
Because of this, we now do the same in the familiar matrix model case,
this can facilitate an understanding of pictures in the next sections.
We use the same {\it colorings} as there.
In the rainbow tensor models, there are several $U(N)$ gauge groups
and fields are charged with respect to different collections
of these groups, thus, colored are the types of indices in the fields
and fields themselves, we call this {\it coloring} as {\it multi-coloring},
preserving the word "color" for the {\it values} of indices inside the
fundamental representation of the particular gauge group.
The multi-coloring could also be called "flavor", or, even better, "techni-flavor",
but we decided to avoid this terminology.
In the Hermitian matrix model, only one type of coloring  remains,
we choose it red. Multi-coloring is also reduced to a single specie:
{\it a pair} of red lines.

The operators $\Tr M^k$ can be depicted as polygons with $k$ angles.
In particular, the ``keystone'' operator $\Tr M^3$ and its first descendant ${\Tr M^4}$ are:

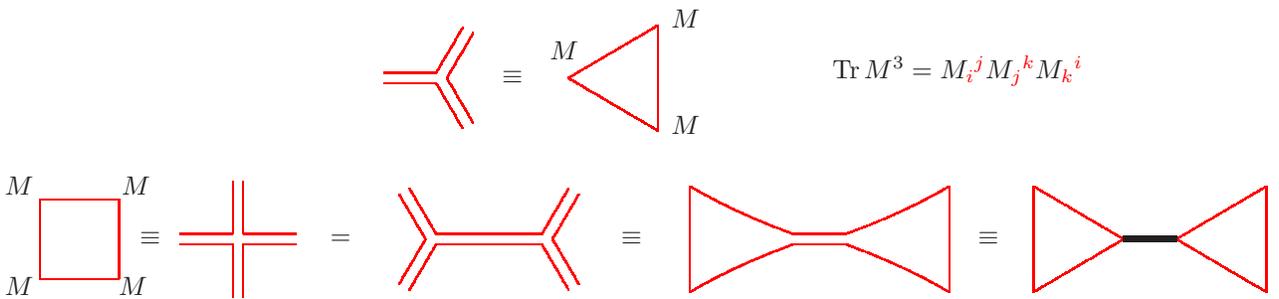
\begin{picture}(300,70)(-150,-25)

\put(0,0){\cre

\put(0,2){\line(-1,0){20}}\qbezier(0,2)(5,10.5)(10,19)
\put(0,-2){\line(-1,0){20}}\qbezier(0,-2)(5,-10.5)(10,-19)
\qbezier(4,0)(9,8.5)(14,17)\qbezier(4,0)(9,-8.5)(14,-17)

\put(50,0){
\qbezier(0,0)(17,10)(34,20)\qbezier(0,0)(17,-10)(34,-20)
\put(34,20){\line(0,-1){40}}
}

}
\put(25,-2){\mbox{$\equiv$}}
\put(150,0){\mbox{$\Tr M^3 = M_{\cre i}^{\cre j} M_{\cre j}^{\cre k} M_{\cre k}^{\cre i}$}}
\put(50,0){
\put(-7,7){\mbox{$M$}}
\put(39,19){\mbox{$M$}}
\put(39,-21){\mbox{$M$}}
}
\end{picture}

\begin{picture}(400,70)(0,-35)

\put(0,0){\cre

\put(75,0){
\put(-2,2){\line(-1,0){20}}\put(-2,2){\line(0,1){20}}
\put(-2,-2){\line(-1,0){20}}\put(-2,-2){\line(0,-1){20}}
\put(2,2){\line(1,0){20}}\put(2,2){\line(0,1){20}}
\put(2,-2){\line(1,0){20}}\put(2,-2){\line(0,-1){20}}
}

\put(0,-15){
\put(0,0){\line(0,1){30}}\put(0,30){\line(1,0){30}}
\put(0,0){\line(1,0){30}}\put(30,0){\line(0,1){30}}
}

\put(100,0){
\put(310,0){
\qbezier(0,0)(-17,10)(-34,20)\qbezier(0,0)(-17,-10)(-34,-20)
\put(-34,20){\line(0,-1){40}}
}
\put(330,0){
\qbezier(0,0)(17,10)(34,20)\qbezier(0,0)(17,-10)(34,-20)
\put(34,20){\line(0,-1){40}}
}

\put(180,0){
\qbezier(5,2)(-17,10)(-34,20)\qbezier(5,-2)(-17,-10)(-34,-20)
\put(-34,20){\line(0,-1){40}}
}
\put(210,0){
\qbezier(-5,2)(17,10)(34,20)\qbezier(-5,-2)(17,-10)(34,-20)
\put(34,20){\line(0,-1){40}}
\put(-5,2){\line(-1,0){20}}  \put(-5,-2){\line(-1,0){20}}
}
}

\put(170,0){
\put(-20,0){
\put(0,2){\line(1,0){20}}\qbezier(0,2)(-5,10.5)(-10,19)
\put(0,-2){\line(1,0){20}}\qbezier(0,-2)(-5,-10.5)(-10,-19)
\qbezier(-4,0)(-9,8.5)(-14,17)\qbezier(-4,0)(-9,-8.5)(-14,-17)
}
\put(20,0){
\put(0,2){\line(-1,0){20}}\qbezier(0,2)(5,10.5)(10,19)
\put(0,-2){\line(-1,0){20}}\qbezier(0,-2)(5,-10.5)(10,-19)
\qbezier(4,0)(9,8.5)(14,17)\qbezier(4,0)(9,-8.5)(14,-17)
}
}

}
\put(38,-2){\mbox{$\equiv$}}
\put(110,-2){\mbox{$=$}}
\put(220,-2){\mbox{$\equiv$}}
\put(355,-2){\mbox{$\equiv$}}

\put(-13,17){\mbox{$M$}}
\put(31,17){\mbox{$M$}}
\put(-13,-21){\mbox{$M$}}
\put(30,-21){\mbox{$M$}}

\linethickness{0.8mm} \put(410,0){\line(1,0){20}}

\end{picture}

\noindent
We remind that the lines in the matrix model pictures are used to describe
the contraction of indices.
Note the interplay between the double and single red lines.
The thick black line denotes the tensor $\delta^i_{i'}\delta_j^{j'}$
or the action of the operator $\Tr \left(\frac{\p}{\p M}\otimes\frac{\p}{\p M}\right)$,
which plays the role of the propagator in the matrix model.
The identity in the picture is manifestation of the relation
\be
\Tr M^4 = \frac{1}{9}\,\frac{\partial \,\Tr M^3 }{\partial M_i^j }\,
\frac{\partial \,\Tr M^3 }{\partial M^j_i}
\nn
\ee
with a combinatorial coefficient omitted.
The parameter $\mu^{-1}$ can be easily included or omitted, as one prefers.
In generic QFT, the propagator contains also a "propagating" (space-time dependent) factor,
which makes the "composite" operator non-local, still a similar formalism
is useful to describe the convolution of indices, it is enough to omit
$\mu$, with all the derivatives it can contain.
This blowing up of interaction vertices (operators) in Feynman diagrams
does not make too much sense in theories with the matrix-valued fields,
like the Yang-Mills theories, however, in the tensor models, where the indices are much less
under control, this formalism becomes very useful.

Clearly, all the operators $\prod_{i=1}^{L+1} \Tr M^{k_i}$ can be
depicted in this blown-up formalism as triangles
connected by thick black lines, moreover, in many different ways.
The single-trace operators with $L+1=1$ emerge in this way from the trees,
while $L$ is the number of loops in the graph with black edges.
In other words, in this formalism of describing the keystone descendants,
the single-traces are the tree operators and the multi-traces are loop operators.
This is a formulation which can be easily extended from matrices to tensors,
where the notion of trace is not very relevant.

Another element of the formalism is an operator $M^m$ with open ends (with no trace),
we denote it by the thick red line (vector):

\begin{picture}(300,40)(-150,-20)

\put(0,0){\cre
\put(0,0){\line(1,0){30}}\put(30,0){\line(1,0){30}}\put(60,0){\line(1,0){10}}
\put(100,0){\line(1,0){10}}\put(110,0){\line(1,0){30}}
\put(30,0){\circle*{4}}\put(60,0){\circle*{4}}\put(110,0){\circle*{4}}
\linethickness{0.8mm} \put(-30,0){\line(-1,0){40}}
\put(-53,10){\mbox{$m$}}
}
\put(-20,-2){\mbox{$\equiv$}}
\put(80,-2){\mbox{$\ldots$}}

\end{picture}

\noindent
Let us note that the colored lines throughout the text are associated not with elements of the Feynman technique, but depict operators, or, more exactly, their color structure: how the concrete operator is constructed from the colored fields. In particular, the loop colored lines (without external ends) denote the invariant operators.

One can use just the same thick line with another label $z$ to denote a sum over $m$,
for example  $(z-M)^{-1}$, then its trace, resolvent will be depicted as
a thick red circle. One can consider also the traces like $\Tr e^{sM}$ etc.
In the next sections, we use only the thick lines and circles with indices $m$.

An important feature of the thick red line is that the thick {\it black} propagator
can be attached to its interior, moreover we have identities like

\begin{picture}(300,100)(-130,-50)

\put(5,-15){\mbox{$m_1$}} \put(30,-15){\mbox{$m_2$}} \put(60,-15){\mbox{$m_3$}}
\put(135,25){\mbox{$m_1+m_3-2$}} \put(130,-17){\mbox{$m_2$}}
\linethickness{0.8mm}

{\cre \put(0,0){\line(1,0){70}}}
\qbezier(20,0)(35,25)(50,0)
\put(30,0){
{\cre
\put(100,20){\line(1,0){50}}
\put(125,-5){\qbezier(-14,0)(-14,14)(0,14) \qbezier(14,0)(14,14)(0,14)
\qbezier(-14,0)(-14,-14)(0,-14)\qbezier(14,0)(14,-14)(0,-14) }
 }
 }

\end{picture}

\noindent
which can be described by the formula
\be
\acontraction{\Big(\Tr}{{\partial\over\partial M}}{{\partial\over\partial M}\Big)\Big[
(M^{m_1-1})_{\cre i}^{\cre m}}{M}
\acontraction[2ex]{\Big(\Tr{\partial\over\partial M}}{{\partial\over\partial M}}{\Big)\Big[
(M^{m_1-1})_{\cre i}^{\cre m}M_{\cre m}^{\cre k}
(M^{m_2})_{\cre k}^{\cre n}}{M}
\Big(\Tr{\partial\over\partial M}{\partial\over\partial M}\Big)\Big[
(M^{m_1-1})_{\cre i}^{\cre m}M_{\cre m}^{\cre k}
(M^{m_2})_{\cre k}^{\cre n}M_{\cre n}^{\cre l}(M^{m_3-1})_{\cre l}^{\cre j}\Big]=\\
=(M^{m_1-1})_{\cre i}^{\cre l}(M^{m_2})_{\cre k}^{\cre k}(M^{m_3-1})_{\cre l}^{\cre j}=(M^{m_1+m_3-2})_{\cre i}^{\cre j}(M^{m_2})_{\cre k}^{\cre k}
\ee
and overbrackets denote the Wick pairing, i.e. the concrete field in the expression that is differentiated.

In application to a thick red circle, this identity converts the thick black propagator
into an operator cutting one circle into two pieces,
then the recursion relation underlying the Virasoro identities is just

\begin{picture}(300,100)(-150,-50)

\put(-105,-2){\mbox{$\mu\ \cdot $}}
\put(-40,-2){\mbox{$= \ \ \sum_{m_1+m_2=m}$}}
\put(-80,20){\mbox{$m$}}
\put(55,20){\mbox{$m_1$}}
\put(55,-25){\mbox{$m_2$}}
\put(200,30){\mbox{$m_1$}}
\put(200,-35){\mbox{$m_2-1$}}

\put(80,-2){\mbox{$= \ \ \sum_{m_1+m_2=m}$}}

\linethickness{0.8mm}

\put(-70,0){\cre
\qbezier(-14,0)(-14,14)(0,14) \qbezier(14,0)(14,14)(0,14)
\qbezier(-14,0)(-14,-14)(0,-14)\qbezier(14,0)(14,-14)(0,-14)
 }

 \put(50,0){\cre
 \qbezier(-14,0)(-14,14)(0,14) \qbezier(14,0)(14,14)(0,14)
\qbezier(-14,0)(-14,-14)(0,-14)\qbezier(14,0)(14,-14)(0,-14)
 }
 \put(36,0){\line(1,0){28}}

\put(180,20){\cre
 \qbezier(-14,0)(-14,14)(0,14) \qbezier(14,0)(14,14)(0,14)
\qbezier(-14,0)(-14,-14)(0,-14)\qbezier(14,0)(14,-14)(0,-14)
 }

\put(180,-20){\cre
 \qbezier(-14,0)(-14,14)(0,14) \qbezier(14,0)(14,14)(0,14)
\qbezier(-14,0)(-14,-14)(0,-14)\qbezier(14,0)(14,-14)(0,-14)
 }

\end{picture}

\subsubsection{Genus expansion, spectral curve and topological recursion}

The simplest way to deal with the Ward identities like (\ref{VirH}) is to rewrite
them in the form of the {\it loop equation}
\be\label{loop}
-\mu z\rho^{(1)} (z)+\mu N+\rho^{(1)}(z)^2+\rho^{(2)}(z,z)+\Big[V'(z)\rho^{(1)}(z)\Big]_-=0
\ee
for (multi-)resolvents (connected components of multi-trace correlators)
\be
\rho^{(n)}(z_1,\ldots,z_n) =  \hat\nabla_{z_1} \ldots \hat\nabla_{z_n}\log {\cal Z}
\ee
Note that we further often refer to $\rho^{(1)}(z)$ as just to $\rho(z)$.
In these formulas, $[...]_-$ means projection onto the negative powers of $z$,
\be
V(z) = \sum_k t_kz^k \ \ \ \ \ \ {\rm  and} \ \ \ \
\hat\nabla_z = \sum_{k\ge 0} \frac{1}{z^{k+1}}\frac{\partial}{\partial z_k}
\ee
and we defined the derivative ${\partial Z\over\partial t_0}\equiv NZ$.

One can take the loop equation (\ref{loop}) at all $t_k=0$,
\be\label{loop1}
-\mu z\rho^{(1)} (z)+\mu N+\rho^{(1)}(z)^2+\rho^{(2)}(z,z)=0
\ee
then, apply the operator $\hat\nabla_{z}$ to (\ref{loop}) and again put all $t_k=0$, which includes $\rho^{(3)}(z)$ etc. This gives a kind of Bogoliubov chain relations. In order to construct an effective recursion, one has to go further and introduce also a parameter $g$ of the quasiclassical, or genus expansion via rescaling $t_k\to {1\over \hbar}t_k$, $\log Z\to {1\over \hbar^2}\log Z$, $N\to {1\over \hbar}N$. Now one can consider the planar limit (leading order in $\hbar$). This reduces (\ref{loop1}) to an algebraic equation
\be
-\mu z\rho_0 (z)+\mu N+\rho_0(z)^2=0
\ee
which solution is
\be\label{plr}
\rho_0(z)={\mu\over 2}\Big(z-\sqrt{z^2-{4N\over \mu}}\Big)
\ee
The imaginary part of $\rho_0(z)$ describes the density of the eigenvalues in the matrix model, and is equal to
\be
y(z)=\mu\sqrt{z^2-{4N\over \mu}}
\ee
This is the notorious semi-circle distribution \cite{sc}, which satisfies the equation
\be
y(z)^2=\mu^2\Big(z^2-{4N\over \mu}\Big)
\ee
and is called the spectral curve (in this concrete case, it is a sphere).

\bigskip

Dealing with the loop equations, one achieves at least two goals:
\begin{itemize}
\item One can promote recursions between particular correlators to those
between their particular generating functions, the best known example
is the genus expansion
with {\it the AMM/EO topological recursion} \cite{AMM1,AMM2,AMM3,AMM/EO} between contributions of different genera
$\rho^{(n|g)}$ of multi-resolvents,
\be
\rho^{(n)}(z_1,\ldots,z_n) =  \sum_{g=0}^\infty  \hbar^{1-g}\rho^{(n)}_g(z_1,\ldots,z_n)
\ee
which are actually meromorphic poly-differentials on
{\it the spectral curve}, which is an equation for the ordinary resolvent
at genus zero.
In the particular case of Hermitian model, the Gaussian planar resolvent
(\ref{plr}) as a function of $1/\mu$
is a generating function of the Catalan numbers.
As pointed out in \cite{AMM/EO}, in abstract form, the topological recursion
is applicable to arbitrary families of Riemann surfaces and
thus works in many examples, where a matrix-model realization is
not yet discovered.
\item One can shift any time-variable $t_k\longrightarrow T_k+t_k$, not only
$t_2 \longrightarrow t_2-\frac{\mu}{2}$,   and to consider $t$ expansions
of ${\cal Z}$ around non-Gaussian points parameterized by the {\it superpotentials}
$W(z) = \sum T_kz^k$.
This leads to the theory of Dijkraaf-Vafa phases \cite{DV,AMM1,AMM2,CM,DVmore,M},
which depend drastically on the power of the polynomial $W(z)$.
\end{itemize}

Despite best studied, the loop equation/resolvent approach has a serious
drawback: even in the simplest case of Hermitian model at the Gaussian point
it does not produce a full answer for correlators:
no multi-resolvent $\rho^{(n)}$ was so far calculated in closed form,
only particular components $\rho^{(n|g)}$ of their genus expansions are known.
To solve {\it this kind} of problems, one can proceed at least in two other ways:
look at the $W$-representations of ${\cal Z}(t)$ and look at less
naive (model-dependent) generating functions rather than at the ordinary
(universal) resolvents.

\subsubsection{$W$-representation}

$W$-representation \cite{ShaW,AMMN2} provides a simple "dual" formula for ${\cal Z}\{t\}$,
expressing it through differentiation rather than integration:
\be\label{Wrep}
{\cal Z}_H\{t\} = e^{\frac{1}{2\mu}\hat W_H } e^{Nt_0}
\ee
where the relevant cut-and-join operator is a $(-2)$-harmonic of the simplest operator
from a big family studied in \cite{MMN1}:
\be
\hat W_H = \sum_{a,b} \left(
abt_at_b\frac{\partial}{\partial t_{a+b-2}} +
(a+b+2)t_{a+b+2} \frac{\partial^2}{\partial t_a\partial t_b}\right)
\label{cjoH}
\ee

\subsubsection{Alternative generating functions and their Fourier transform}

Instead of resolvents, one can consider various other generating functions of the correlators in the matrix model. For instance, one can use the Wilson loops \cite{expgen}
\be
\Tr e^{sM}=\sum_{k} \frac{s^k}{k!}\,\Tr M^k
\ee
and, especially interesting, the Harer-Zagier generating function  \cite{HZ,MorShaHZ}
\be
\sum_{k=0}^\infty \frac{z^{k}}{(2k-1)!!}\Big<{\rm Tr}_{_{N\times N}} M^{2k}\Big>
= \frac{\mu}{2z}\left(\left(\frac{\mu+z}{\mu-z}\right)^N-1\right)
\label{HZ1pH}
\ee
and its Fourier transform (FT)
in the matrix size \cite{AMM1}
\be
\sum_{N,k} \frac{\lambda^N z^k}{(2k-1)!!}\Big<{\rm Tr}_{_{N\times N}} M^{2k}\Big>
= \frac{ \lambda}{(1-\lambda)\Big(1-\lambda-(1+\lambda)z/\mu\Big)}
\ee
This FT generating function leads to far more explicit expressions for matrix model averages.
In result,
\be
{\rm FT}_{[k]} =
\sum_{\lambda} \lambda^N \Big<{\rm Tr}_{_{N\times N}} M^{2k}\Big> =
\frac{\lambda(1+\lambda)^k}{\mu^k(1-\lambda)^{k+2}}\cdot (2k-1)!!
\label{HZ1pHl}
\ee
which one can easily use with the help of binomial expansion
\be
\frac{1}{(1-\lambda)^{k+2}} = \sum_N \lambda^N \,\frac{(N+k+1)!}{(k+1)!\,N!}
\ee
In particular,
\be
\mu\Big< \Tr M^2\Big> = N^2, \ \  \ \ \mu^2\Big< \Tr M^4\Big> = 2N^3+N, \ \ \ \
\mu^3\Big< \Tr M^6\Big> = 5N^4+10N^2, \ \ \ \ \mu^4\Big< \Tr M^8\Big> = 14N^5+70N^3+21N, \nn \\
\!\!\!\!\!\!\!\! \mu^5\Big< \Tr M^{10}\Big> = 42N^6+420N^4+483N^2, \ \ \ \
\mu^6\Big< \Tr M^{12}\Big> = 132N^7+2310N^5+6468N^3+1485 N, \ \ \ldots \ \ \ \ \ \ \ \
\ee
Similar generating functions for the exact Gaussian correlators are also available from
\cite{MorShaHZ} for the double- and triple-trace averages
$\Big< \Tr M^{k_1}\Tr M^{k_2}\Big>$ and $\Big< \Tr M^{k_1}\Tr M^{k_2}\Tr M^{k_3}\Big>$.

\bigskip

These Gaussian correlators satisfy amusing sum rules, for example:
\be
\pm \Big<\Tr M^2\Big> + \Big<(\Tr M)^2\Big> = \pm N(N\pm 1),
\nn\\ \nn \\
\pm 6\Big<\Tr M^4\Big>  + 8\Big<\Tr M^3\, Tr M\Big> + 3\Big<(\Tr M^2)^2\Big> \pm
6\Big<\Tr M^2\,(\Tr M)^2\Big> + \Big<(\Tr M)^4\Big> =  3N(N\pm 1)(N\pm 2)(N\pm 3),
\nn\\
\mp 2\Big<\Tr M^4\Big>    - \Big<(\Tr M^2)^2\Big> \pm
2\Big<\Tr M^2\,(\Tr M)^2\Big> + \Big<(\Tr M)^4\Big> =  -(N+1)N(N-1)(N\pm 2),
\nn\\
 -4 \Big<\Tr M^3\, Tr M\Big> + 3\Big<(\Tr M^2)^2\Big>   + \Big<(\Tr M)^4\Big> =
 3(N+1)N^2(N-1),
\nn\\ \nn \\
\ldots
\label{sum1}
\ee
The coefficients at the l.h.s. are actually the properly normalized symmetric
group characters $\varphi_R(\Lambda)$ from
\cite{MMN1},
so that, in general, the sum rules are
\be\label{sum}
\left.\frac{1}{d_R\cdot {\cal Z}_H}\cdot
\chi_R\left\{\frac{1}{n}\frac{\p}{\p t_n}\right\}{\cal Z}_H\right|_{t=0}
= \sum_{\Lambda\,\vdash |R|}\varphi_R(\Lambda)\cdot {\cal O}_{\Lambda}
= c_R  \cdot \frac{D_R(N)}{d_R}
\ee
for all Young diagrams $R$ of even size (number of boxes) $|R|$.
Here $\chi_R$ and $D_R(N)$ are respectively the Schur polynomials and the dimensions of representation $R$ of the linear group
$GL(N)$, the factor $d_R=\chi_R(t_n=\delta_{n,1})$
is the dimension of representation $R$ of the symmetric group $S_{|R|}$ divided by $|R|!$
\cite{Fulton}.
The coefficients $c_R$ are occasionally equal to $\varphi_R([2])$,   $\varphi_R([2,2])$
and $\varphi_R([2,2,2])$
for $|R|=2$, $|R|=4$ and $|R|=6$ respectively, with an obvious implication for the general case.
These sum rules allow one to express all averages ${\cal O}_{\Lambda}$ through those for the
single-line Young diagrams ${\cal O}_{[\,|\Lambda|\,]}$,
which are fully described by (\ref{HZ1pHl}).
They also provide a simple formula for the character expansion of the partition function:
\be\label{ZH}
\boxed{
{\cal Z}_H\{t\} = \sum_{{\rm even\ size}\ R}\varphi_R\Big(\underbrace{[2,\ldots,2]}_{|R|/2}\Big)\cdot D_R(N)\cdot \chi_R\{t\}=
\sum_{{\rm even\ size}\ R} \varphi_R\Big(\underbrace{[2,\ldots,2]}_{|R|/2}\Big)\cdot \chi_R\left\{t_n=\frac{N}{n}\right\}\cdot \chi_R\{t\}
}
\ee
Existence of simple formulas like (\ref{HZ1pH}),
and thus of their far-going generalization (\ref{sum})
for the Hermitian matrix model seems
to reflect \cite{AMM1,MorShaHZ} its KP/Toda integrability \cite{GMMOZ,versus,UFN23},
i.e. a somewhat deeper structure than just the Ward identities.
In particular, integrability requires the coefficients $c_R$ to be made from the
Casimir exponentials \cite{GKM2,AMMN,Charint}.
Like the Virasoro recursion these relations are {\it linear} in correlators
and like the Hirota bilinear identities they preserve the grading:
hence, they combine the advantages of these both.
They are sufficient to obtain {\it any} Gaussian correlator: using the orthogonality relation
\be
\sum_R d_R^2\varphi_R(\Lambda)\varphi_R(\Lambda')={1\over z_\Lambda} \delta_{\Lambda,\Lambda'}
\ee
where $z_{\Lambda}$ is the standard symmetric factor of the Young diagram (order of the automorphism)
\cite{Fulton}, one can obtain from (\ref{sum})
\be\label{completesolH}\boxed{
{\cal O}_{\Lambda}=z_\Lambda\sum_{R\,\vdash |\Lambda|}
c_R\cdot d_R\cdot D_R(N)\cdot \varphi_R(\Lambda)
= \sum_{R\,\vdash |\Lambda|}
\varphi_R\Big(\underbrace{[2,\ldots,2]}_{|R|/2}\Big)\cdot D_R(N)\cdot \psi_R(\Lambda)
}
\ee
where $\psi_R(\Lambda) =  z_\Lambda\, d_R\,  \varphi_R(\Lambda)$
are the conventionally normalized characters \cite{Fulton} called by the command $Chi(R,\Lambda)$
in MAPLE.
Since all the quantities $\varphi_R(\Lambda)$, $d_R$, $z_\Lambda$ and $D_R(N)$ are
well-known from the basic representation theory,
these formulas provide a long-looked-for {\it complete} perturbative {\it solution}
to the Hermitian matrix model
(perturbative means that it is still restricted to the Gaussian point,
while the non-perturbative analysis of the Dijkraaf-Vafa phases still requires the
use of Virasoro constraints {\it a la} \cite{AMM1}).
It would be very interesting to find a counterpart of this phenomenon
and these formulas for the tensor model, see s.\ref{solvten} below for a first step
in this direction.

\bigskip

One more important property of the Gaussian correlators
is much simpler: it just reflects the fact that $-\frac{\mu}{2}$  is
a background value of the time variable $t_2$.
Because of this, the very special kind of averages gets factorized:
\be
{\cal O}_{[\Lambda,2^n]}
= {\mu^{N^2/2}}\! \left.\left(\frac{\partial}{\partial t_2}\right)^{\!n}\!\!
\left(\frac{1}{\mu^{N^2/2}} {\cal O}_{\Lambda}\right) \right|_{t=0} =
 {\mu^{N^2/2}}\! \left(-2\frac{\partial}{\partial \mu}\right)^{\!n}\!\!
 \left(\frac{1}{\mu^{N^2/2}}{\cal O}_{\Lambda}\right)
= \frac{1}{\mu^{ n}} {\cal O}_{\Lambda} \cdot
\prod_{i=0}^{n-1} \left(N^2+|\Lambda|+2i\right)
\label{factHe}
\ee
where we took into account the obvious fact that ${\cal O}_\Lambda \sim \mu^{-|\Lambda|/2}$.

\subsubsection{Kontsevich representation of Hermitian model}

Integrability of matrix model (see s.\ref{int}) inspires a highly non-trivial transform of the partition functions called Miwa transformation which expresses
the time-variables $t_k$ in terms of a matrix-valued background field $A$,
\be
t_k = -\frac{1}{k}\Tr A^{-k}
\ee
At the particular values of the matrix size $N$, this would be a counterpart of the {\it topological locus} in Chern-Simons/knot theory
\cite{RJ,DMMSS,MMM} describing particular distinguished slices in the space of time-variables, however, it should be considered at arbitrary large $N$.
For the Hermitian model in the Gaussian phase, this transformation was first described
in \cite{CheGHK,versus} (see also \cite{AMM3}) and looks like:
\be\label{Kont}
\left.{\cal Z}_H\right|_{t_k =- \frac{1}{k}\Tr A^{-k}}
=
{\cal Z}_{GHK}(A) \sim \int dM \det M^N \exp\left(-\frac{1}{2\mu} \Tr M^2 -i\Tr MA\right)
\ee
The Virasoro constraints are now straightforward consequences of the equations of motion,
and the cut-and-join operator generating the $W$-representation (\ref{Wrep}) is just a Laplacian
\be
\hat W = \tr \left(\frac{\partial^2}{\partial A^2}-{N\over A}\right)
\ee
in the Miwa variables.

A counterpart of the Kontsevich transform in non-Gaussian DF phases has been never
worked out.

\subsubsection{Integrability\label{int}}

The Gaussian Hermitian matrix model describes an integrable system: the partition function (\ref{HMM}) is a $\tau$-function of the (forced) Toda chain \cite{GMMOZ,versus} (see \cite{MMZ} for a discussion of integrability in non-Gaussian phases). This means that it satisfies the equation w.r.t. the size of matrix $N$,
\be
{\cal Z}_H\{t|N\}\frac{\p^2 {\cal Z}_H\{t|N\}}{\partial t_1^2}
-\left(\frac{\p {\cal Z}_H\{t|N\}}{\partial t_1^2}\right)^2
={\cal Z}_H\{t|N+1\}{\cal Z}_H\{t|N-1\}
\ee
or, in terms of the resolvent,
\be
\rho (z|N+1)+\rho (z|N-1)-2\rho (z|N)={1\over N}\p^2_z\rho (z|N)
\ee
The latter equation is easily transformed into formulas (\ref{HZ1pH})-(\ref{HZ1pHl}), \cite[Part IV]{AMM1}, see also \cite{MorShaHZ}.
Explicit solution of the Toda chain that describes this concrete matrix model
is distinguished either by the string equation \cite{GMMOZ,versus},
or by the determinant representation explicitly
\be\label{detrep}
{\cal Z}_H\{t|N\}=\det_{0\le i,j\le N} C_{i+j},\ \ \ \ \ \ \ \ C_k\equiv \int dx\ x^k \exp\left(-{\mu x^2\over 2}+\sum_k t_kx^k\right)
\ee

However, there is another possibility to relate the Gaussian Hermitian model with the Toda lattice $\tau$-function \cite{MMi}: one can note that it is equal to the concrete model from the big family of Hurwitz partition functions considered in \cite{AMMN2},
\be
Z_{(k,m)}\Big\{\mu,N_1,\ldots,N_m\,|\,t^{(i)}\Big\} =
\sum_R \mu^{-|R|} d_R^{2-k-m}  \left( \prod_{i=1}^k \chi_R\{t^{(i)}\}\right)
\left( \prod_{i=1}^m D_R(N_i)\right)
\label{Zotau}
\ee
Indeed, from (\ref{ZH}) one observes that the Gaussian Hermitian model is equivalent to the Toda lattice $\tau$-function $Z_{(2,1)}$ at some special point in the second set of time variables $\{\bar t\}$:
\be
{\cal Z}_H\{t|N\}=Z_{(2,1)}\left\{\mu,N\,\Big|\,\bar t_k={1\over 2}\delta_{k,2},\  t_k\right\}
\ee
This explains emerging the strange group character factor $\varphi_R\Big(\underbrace{[2,\ldots,2]}_{|R|/2}\Big)$ in (\ref{completesolH}) and the sum over the Young diagrams of even sizes, and can be directly obtained from the matrix model of the Kontsevich type describing $Z_{(2,1)}$, \cite[eq.(64)]{AMMN2}: one can note that this matrix model reduces to (\ref{Kont}) upon putting $\bar t_k={1\over 2}\delta_{k,2}$, the latter being equivalent to the Gaussian Hermitian model, as was explained in the previous subsection.

\subsection{Complex matrix model of \cite{CMM,AMM3}}

The complex matrix model is an integral over complex $N\times N$ matrices $M$
with the Gaussian kinetic term $\Tr MM^\dagger=\Tr M^\dagger M$.
In what follows, we often denote $M^\dagger\equiv\bar M$ to simplify formulas, i.e. $\bar M$ denotes the Hermitian, not just complex conjugation.
The kinetic term can be perturbed in two essentially different ways:
\be
\int d^2M \exp\Big(-\mu\,\Tr MM^\dagger + \Tr M^m + \Tr {M^\dagger}^m\Big)
\ee
with $m=3$ or $m=4$
or
\be
\int d^2M \exp\Big(-\mu\,\Tr MM^\dagger + \Tr (MM^\dagger)^2  \Big)
\ee
These different choices of keystone operators lead to RG-completions with
essentially different symmetries: $U(N)$ and $U(N)\otimes U(N)$.
In the latter case, "gauge" invariant are only the operators made from $\Tr (M\bar M)^k$,
while, in the former case, one can take traces of arbitrary matrix products.
We mostly consider the latter model with the extended symmetry in this paper,
though the former one is also used in some examples.

The extended partition function with $U(N)\otimes U(N)$ symmetry is defined by the integral
\be
{ {\cal Z}_C}\{{ t}\} = \int d^2M \exp\Big(-\mu\,\Tr MM^\dagger +
\sum_k {   t_k\Tr (MM^\dagger)^k}\Big)
\label{ZC}
\ee
Its significant difference from the Hermitian model is that the odd powers of $M$
can not appear in the action, and, therefore,
this time the $\hat L_{-1}$ constraint is absent:
\be
\left(-\mu\frac{\partial}{\partial { t_{n+1}}} +
\sum_k k{ t_k} \frac{\partial}{\partial { t_{k+n}}} +
\sum_{a=1}^{n-1} \frac{\partial^2}{\partial { t_a}\partial { t_{n-a}}}
+ 2N\cdot(1-\delta_{n,0})\frac{\partial}{\partial { t_{n}}}
+ N^2\cdot \delta_{n,0} \right)
{{\cal Z}_C\{t\}} = 0\ ,\ \ \ \ \ \ \ n\ge 0
\label{VirCR}
\ee
Instead, the first term with the coefficient $\mu$ contains $t_{n+1}$ rather than $t_{n+2}$,
what makes the recursive extraction of correlators well defined.
The first few examples are
(they are particular cases of  (\ref{corrZpertrN}) below with $\alpha=2N$ and $\beta=N^2$):
\be
{\cal O}_{[1]} =  {\frac{N^2}{\mu}}
\nn
\ee
\hrule
\be
{\cal O}_{[2]} =   {\frac{2N^3}{\mu^2}}
\ \ \ \ {\cal O}_{[1,1]} =  \frac{N^2(N^2+1)}{\mu^2} \nn
\ee
\hrule
\be
{\cal O}_{[3]} =   \frac{N^2(5N^2+1)}{\mu^3}
\ \ \ \ {\cal O}_{[2,1]} =   \frac{2N^3(N^2+2)}{\mu^3}
\ \ \ \ {\cal O}_{[1,1,1]} = {N^2(N^2+1)(N^2+2)\over\mu^3}
\nn
\ee
\hrule
\be
{\cal O}_{[4]} = \frac{2N^3(7N^2+5)}{\mu^4}
\ \ \ \ {\cal O}_{[3,1]} = \frac{N^2(N^2+3)(5N^2+1)}{\mu^4}
\ \ \ \ {\cal O}_{[2,2]} = \frac{N^2(2N^4+9N^2+1)}{\mu^4}
\\ {\cal O}_{[2,1,1]} ={2N^3(N^2+2)(N^2+3)\over\mu^4}  \ \ \ \
{\cal O}_{[1,1,1,1]} =  {N^2(N^2+1)(N^2+2)(N^2+3)\over\mu^4}
\nn
\ee
\hrule
\be
{\cal O}_{[5]} = \frac{2N^2( 21N^4 +35N^2+4)}{\mu^5}
\ \ \ \ {\cal O}_{[4,1]} = \frac{2N^3(N^2+4)(7N^2+5)}{\mu^5}
\ \ \ \ {\cal O}_{[3,2]} = \frac{2N^3(9N^4 +37N^2+18)}{\mu^5}\nn\\
{\cal O}_{[3,1,1]} = {N^2(5N^2+1)(N^2+3)(N^2+4)\over\mu^5}\ \ \ \
{\cal O}_{[2,2,1]} = {2N^2(11N^2+1)(N^2+4)\over\mu^5}\nn\\
{\cal O}_{[2,1,1,1]}={2N^3(N^2+2)(N^2+3)(N^2+4)\over\mu^5} \ \ \ \ {\cal O}_{[1,1,1,1,1]} = {N^2(N^2+1)(N^2+2)(N^2+3)(N^2+4)\over\mu^5}
\nn
\ee
\hrule
\be
\ldots\nn
\label{corrZpertrN}
\ee

So far the generating functions like (\ref{HZ1pH}) and (\ref{HZ1pHl})
were not  available for these correlators.
Moreover, in this respect the situation may look somewhat hopeless:
\be
\sum_N \lambda^N {\cal O}_{[1]}^{N\times N}
=  \frac{\lambda(\lambda+1)}{\mu(1-\lambda)^3}\nn \\
\sum_N \lambda^N {\cal O}_{[2]}^{N\times N}
=  \frac{2\lambda(\lambda^2+4\lambda+1)}{\mu^2(1-\lambda)^4} \nn \\
\sum_N \lambda^N {\cal O}_{[3]}^{N\times N}
=  \frac{6\lambda(\lambda+1)(\lambda^2+8\lambda+1)}{\mu^3(1-\lambda)^5}
\nn \\
\sum_N \lambda^N {\cal O}_{[4]}^{N\times N}
=   \frac{24\lambda(\lambda^4+16\lambda^2+36\lambda^2+16\lambda+1)}{\mu^4(1-\lambda)^6} \nn \\
\sum_N \lambda^N {\cal O}_{[5]}^{N\times N}
= \frac{120\lambda(\lambda+1)(\lambda^4+24\lambda^3+76\lambda^2+24\lambda+1)}{\mu^5(1-\lambda)^7} \nn \\
\ldots
\label{HZCM}
\ee
The reason for this is, however, the unjustified restriction to square matrices,
see s.(\ref{HZRCM}) below.

Instead seen from the table is factorization of the averages
for the Young diagrams with the single-line tails:
\be
\boxed{
{\cal O}_{[\Lambda,1^k]} = \frac{1}{\mu^k}\,{\cal O}_{[\Lambda]}\cdot \prod_{i=0}^{k-1} (N^2+k+i)
}
\label{factCM}
\ee
This is somewhat similar to the property of extended symmetric group characters
$\varphi$ in \cite{MMN1}.
It is a direct counterpart of (\ref{factHe}) for the complex matrix model:
a corollary of the fact that $-\mu$ is the background
value of the first time-variable, and everything is invariant under simultaneous
shift of $\mu$ and $t_1$.

\subsection{Rectangular complex matrix model
\label{RCM}}

\subsubsection{Partition function and Ward identities}

In fact, there is no need for the matrix $M$ to be square, it can
be arbitrary rectangular matrix $N_1\times N_2$, so that
square are the matrices $MM^\dagger$ and $M^\dagger M$.
There is an evident duality between $N_1$ and $N_2$ in the matrix integral
\be
{ {\cal Z}_C}\{{ t}\} = \int d^2M \exp\Big(-\mu\,\Tr MM^\dagger +
\sum_k {   t_k\Tr (MM^\dagger)^k}\Big)
\label{ZCR}
\ee
Considering the deformation
\be
\delta M =  (MM^\dagger)^n M
\ee
of the integration variable in this integral,
one deduces that the partition function
satisfies the same Virasoro constraints (with $n\geq 0$) as in the square case:
\be
\left(-\mu\frac{\partial}{\partial { t_{n+1}}} +
\sum_k k{ t_k} \frac{\partial}{\partial { t_{k+n}}} +
\sum_{a=1}^{n-1} \frac{\partial^2}{\partial { t_a}\partial { t_{n-a}}}
+ \underbrace{(N_1+N_2)}_{\alpha}\cdot\,(1-\delta_{n,0})\,\frac{\partial}{\partial { t_{n}}}
+ \underbrace{N_1N_2}_\beta\cdot \,\delta_{n,0} \right)
{{\cal Z}_C\{t\}} = 0
\ee
only the parameters $\alpha$ and $\beta$ are now independent.

\subsubsection{The simplest averages from Virasoro recursion
\label{averc}}

From the Virasoro relations one can recursively deduce the
Gaussian correlators in the rectangular model:

{\footnotesize
\be
\!\!\!\!\!\!\!\!\!\!\!\!\!\!\!\!\!\!\!\!\!\!\!\!
\begin{array}{ll }
{\cal O}_{[1]}=\Big<\Tr M\bar M \Big> =  {\frac{N_1N_2}{\mu}}  \\ \\
{\cal O}_{[2]} =\Big<\Tr (M\bar M)^2 \Big>=   {\frac{N_1N_2(N_1+N_2)}{\mu^2}}
& {\cal O}_{[1,1]}=\Big<(\Tr M\bar M)^2\Big>
=  \frac{N_1N_2(N_1N_2+1)}{\mu^2} \\ \\
{\cal O}_{[3]} =\Big<\Tr (M\bar M)^3\Big>=
\frac{N_1N_2\Big(N_1^2+3N_1N_2+N_2^2+1\Big)}{\mu^3}
& {\cal O}_{[2,1]}=\Big<\Tr (M\bar M)^2 \ \Tr M\bar M \Big>
=   \frac{N_1N_2(N_1+N_2)(N_1N_2+2)}{\mu^3} \\ \\
& {\cal O}_{[1,1,1]} = \Big<(\Tr M\bar M)^3\Big> =
\frac{N_1N_2(N_1N_2+1)(N_1N_2+2)}{\mu^3} \\ \\
{\cal O}_{[4]} = \frac{N_1N_2(N_1+N_2) \Big(N_1^2+5N_1N_2+N_2^2+5\Big)}{\mu^4}
& {\cal O}_{[3,1]} = \frac{N_1N_2(N_1N_2+3)\Big(N_1^2+3N_1N_2+N_2^2 +1\Big)}{\mu^4} \\ \\
&  { {\cal O}_{[2,2]}
= \frac{N_1N_2\Big((N_1+N_2)^2N_1N_2+4N_1^2+10N_1N_2+4N_2^2+2\Big)}{\mu^4} }\\ \\
& {\cal O}_{[2,1,1]} = \frac{N_1N_2(N_1+N_2)(N_1N_2+2)(N_1N_2+3)}{\mu^4} \\ \\
& {\cal O}_{[1,1,1,1]} = \frac{N_1N_2(N_1N_2+1)(N_1N_2+2)(N_1N_2+3)}{\mu^4}  \\ \\
{\cal O}_{[5]} = \frac{N_1N_2\Big( N_1^4+10N_1^3N_2+20N_1^2N_2^2+10N_1N_2^3
+N_2^4+15N_1^2+40N_1N_2+15N_2^2+8\Big)}{\mu^5}
& {\cal O}_{[4,1]} = \frac{N_1N_2(N_1+N_2)(N_1N_2+4)\Big(N_1^2+5N_1N_2+N_2^2+5\Big)}{\mu^5} \\ \\
&  {{\cal O}_{[3,2]} = \frac{N_1N_2(N_1+N_2)\Big((N_1^2 +3N_1N_2+N_2^2)N_1N_2
+6N_1^2+25N_1N_2+6N_2^2 +18\Big)}{\mu^5}} \\ \\
& {\cal O}_{[3,1,1]} = \frac{N_1N_2(N_1N_2+3)(N_1N_2+4)\Big(N_1^2+3N_1N_2+N_2^2+1\Big)}{\mu^5} \\ \\
& {\cal O}_{[2,2,1]} =
\frac{N_1N_2(N_1N_2+4)\Big((N_1+N_2)^2N_1N_2+4N_1^2+10N_1N_2+4N_2^2 +2\Big)}{\mu^5}\\ \\
 & {\cal O}_{[2,1,1,1]}=\frac{(N_1+N_2)N_1N_2(N_1N_2+2)(N_1N_2+3)(N_1N_2+4)}{\mu^5} \\ \\
 & {\cal O}_{[1,1,1,1,1]} = \frac{ N_1N_2(N_1N_2+1)(N_1N_2+2)(N_1N_2+3)(N_1N_2+4)}{\mu^5} \\ \\
{\cal O}_{[6]} = \frac{ N_1N_2(N_1+N_2)
\Big(N_1^4+14N_1^3N_2+36N_1^2N_2^2+14N_1N_2^3+N_2^4+35N_1^2+210N_1N_2+35N_2^2+84\Big)}{\mu^6}
\!\!\!\!\!\!\!\!\!\!\!\!\!\!\!\!\!\!   \\ \\
\ \ \ \ \ \ \ \ \ \ \ \ldots
\end{array}
\nn
\ee
}

The FT formula should now include the Fourier transforms in the both variables
$N_1$ and $N_2$
\be
{\rm FT}_{\Lambda}=\mu^{|R|}\sum_{N_1,N_2} \lambda_1^{N_1}\lambda_2^{N_2} {\cal O}_{\Lambda}^{N_1\times N_2}
\ee
and this immediately provides a simple formula,
which substitutes the ugly set (\ref{HZCM}):
\be
\boxed{
{\rm FT}_{[m]}=  m!\cdot \frac{\lambda_1\lambda_2\,(1-\lambda_1\lambda_2)^{m-1}}
{(1-\lambda_1)^{m+1}\,(1-\lambda_2)^{m+1}}
}
\label{HZRCM}
\ee
or
\be
\sum_m{\rm FT}_{[m]}
\cdot\frac{z^m}{m!} =
\frac{\lambda_1\lambda_2}{1-\lambda_1\lambda_2}\cdot
\frac{1}{(1-\lambda_1)(1-\lambda_2) - z\,(1-\lambda_1\lambda_2)}
\ee
The FT functions for other Young diagrams are a little more involved:
\be
{\rm FT}_{[1,1]} = \frac{2 \lambda_1 \lambda_2 (\lambda_1 \lambda_2+1)}
{(1-\lambda_1)^3 (1-\lambda_2)^3}
\nn\\
\nn\\
{\rm FT}_{[2,1 ]} = \frac{   6 \lambda_1 \lambda_2 (1-\lambda_1 \lambda_2 ) (\lambda_1 \lambda_2+1)}
{(1-\lambda_1)^4 (1-\lambda_2)^4)}
\nn \\
{\rm FT}_{[1,1,1 ]} = \frac{  6 \lambda_1 \lambda_2 (\lambda_1^2 \lambda_2^2+4 \lambda_1 \lambda_2+1)}
{(1-\lambda_1)^4 (1-\lambda_2)^4)}
\nn \\
\nn \\
{\rm FT}_{[3,1 ]} = \frac{  12 \lambda_1 \lambda_2 (2 \lambda_1^3 \lambda_2^3
-\lambda_1^2 \lambda_2^2-\lambda_1^2 \lambda_2-\lambda_1 \lambda_2^2-\lambda_1 \lambda_2+2)}
{(1-\lambda_1)^5 (1-\lambda_2)^5}
\nn\\
{\rm FT}_{[2,2 ]} = \frac{  24 \lambda_1 \lambda_2 (\lambda_1^3 \lambda_2^3-2 \lambda_1^2 \lambda_2^2
+\lambda_1^2 \lambda_2+\lambda_1 \lambda_2^2-2 \lambda_1 \lambda_2+1)}
{(1-\lambda_1)^5 (1-\lambda_2)^5}
\nn\\
{\rm FT}_{[2,1,1 ]} = \frac{   24 \lambda_1 \lambda_2 (1-\lambda_1 \lambda_2)
(\lambda_1^2 \lambda_2^2+4 \lambda_1 \lambda_2+1)}{(1-\lambda_1)^5 (1-\lambda_2)^5}
\nn\\
{\rm FT}_{[1,1,1,1 ]} = \frac{  24 \lambda_1 \lambda_2 (\lambda_1^3 \lambda_2^3
+10 \lambda_1^2 \lambda_2^2+\lambda_1^2 \lambda_2+\lambda_1 \lambda_2^2
+10 \lambda_1 \lambda_2+1)}{(1-\lambda_1)^5 (1-\lambda_2)^5}
\nn\\
\nn \\
{\rm FT}_{[4,1 ]} = \frac{  24 \lambda_1 \lambda_2 (1-\lambda_1 \lambda_2)
(5 \lambda_1^3 \lambda_2^3+\lambda_1^2 \lambda_2^2-6 \lambda_1^2 \lambda_2
-6 \lambda_1 \lambda_2^2+\lambda_1 \lambda_2+5)}{(1-\lambda_1)^6 (1-\lambda_2)^6}
\nn\\
{\rm FT}_{[3,2 ]} = \frac{  24 \lambda_1 \lambda_2 (1-\lambda_1 \lambda_2)
(5 \lambda_1^3 \lambda_2^3-11 \lambda_1^2 \lambda_2^2+6 \lambda_1^2 \lambda_2
+6 \lambda_1 \lambda_2^2-11 \lambda_1 \lambda_2+5)}{(1-\lambda_1)^6 (1-\lambda_2)^6}
\nn\\
{\rm FT}_{[3,1,1 ]} = \frac{  24 \lambda_1 \lambda_2
(5 \lambda_1^4 \lambda_2^4+16 \lambda_1^3 \lambda_2^3-6 \lambda_1^3 \lambda_2^2
-6 \lambda_1^2 \lambda_2^3-18 \lambda_1^2 \lambda_2^2-6 \lambda_1^2 \lambda_2
-6 \lambda_1 \lambda_2^2+16 \lambda_1 \lambda_2+5)}{(1-\lambda_1)^6 (1-\lambda_2)^6}
\nn \\
{\rm FT}_{[2,2,1 ]} = \frac{  24 \lambda_1 \lambda_2
(5 \lambda_1^4 \lambda_2^4+4 \lambda_1^3 \lambda_2^3+6 \lambda_1^3 \lambda_2^2
+6 \lambda_1^2 \lambda_2^3-42 \lambda_1^2 \lambda_2^2+6 \lambda_1^2 \lambda_2
+6 \lambda_1 \lambda_2^2+4 \lambda_1 \lambda_2+5)}{(1-\lambda_1)^6 (1-\lambda_2)^6}
\nn\\
{\rm FT}_{[2,1,1,1 ]} = \frac{  24 \lambda_1 \lambda_2
(1-\lambda_1 \lambda_2) (5 \lambda_1^3 \lambda_2^3+49 \lambda_1^2 \lambda_2^2
+6 \lambda_1^2 \lambda_2+6 \lambda_1 \lambda_2^2+49 \lambda_1 \lambda_2+5)}
{(1-\lambda_1)^6 (1-\lambda_2)^6}
\nn\\
{\rm FT}_{[1,1,1,1,1 ]} = \frac{  120 \lambda_1 \lambda_2
(\lambda_1^4 \lambda_2^4+20 \lambda_1^3 \lambda_2^3+6 \lambda_1^3 \lambda_2^2
+6 \lambda_1^2 \lambda_2^3+54 \lambda_1^2 \lambda_2^2+6 \lambda_1^2 \lambda_2
+6 \lambda_1 \lambda_2^2+20 \lambda_1 \lambda_2+1)}{(1-\lambda_1)^6 (1-\lambda_2)^6}
\label{51}
\ee
However, they satisfy elegant sum rules, the analogue of (\ref{sum1}) and (\ref{sum}):
\be
{\rm FT}_{[1]} =
\frac{\lambda_1}{(1-\lambda_1)^2}\cdot \frac{\lambda_2}{(1-\lambda_2)^2},
\nn \\
{\rm FT}_{[2]} + {\rm FT}_{[1,1]} =
\frac{2\lambda_1}{(1-\lambda_1)^3}\cdot \frac{2\lambda_2}{(1-\lambda_2)^3},
\nn \\
2\cdot{\rm FT}_{[3]} + 3\cdot{\rm FT}_{[2,1]} + {\rm FT}_{[1,1,1]}  =
\frac{6\lambda_1}{(1-\lambda_1)^4}\cdot \frac{6\lambda_2}{(1-\lambda_2)^4},
\nn \\
 6\cdot{\rm FT}_{[4]}
+  8\cdot{\rm FT}_{[3,1]}+  3\cdot{\rm FT}_{[2,2]}
+  6\cdot{\rm FT}_{[2,1,1]}+{\rm FT}_{[1,1,1,1]} =
\frac{24\lambda_1}{(1-\lambda_1)^5}\cdot \frac{24\lambda_2}{(1-\lambda_2)^5},
 \nn \\
\!\!\!\!\!\!\!\!\!\!\!\!\!\!\!\!\!\!\!\!\!\!\!
24\cdot{\rm FT}_{[5]}
+  30\cdot{\rm FT}_{[4,1]}+ 20\cdot{\rm FT}_{[3,2]}
+ 20\cdot{\rm FT}_{[3,1,1]}+ 15\cdot{\rm FT}_{[2,2,1]}
+ 10\cdot{\rm FT}_{[2,1,1,1]}+ {\rm FT}_{[1,1,1,1,1]} =
\frac{120\lambda_1}{(1-\lambda_1)^6}\cdot \frac{120\lambda_2}{(1-\lambda_2)^6}, \nn \\
\ldots
\nn
\ee
As in the case of Hermitian matrix model, one easily recognizes  in the coefficients
here the appropriately normalized
symmetric group characters $\varphi_R(\Lambda)$ from \cite{MMN1}.
Hence, one immediately obtains the general formula
\be
\sum_{\Lambda\vdash k} \varphi_{[k]}(\Lambda)\cdot{\rm FT}_{\Lambda}=\frac{k!\lambda_1}{(1-\lambda_1)^{k+1}}\cdot \frac{k!\lambda_2}{(1-\lambda_2)^{k+1}}
\ee
and, for $R=[k]$,
\be
\left.\frac{1}{d_R}\chi_R\left\{\frac{1}{n}\frac{\p}{\p t_n}\right\}\log{\cal Z}_C\right|_{t=0} =
\sum_{\Lambda\vdash k} \varphi_{[k]}(\Lambda)\cdot {\cal O}_{[\Lambda]}^{N_1\times N_2}
= {1\over \mu^k}\frac{\Gamma(N_1+k)}{\Gamma(N_1)}\frac{\Gamma(N_2+k)}{\Gamma(N_2)}
\label{surN1N2}
\ee
Moreover, the factorization persists for an arbitrary $R$ and it is especially simple
for the single-hook diagrams $R=[k,1^{l-1}]$:
\be
\sum_{\Lambda\vdash |R|} \varphi_{[R]}(\Lambda)\cdot {\rm FT}_{[\Lambda]} =
\sum_{\Lambda\vdash |R|} \varphi_{[R]}(\Lambda)\cdot
\left(\sum_{N_1,N_2} \lambda_1^{N_1}\lambda_2^{N_2}\cdot
{\cal O}_{[\Lambda]}^{N_1\times N_2}\right)
= \frac{|R|!\cdot\lambda_1^l}{(1-\lambda_1)^{|R|+1}}\cdot
\frac{|R|!\cdot\lambda_2^l}{(1-\lambda_2)^{|R|+1}}
\label{suruN1N2}
\ee
where $l$ is the number of lines in $R$.
For more complicated diagrams $R$, there are simple factors in the numerator,
e.g. for $R=[3,2]$ the factorial $5!$ gets substituted by $4!\cdot(3\lambda+2)$,
while the transposition to $R=[2,2,1]$ changes it for $4!\cdot(3+2\lambda)$.
Similarly, for $R=[2,2]$ the factorial $4!$ changes for $3!\cdot (2\lambda+2)$.
We discuss the origins and implications of these formulas elsewhere.

{\bf The complete  perturbative solution to the rectangular complex model}, i.e. an explicit formula
for arbitrary Gaussian correlator is now provided by
a somewhat simpler counterpart of (\ref{completesolH}):
\be\boxed{
{\cal O}_{\Lambda} = \frac{1}{\mu^{|\Lambda|}}\sum_{ R\,\vdash |\Lambda|}
\frac{ D_R(N_1)D_R(N_2)}{d_R}\cdot \psi_R(\Lambda)
}
\label{completesolRCM}
\ee
and similarly for the partition function
\be
\boxed{
{\cal Z}_C\{t\} = \sum_{R} \frac{1}{\mu^{|R|}}\frac{D_R(N_1)\,D_R(N_2)}{d_R} \cdot \chi_R\{t\}
}
\label{charexpC}
\ee
One can immediately associate this partition function with a partition function from the family (\ref{Zotau}):
\be\label{CZotau}
{\cal Z}_C\{t\} =Z_{(1,2)}\Big\{\mu,N_1,N_2|t_k\Big\}=Z_{(2,2)}\Big\{\mu,N\,\Big|\,\bar t_k=\delta_{k,1},\ t_k\Big\}
\ee
its complex matrix model representation found in \cite{AMMN2} being slightly different from ${\cal Z}_C\{t\}$.

\bigskip

The factorization property (\ref{factCM}) also survives, with a simple modification:
\be
\boxed{
{\cal O}_{[\Lambda,1^k]} = \frac{1}{\mu^k}\,{\cal O}_{[\Lambda]}\cdot \prod_{i=0}^{k-1} (N_1N_2+k+i)
}
\label{factRCM1}
\ee
A relative complexity of the FT formulas for the averages (\ref{51}) can be attributed
to dependence of the factor on the product of two $N_1N_2$, which can be modeled by action
of the Casimir-type operator $\lambda_1\lambda_2\frac{\p^2}{\partial \lambda_1\partial \lambda_2}$
on $FT_{[\Lambda]}$ and can not
be reduced in any way  to just a shift of variables (what could be achieved by an
adequate integral transform if it was the action of just $\lambda\frac{\p}{\p\lambda}$).

\bigskip

For $N_2=1$, i.e. for the {\it vector} model with $N_1=N$,
this factorization extends to {\it all} Gaussian correlators:
\be
{\cal O}_{[\Lambda]}^{N\times 1} =
\frac{1}{\mu^{|\Lambda|}}  \prod_{i=0}^{|\Lambda|-1} (N +i)
=  \frac{\Gamma(N+|\Lambda|)}{\mu^{|\Lambda|}\,\Gamma(N)}
\label{factvect}
\ee
in particular, the generating function (\ref{HZRCM}) of single-traced averages
in the case of vector model reduces to
\be
\sum_{N=0}^\infty {\cal O}_{[m]}^{N \times 1}\cdot {\lambda^N}
= \sum_{N=0}^\infty \frac{(N+m-1)!}{(N-1)!}\cdot \lambda^N
= \sum_{N=0}^\infty\frac{\Gamma(N+m)}{\Gamma(N)}\cdot\lambda^N
= m!\cdot\frac{ \lambda}{(1-\lambda)^{m+1}}
\ee
Moreover, the operators and thus their averages for all other
Young diagrams $\Lambda=\{m_1\geq m_2\geq\ldots \geq 0\}$
depend only on their sizes $|\Lambda|=\sum_k m_k$, thus the above answer for
the single-line diagrams $\Lambda=[\,|\Lambda|\,]=\left[\sum_k m_k\right]$ is exhaustive
in this case:
\be
{\cal O}_{[\Lambda]}^{N\times 1} = \left<
\prod_k \left(\sum_{i=1}^{N} M_i\bar M_i\right)^{m_k} \right>
= \mu^{N}  \prod_{i=1}^N \int dM_id\bar M_i e^{-\mu M_i\bar M_i}
 \left(\sum_{i=1}^{N} M_i\bar M_i\right)^{|\Lambda|}
= {\cal O}_{[\,|\Lambda|\,]}^{N\times 1}
\ee

\subsubsection{$W$-representation}

The $W$-representation for the rectangular complex model can be read off from formulas of \cite{AMMN2} upon its identification with $Z_{(1,2)}$:
\be\label{CMMW}
{\cal Z}_C\{t\} =
\exp\left\{{1\over\mu}\Big(N_1N_2t_1+(N_1+N_2)\hat L_{1} + \hat W_{1}\Big)\right\}\cdot 1
\ee
with
\be
\hat L_1 =\sum_m (m+1)t_{m+1}\frac{\p}{\p t_m},  \\
\hat W_1 =\sum_{a,b} abt_at_b \frac{\p}{\p t_{a+b-1}} + (a+b+1) t_{a+b+1}\frac{\p^2}{\p t_a\p t_b}
\ee
In variance with (\ref{Wrep})-(\ref{cjoH}), when the $W$-operator has the grading +2, these operators have the grading +1, which is related with the fact that the bare action is given by the shift of the first time, i.e. $-\mu$ is the background value of $t_1$, while, in the Gaussian Hermitian case, it is $t_2$ whose background value is equal to $-\mu/2$.

\section{On the universal structure of Virasoro-like constraints
\label{BZrem}}

In fact, in many different models the construction of
Ward identities follows one and the same line.
The principal player in the game is the special set of operators
originating from those in the bare action.
We call the non-bilinear operators in the bare action {\it keystone},
and the set of interest is built from them by various kinds of contractions
leading to {\it tree operators} and {\it loop operators}.
These are the only ones needed for the RG-completion of the theory,
and they do not necessarily include all possible operators
allowed by symmetries.
Instead, these are exactly the operators emerging in the derivation of
Ward identities along the lines of \cite{virrel}.

\subsection{Keystone operators and their RG-descendants}

Usually in theoretical physics, one begins from the study of QFT models
at some intermediate energy scale, and describes them as a collection
of certain degrees of freedom (say, moving (quasi)particles, or spins
at fixed positions, etc), which can interact with each other.
Accordingly, we write down an action consisting of kinetic terms
which are quadratic in fields, and certain interaction, which, within the context
of the present paper, we call non-quadratic {\it keystone} operators.
In the case of Hermitian matrix model, this starting action is
\be
-\frac{\mu}{2}\,\Tr M^2 + \Tr M^3
\label{HMkeystone}
\ee
The main feature of QFT is that in general such an action turns out to be
drastically changed by quantum corrections modulo a few notable exceptions,
which include the fundamental theory of nature, the Standard Model of
elementary particles, and the starting action gets "dressed" and acquires
an absolutely different form.
New interaction terms are immediately generated, and the resulting action
has many operators with the entire variety of couplings.
In the Hermitian matrix model, this corresponds to switching from (\ref{HMkeystone}) to
\be
-\frac{\mu}{2}\,\Tr M^2 + \sum_{k=1}^{\infty} t_k\Tr M^k
\ee
Usually this dressing process is described in terms of the renormalization group (RG)
flows in the moduli space of couplings (time-variables) $\{t_k\}$,
and the resulting action is the one which is {\it RG-complete}:
no more operators are needed to describe any correlator that is non-vanishing.
One of the basic problems in QFT is to find the RG-completion of the given
starting action, i.e. to identify all the RG-descendants of the given keystone
operators.
The thing is that this set can actually be smaller than all the operators
which are allowed by symmetries, this phenomenon is well known in the conventional
QFT as the existence of UV- or IR-renormalizable models.
There, however, one usually deals with theories that possess the space-time, where one
can additionally distinguish between local and non-local operators,
and often only local operators are included into the RG considerations,
at least, in the UV region.
The standard renormalizability in the UV region is then usually restricted by various types of unitarity constraints and requires the RG-completion by local operators.
In matrix models as well as in general in string theory, there is no
space time, locality does not play any special role and unitarity is present
by the construction.
Criteria for the RG-completeness are instead related to existence of rich
Ward identities, known in matrix models under the name of Virasoro/W-constrains
(because these are the algebras to which they belong, as Borel subalgebras,
in the simplest matrix models).
In general QFT, these Ward identities underlying the theory of RG flows
are representations of the peculiar algebra of rooted trees,
the corresponding construction is known as Bogoliubov-Zimmermann theory
and we are going to briefly review it in this section.
Since our main task in this paper is lifting the matrix model theory to the tensor models,
we rely upon the BZ-formalism in the presentation of \cite{GMS} (see also \cite{Riv} for an interesting related issue).
Similar considerations can be also found in \cite{GurVir,RA,ART}.

\bigskip

Though we do not go that far in this paper, the first really interesting
tensor model to analyze within this context is the {\it rainbow} model of
\cite{rainbow} (see also \cite{Gur} for earlier works).
In the rainbow model, each index of the rank-$r$ tensor field belongs
to the representation of its own unitary group and, as a consequence,
all the $r+1$ fields merging at the hyper-tetrahedron (simplex) vertex are
different. In result, there are $r+1$ different propagators,
each being a tube/cable with $r$ lines of different coloring,
and the total number of different colorings is $\frac{r(r+1)}{2}$.
For the simplest non-trivial case of $r$ this is $6$, hence, the
name "rainbow".
The keystone operators are provided by the tetrahedron vertices (since the vertex is tetrahedron in the first non-trivial (tensor) case of $r=3$, for the sake of simplicity, we always call it just tetrahedron),
which can be depicted as follows:

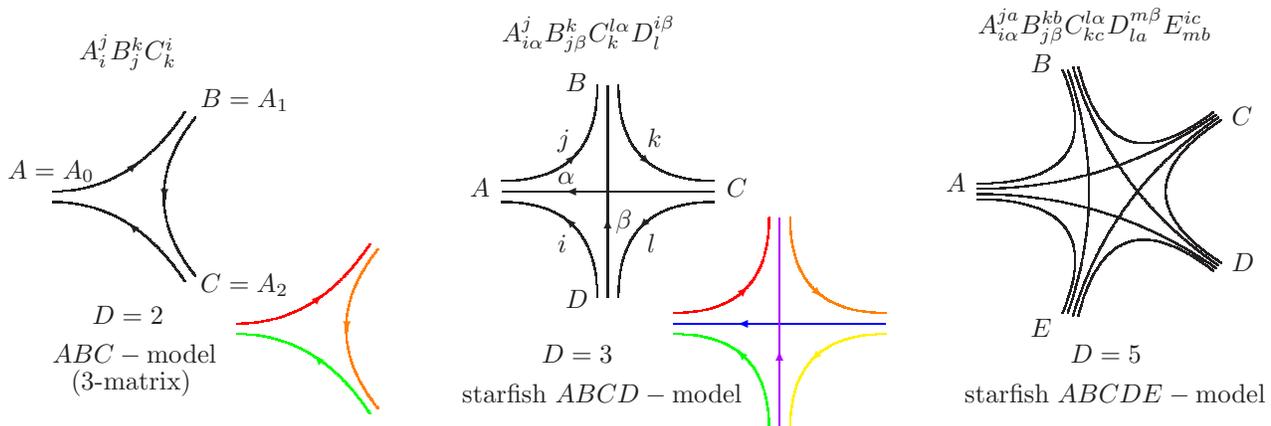
\begin{picture}(300,200)(0,-125)
\qbezier(0,0)(30,0)(50,30)\put(30,9){\vector(1,1){2}}
\qbezier(0,-4)(30,-4)(50,-34) \put(42.4,-1){\vector(0,-1){2}}
\qbezier(54,28)(30,-2)(54,-32) \put(31,-13.8){\vector(-1,1){2}}
\put(70,-50){
\put(0,0){{\color{red}\qbezier(0,0)(30,0)(50,30) \put(30,9){\vector(1,1){2}}} }
{\color{green}\qbezier(0,-4)(30,-4)(50,-34) \put(31,-13.8){\vector(-1,1){2}}}
\put(-1,0){{\co\qbezier(54,28)(30,-2)(54,-32) \put(42.4,-1){\vector(0,-1){2}} } }
}
\put(-17,5){\mbox{$A=A_0$}}
\put(56,32){\mbox{$B=A_1$}}
\put(56,-38){\mbox{$C=A_2$}}
\put(10,50){\mbox{$A_{i}^j B_{j}^k C_k^i$}}
\put(15,-50){\mbox{$D=2$}}
\put(0,-65){\mbox{$ABC-{\rm model}$}}
\put(7,-75){\mbox{(3-matrix)}}
\put(50,0){
\qbezier(120,4)(156,4)(156,40)\put(146,12){\vector(1,1){2}}
\qbezier(120,-4)(156,-4)(156,-40)\put(146,-12){\vector(-1,1){2}}
\qbezier(200,4)(164,4)(164,40)\put(174,12){\vector(1,-1){2}}
\qbezier(200,-4)(164,-4)(164,-40)\put(174,-12){\vector(-1,-1){2}}
\put(120,0){\line(1,0){80}}\put(146,0){\vector(-1,0){2}}
\put(160,40){\line(0,-1){80}}\put(160,-12){\vector(0,1){2}}
\put(65,-50){
\put(0,0){{\cre\qbezier(120,4)(156,4)(156,40)\put(146,12){\vector(1,1){2}}}}
{\cg\qbezier(120,-4)(156,-4)(156,-40)\put(146,-12){\vector(-1,1){2}}}
\put(0,0){{\co\qbezier(200,4)(164,4)(164,40)\put(174,12){\vector(1,-1){2}}}}
{\cy\qbezier(200,-4)(164,-4)(164,-40)\put(174,-12){\vector(-1,-1){2}}}
\put(0,0){\cb\put(120,0){\line(1,0){80}}\put(146,0){\vector(-1,0){2}}}
\put(0,0){\cv\put(160,40){\line(0,-1){80}}\put(160,-12){\vector(0,1){2}}}
}
\put(108,-2){\mbox{$A$}}
\put(144,38){\mbox{$B$}}
\put(205,-2){\mbox{$C$}}
\put(144,-44){\mbox{$D$}}
\put(120,57){\mbox{$A_{i\alpha}^j B_{j\beta}^k C_k^{l\alpha} D_l^{i\beta}$}}
\put(135,-65){\mbox{$D=3$}}
\put(105,-80){\mbox{starfish $ABCD-{\rm model}$}}
\put(141,17){\mbox{$j$}} \put(175,17){\mbox{$k$}}
\put(141,-23){\mbox{$i$}} \put(175,-23){\mbox{$l$}}
\put(141,3){\mbox{$\alpha$}} \put(163,-14){\mbox{$\beta$}}
}
\put(400,0){
%

\qbezier(-50,3)(0,3)(-18,46)
\qbezier(-12,47)(0,0)(39,30)
\qbezier(42,27)(0,0)(42,-27)
\qbezier(-12,-47)(0,0)(39,-30)
\qbezier(-50,-3)(0,-3)(-18,-46)
\qbezier(-50,1)(0,1)(40,29)
\qbezier(-50,-1)(0,-1)(40,-29)
\qbezier(-16,46)(0,0)(-16,-46)
\qbezier(-14,47)(0,0)(41,-28)
\qbezier(-14,-47)(0,0)(41,28)
\put(-50,60){\mbox{$A_{i\alpha}^{ja} B_{j\beta}^{kb} C_{kc}^{l\alpha} D_{la}^{m\beta} E_{mb}^{ic}$}}
\put(-15,-65){\mbox{$D=5$}}
\put(-55,-80){\mbox{starfish $ABCDE-{\rm model}$}}
\put(-62,0){\mbox{$A$}}
\put(-30,45){\mbox{$B$}}
\put(46,25){\mbox{$C$}}
\put(46,-30){\mbox{$D$}}
\put(-30,-55){\mbox{$E$}}
 %


%
}
\put(0,-100){\mbox{
\footnotesize{In general, the indices here belong to different groups
(tensors are "rectangular"): \
 $i=1,\ldots,N_{{\rm green}}$, $j=1,\ldots,N_{{\rm red}}$,}}}
\put(0,-110){\mbox{
\footnotesize{
 $k=1,\ldots,N_{{\rm orange}}$, $l=1,\ldots,N_{{\rm yellow}}$, $\alpha=1,\ldots, N_{{\rm blue}}$,
$\beta=1,\ldots,N_{{\rm violet}}$, $a=1,\ldots, N_{{\rm brown}}$, $b=1,\ldots, N_{{\rm pink}}$,
$\ldots$
}}}

\end{picture}

\noindent
The first task in the study of this model is to build the RG-descendants of these
keystone operators and describe this emerging set in some efficient way.
The first step on this way i.e. in constructing the tree and loop operators
from the keystones is, in fact, universal, while the relation of the loop and
tree operators is model-dependent and its investigation is still a piece of art.
In this section, we describe the universal part of the story,
while in the following two we use much simpler tensor models to illustrate
a possibility of artistic steps.
Lifting these considerations to the rainbow models themselves remains for the future.

As to the Ward identities, they can be formulated at two different levels.
The easy and universal step to be actually described below
is constructing {\it recursion relations} between particular {\it Gaussian} averages,
which can allow one to build them one after another.
Usually this recursion is just in the power (the number) of fields in the operator.
A more artistic step is to collect these recursions into equations in terms
of generating functions.
As we saw in section 2, this can be actually done in different ways,
useful for different purposes.
What is important, at the level of generating functions, one can actually
move away from the Gaussian point and consider other phases.
Once equations for the generating functions are known,
this non-perturbative treatment is provided just
by a shift of the time-variables $t\longrightarrow T+t$.
As soon as such a description of the rainbow models is worked out
(not in the present paper, yet), one is able to treat the
tetrahedron vertices non-perturbatively, as lifting of the theory of Dijkgraaf-Vafa
phases from the matrix to rainbow tensor models.

\subsection{Tree operators as the {\it base} of RG-complete set}

We now remind the first steps of the RG-completion of the given keystone interaction.
They are absolutely universal and applicable to any QFT model.
We will be illustrating this general construction by
two examples, relevant for the purpose of this paper:
the rectangular complex and rainbow $ABCD$ models.

\bigskip

1) Specify integration variables (fields) and the kinetic term
(Gaussian weight), e.g.
\be\label{31}
\int d^2M\ e^{-\mu\Tr MM^\dagger}, \ \ \  {\rm or} \ \ \
\int d^2\!A d^2B d^2C d^2D\  e^{-\mu \Tr A\bar A + \Tr B\bar B + \Tr C\bar C + \Tr D\bar D}
\ee

\bigskip

2) Select a {\it keystone} operator or a pair of these, e.g.
\be
{\cal K} = \Tr\Big(MM^\dagger\Big)^2
\ee
or, in the square-matrix case,
\be\label{var2}
\Big({\cal K} = \Tr M^4\Big) \oplus \Big({\cal K} = \Tr \bar M^4\Big)
\ee
for the first matrix model in (\ref{31}), and, for the second model, in (\ref{31}),
\be
\Big({\cal K} = [ABCD]\Big) \oplus \Big(\bar{\cal K} = [\bar D\bar C\bar B\bar A]\Big)
\ee
which we depict as fat points (black $\oplus$ white) with four (in these examples)
thick ($r$-fat) external lines.
The tin lines will be used to describe the {\it internal} structure of propagators
and vertices: in matrix models, the thick lines are called {\it fat} and made from a pair
of think lines.
For rank-$r$ tensors, the thick lines are tubes/cables containing $r$ thin lines,
which, for the rainbow models, are all of different colors.
Moreover, the cables can contain different (but not arbitrary) combinations of $r$ colors
and thus are themselves {\it multicolored}.
The fat points, where different cables merge, can have a complicated internal structure
describing reshuffling of the thin lines between the cables, and they can be very
different.
To handle this variety, we agree to denote by the thick points only the keystone vertices,
while all other types of cable mergers will be induced from them, actually, by the
Feynman diagrams.

Indeed, the thick points and lines are the ones describing vertices and propagators
in the ordinary Feynman diagrams for the keystone interaction.
In fact, these Feynman diagrams generate new operators.
In conventional QFT, we do not pay too much attention to this, because these
new operators are usually non-local, and only some of them contain essentially local
contributions (like, say, the tadpoles or the UV-divergent diagrams).
However, in theories where one does not care about the space-time and locality,
like in the case of matrix models, all operators arising from the Feynman diagrams
are relevant.

\bigskip

3) Construct new {\it connected}
operators from these by connecting some of the thick lines,
i.e. by applying the operations
\be\label{oper}
\Tr \frac{\partial}{\partial M^\dagger} \otimes \frac{\partial}{\partial M}
\ \ \ \ \ {\rm or} \ \ \ \
 \Tr \frac{\partial}{\partial \bar A} \otimes \frac{\partial}{\partial A}
 +  \Tr \frac{\partial}{\partial \bar B} \otimes \frac{\partial}{\partial B}
 + \Tr \frac{\partial}{\partial \bar C} \otimes \frac{\partial}{\partial C}
 + \Tr \frac{\partial}{\partial \bar D} \otimes \frac{\partial}{\partial D}
 \ee

Let us consider the second keystone operator (\ref{var2}). If applied once to a pair of points, the operation (\ref{oper}) provides an operator with six external legs $\Tr M^3\bar M^3$. In our notations of s.\ref{mmp}, this is depicted as

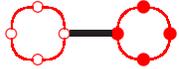
\begin{picture}(300,90)(-100,-55)
\put(-5,-40){\mbox{$\Tr M^3\bar M^3$}}


\put(0,0){
\put(0,0){\cre
\qbezier(-10,2)(-10,10)(-2,10)\qbezier(10,2)(10,10)(2,10)
\qbezier(-10,-2)(-10,-10)(-2,-10)\qbezier(10,-2)(10,-10)(2,-10)
\put(-10,0){\circle{4}}\put(10,0){\circle{4}}
\put(0,-10){\circle{4}}\put(0,10){\circle{4}}
}
\put(40,0){\cre
\qbezier(-10,0)(-10,10)(0,10)\qbezier(10,0)(10,10)(0,10)
\qbezier(-10,0)(-10,-10)(0,-10)\qbezier(10,0)(10,-10)(0,-10)
\put(-10,0){\circle*{5}}\put(10,0){\circle*{5}}
\put(0,-10){\circle*{5}}\put(0,10){\circle*{5}}
}
}

\linethickness{0.8mm}
\put(12,0){\line(1,0){16}}

\end{picture}

\noindent
However, for illustrative purposes, in this paragraph we temporarily return
to the standard Feynman graph notation,
though it will be used to enumerate the local operators.
Then, the Feynman diagrams with six ($\Tr M^3\bar M^3$ drawn above)
and eight external legs (if the operation is applied twice to a set of six points)
look like

\begin{picture}(300,90)(-100,-55)
\put(0,0){\circle{6}}\put(30,0){\circle*{6}}
\put(-5,-40){\mbox{$\Tr M^3\bar M^3$}}
\linethickness{0.8mm}
\put(0,0){\line(1,0){30}}
\put(0,0){\line(-1,0){15}}\put(0,0){\line(0,1){15}}\put(0,0){\line(0,-1){15}}
\put(30,0){\line(1,0){15}}\put(30,0){\line(0,1){15}}\put(30,0){\line(0,-1){15}}
\put(110,0){
\put(0,0){\circle{6}}\put(30,0){\circle*{6}}\put(60,0){\circle{6}}
\put(-5,-40){\mbox{$\Tr M^3\bar M M^3\bar M$}}
\linethickness{0.8mm}
\put(0,0){\line(1,0){60}}
\put(0,0){\line(-1,0){15}}\put(0,0){\line(0,1){15}}\put(0,0){\line(0,-1){15}}
\put(30,0){\line(0,1){15}}\put(30,0){\line(0,-1){15}}
\put(60,0){\line(1,0){15}}\put(60,0){\line(0,1){15}}\put(60,0){\line(0,-1){15}}
}
\put(250,-10){
\put(0,0){\circle{6}}\put(30,0){\circle*{6}}\put(30,30){\circle{6}}
\put(-5,-40){\mbox{$\Tr M^6\bar M^2$}}
\linethickness{0.8mm}
\put(0,0){\line(1,0){30}}
\put(0,0){\line(-1,0){15}}\put(0,0){\line(0,1){15}}\put(0,0){\line(0,-1){15}}
\put(30,0){\line(1,0){15}}\put(30,0){\line(0,1){30}}\put(30,0){\line(0,-1){15}}
\put(30,30){\line(-1,0){15}}\put(30,30){\line(0,1){15}}\put(30,30){\line(1,0){15}}
}
\end{picture}

\noindent
and so on.
In this picture, we show an example of the square matrix model,
for the rectangular case there are no chiral operators for the role
of keystone ones, only ${\cal K}=\Tr (MM^\dagger)^2$,
thus, all vertices will be the same (no black and white), and the
the emerging operators will be just two instead of three:
$\Tr (MM^\dagger)^3$ and $\Tr (MM^\dagger)^4$,
according to the higher symmetry of the model.
For the rainbow model, the pictures will remain the same, but the internal
structure of emerging operators (contraction of indices) will be a little
more involved, and can be easily depicted in terms of the thin-line diagrams.
The total number of emerging operators will not actually
increase too much, because the growth of the number of fields ($A,B,C,D$ instead
of a single $M$) will be compensated by the increased symmetry:
the $r$-colorings of propagator tubes/cables will not be arbitrary and
there will be at most four options per each thick line,
with additional constraints that all the thick lines in each tetrahedron vertex
are different.

One can also apply the same operation twice to just two points, giving rise
to operators with four external legs:

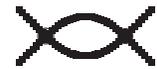
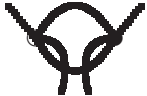
\begin{picture}(300,90)(-100,-55)
\put(0,0){\circle{6}}\put(30,0){\circle*{6}}
\put(-15,-40){\mbox{$N\cdot \Tr M^2\bar M^2$}}
\linethickness{0.8mm}
\qbezier(0,0)(15,15)(30,0) \qbezier(0,0)(15,-15)(30,0)
\qbezier(0,0)(-5,5)(-10,10)\qbezier(0,0)(-5,-5)(-10,-10)
\qbezier(30,0)(35,5)(40,10)\qbezier(30,0)(35,-5)(40,-10)
\put(150,0){
\put(0,0){\circle{6}}\put(30,0){\circle*{6}}
\put(-10,-40){\mbox{$\Big(\Tr M\bar M\Big)^2$}}
\linethickness{0.8mm}
\qbezier(0,0)(15,25)(30,0) \qbezier(0,0)(15,-25)(30,0)
\qbezier(0,0)(-5,6)(-10,12)\qbezier(0,0)(15,0)(10,-20)
\qbezier(30,0)(35,6)(40,12)\qbezier(30,0)(15,0)(20,-20)
}
\end{picture}

\noindent
but these will be "loop" rather than "tree" operators.
If we look at these operators in the thin-line representation,
then, for the matrix models, it gets clear that all tree operators
are just the ordinary single-trace operators, while
the loop operators are either single- or multi-trace operators:

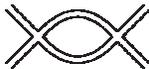
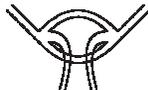
\begin{picture}(300,90)(-100,-55)
\put(-15,-40){\mbox{$N\times \Tr M^2\bar M^2$}}
\qbezier(0,2)(15,18)(30,2) \qbezier(0,-2)(15,-18)(30,-2)
\qbezier(2,0)(15,15)(28,0) \qbezier(2,0)(15,-15)(28,0)
\qbezier(-2,0)(-7,5)(-12,10)\qbezier(-2,0)(-7,-5)(-12,-10)
\qbezier(32,0)(37,5)(42,10)\qbezier(32,0)(37,-5)(42,-10)
\qbezier(0,2)(-5,7)(-10,12)\qbezier(0,-2)(-5,-7)(-10,-12)
\qbezier(30,2)(35,7)(40,12)\qbezier(30,-2)(35,-7)(40,-12)
\put(150,0){
\put(-10,-40){\mbox{$\Big(\Tr M\bar M\Big)^2$}}
\qbezier(0,2)(15,18)(30,2) \qbezier(0,-2)(15,-18)(30,-2)
\qbezier(4,2)(15,13)(26,2) \qbezier(4,-2)(15,-13)(26,-2)
\qbezier(0,-2)(-7,5)(-12,10)
\qbezier(30,-2)(37,5)(42,10)
\qbezier(0,2)(-5,7)(-10,12)
\qbezier(30,2)(35,7)(40,12)
\qbezier(4,2)(16,2)(10,-20)\qbezier(26,2)(14,2)(20,-20)
\qbezier(4,-2)(13,-2)(8,-20)\qbezier(26,-2)(17,-2)(22,-20)
}
\end{picture}

\noindent
However, in the tensor case, there is no better
term than {\it tree} and {\it loop} operators for the substitutes of
the matrix model single- and multi-traces.

Note that, in the case of matrix models, all planar diagrams actually give rise
to the single-trace operators (times traces of unity, which are just powers of $N$),
while the true multi-trace operators emerge only from the non-planar diagrams.
The number of traces is related to the degree of non-planarity (that is, to the
genus of the surface obtained by putting all external lines together in a cyclic order).
The counterpart of this feature for the tensor models depends on the choice
of keystone operators and also plays a role in structure of the
extended partition functions and the Ward identities.

\bigskip

4) Define the extended partition function by putting all {\it tree} operators
in the action:
\be
{\cal Z}(t) = d^2{\cal M} \exp\Big(-[{\cal M}\bar{\cal M}] + t{\cal K} + \bar t\bar {\cal K}
+ \sum_{trees} t_{tree} {\cal K}_{tree}\Big)
\label{genZtree}
\ee
${\cal M}$ is a symbolic unifying notation for the dynamical field.
The keystone operators can be considered as associated with the simplest tree
consisting of one vertex (black or white, in the chiral case).

\bigskip

5) Virasoro like constraints reflect the invariance of extended partition function
under the changes of integration variable, generated by a gradient of any operator in
the action,
\be
\delta_{tree} {\cal M} = \frac{\partial}{\partial \bar {\cal M}} {\cal K}_{tree}
\label{treenytreetransform}
\ee
(the exact correspondence between the indices in ${\cal M}$ and $\bar{\cal M}$
is dictated by the kinetic term).
In other words, as any Ward identities, they are essentially
averages of the equations of motion.
This transformation changes any term in the extended action by
\be
t_{tree'}\frac{\partial}{\partial \bar {\cal M}} {\cal K}_{tree}
\frac{\partial}{\partial \bar {M}} {\cal K}_{tree'}
\ee
which is by definition again a tree operator,
this produces a term like
\be
\sum_{T'}t_T'\frac{\partial}{\partial t_{T\circ T'}}{\cal Z}
\ee
in the Virasoro constraints with a clear notion of tree composition $T\circ T'$
(the tree $T$ is attached by some of its vertices to the tree $T'$ at some of its vertices,
and all possible choices are summed up).

Actually the trees are {\it rooted} and it is also convenient to consider
variation, where ${\cal M}$-gradient is taken w.r.t. the fields in the root vertex
only, then the composition operation $\circ$ of the rooted trees becomes even simpler:
one attaches a root of one tree to any vertex of another.

\bigskip

6) However, the Jacobian of the transformation (\ref{treenytreetransform})
contains loop operators, i.e. is no longer expanded in the trees.
Moreover, there is no reason for it to be expressed via any number
of derivatives w.r.t. the tree time-variables $t_{tree}$,
as it happens in the matrix models.
In other words, the extended partition function (\ref{genZtree})
looks to be not RG-complete.

From this point, we have two obvious ways to proceed:
introduce more terms into the extended action to make it RG complete
or to look for a factorization of loop operator averages at large $N$
and to get a closed set of constraints, at least, in this limit
(i.e. to construct a counterpart of the spectral curve with the hope to
build further a counterpart of the AMM/EO topological recursion over it).

\subsection{The simplest recursions}

If we begin with the action $-\mu[{\cal M}\bar{\cal M}]+t_\circ {\cal K} +
t_\bullet \bar{\cal K}$, with ${\cal K}$ being of the forth power in ${\cal M}$,
 and consider its variation $\delta\bar{\cal M} = d_M{\cal K}$,
then we get
\be
-\mu [Md_M{\cal K}] + t_\bullet d_{\bar M}\bar{\cal K} d_M{\cal K} =
-4\mu {\cal K} + 16 t_\bullet {\cal K}_{1,\bar 1}
\ee
or, pictorially (with combinatorial factors omitted),

\begin{picture}(300,30)(-150,-15)

\qbezier(-25,-10)(-33,0)(-25,10)
\qbezier(117,-10)(125,0)(117,10)
\put(125,-2){\mbox{$\cdot \ \
{\Large:e^{t_{\bullet} \ \circ} \, \cdot  \ e^{t_\circ \, \bullet}: } \ \ \cong \ \ 0$}}
\put(0,0){\circle{6}}
\put(-10,0){
\put(50,0){\circle*{6}}\put(73,0){\circle{6}}
\linethickness{0.8mm} \put(50,0){\line(1,0){20}}
}
\put(-20,-2){\mbox{$-\mu$}}
\put(10,-2){\mbox{$+ \ \ t_\bullet$}}
\put(75,-2){\mbox{$+$}}
\put(90,0){\circle{6}}
\put(100,0){\circle{19}}
\put(100,0){\circle{21}}

\end{picture}

\noindent
The last term comes from the Jacobian $d_M\delta \bar M = d_Md_M{\cal K}$.
The Ward identity says that the Gaussian {\it average} of this sum should vanish
(the fact that vanishing takes place only after averaging is expressed by the sign
$\cong$ instead of equality).
The Gaussian averages of the two chiral operators in the sum are zero, but
one should take into account the contributions proportional to $t_\bullet$ and
coming from the exponentials.
This is similar to the usual story: in
$$
-\mu \frac{\partial}{\partial t_n} + \sum_k kt_k\frac{\partial}{\partial t_{k+n}}
$$
one should either put $t$ to zero and stay with just
$$
-\mu \frac{\partial}{\partial t_n}
$$
or first differentiate over $t_k$ to stay with
$$
-\mu \frac{\partial^2}{\partial t_k\partial t_n} + k\frac{\partial}{\partial t_{k+n}}
$$
In result, we get in the first non-vanishing order in $t$

\begin{picture}(300,30)(-150,-15)

\put(0,0){\circle{6}}\put(-10,0){\circle*{6}}
\put(-10,0){
\put(50,0){\circle*{6}}\put(73,0){\circle{6}}
\linethickness{0.8mm} \put(50,0){\line(1,0){20}}
}
\put(-25,-2){\mbox{$\mu$}}
\put(14,-2){\mbox{$\cong  $}}
\put(75,-2){\mbox{$+$}}
\put(92,0){\circle*{6}}
\put(12,0){
\put(90,0){\circle{6}}
\put(100,0){\circle{19}}
\put(100,0){\circle{21}}
}

\end{picture}

\noindent
In application to the particular model,
one should also insert combinatorial factors
and put the normal ordering around the operator $\bullet$.
Taking all this into account, together with vanishing of the
Gaussian averages of chiral operators, which explains the
elimination of disconnected averages, one recognizes a trivial
identity.
Not quite trivial is only the matching of combinatorics
at all orders in $\mu^{-1}$.

Similarly one can {\it draw} a generic tree Virasoro constraint,
with one tree attached to all vertices of another, in the above example
each of the two trees consists of a single vertex.

\subsection{BZ exponential and rooted trees }

As already mentioned in \cite{GMS} and \cite{rainbow},
one of the ways to construct generating functions of trees
is provided by the Bogoliubov-Zimmermann forest formula
\be
e^{\hat V}
= 1+\sum_{{\rm forests}\ {\cal F}} \frac{1}{{\rm Tree}({\cal F})!}
\prod_{{\rm trees}\ {\cal T}\in{\cal F}} \frac{\hat V_{{\cal T}}}{\sigma_{{\cal T}}{\cal T}!}
\label{BZf}
\ee
for the expansion of the exponentiated vector field
\be
\hat V = \sum_{\alpha=1}^P v^\alpha\partial_\alpha
\ee

\noindent
Notation in this expansion is best explained pictorially:
\be
e^{\hat V} = \ \sum_{n=0}^\infty\ \frac{\hat V^n}{n!}\ = \
1 \  + \  v^\alpha\p_\alpha + \frac{1}{2}v^\gamma \p_\gamma v^\alpha\p_\alpha
+ \frac{1}{6}v^\gamma \p_\gamma v^\beta\p_\beta v^\alpha\p_\alpha \ + \ \ldots\  = \nn \\
\ee

\bigskip

\begin{picture}(300,80)(-140,-20)
\put(-75,-2){\circle{6}}
\put(-45,0){
\put(0,-2){\circle{6}}
\put(0,20){\vector(0,-1){20}}
\put(0,20){\circle*{4}}
}
\put(-10,0){
\put(0,-2){\circle{6}}
\put(0,20){\vector(0,-1){20}}
\put(0,20){\circle*{4}}
\put(0,40){\vector(0,-1){20}}
\put(0,40){\circle*{4}}
}
\put(60,0){
\put(0,-2){\circle{6}}
\put(0,20){\vector(0,-1){20}}
\put(0,20){\circle*{4}}
\put(0,40){\vector(0,-1){20}}
\put(0,40){\circle*{4}}
\put(0,60){\vector(0,-1){20}}
\put(0,60){\circle*{4}}
}
\put(160,0){
\put(0,-2){\circle{6}}
\put(0,20){\vector(0,-1){20}}
\put(0,20){\circle*{4}}
\put(20,40){\vector(-1,-1){20}}
\put(20,40){\circle*{4}}
\put(-20,40){\vector(1,-1){20}}
\put(-20,40){\circle*{4}}
}
\end{picture}
\be
= 1 +\boxed{\left(v^\alpha+\frac{1}{2}v^\gamma(\p_\gamma v^\alpha) +
\frac{1}{6}v^\gamma(\p_\gamma v^\beta)(\p_\beta v^\alpha) +
\frac{1}{6}v^\beta v^\gamma (\p_\beta\p_\gamma v^\alpha) + \ldots\right)\p_\alpha} \ + \nn
\ee

\begin{picture}(300,100)(-155,-20)
\put(-75,0){
\put(0,-2){\circle{6}}
\put(0,20){\vector(0,-1){20}}
\put(0,20){\circle*{4}}
}
\put(-60,0){
\put(0,-2){\circle{6}}
\put(0,20){\vector(0,-1){20}}
\put(0,20){\circle*{4}}
}
\put(0,0){
\put(0,-2){\circle{6}}
\put(0,20){\vector(0,-1){20}}
\put(0,20){\circle*{4}}
\put(0,40){\vector(0,-1){20}}
\put(0,40){\circle*{4}}
}
\put(15,0){
\put(0,-2){\circle{6}}
\put(0,20){\vector(0,-1){20}}
\put(0,20){\circle*{4}}
}
\put(60,0){
\put(0,-2){\circle{6}}
\put(0,20){\vector(0,-1){20}}
\put(0,20){\circle*{4}}
}
\put(75,0){
\put(0,-2){\circle{6}}
\put(0,20){\vector(0,-1){20}}
\put(0,20){\circle*{4}}
\put(0,40){\vector(0,-1){20}}
\put(0,40){\circle*{4}}
}
\put(140,0){
\put(0,-2){\circle{6}}
\put(0,20){\vector(0,-1){20}}
\put(0,20){\circle*{4}}
\put(0,40){\vector(0,-1){20}}
\put(0,40){\circle*{4}}
}
\put(155,0){
\put(0,-2){\circle{6}}
\put(0,20){\vector(0,-1){20}}
\put(0,20){\circle*{4}}
\put(0,40){\vector(0,-1){20}}
\put(0,40){\circle*{4}}
}
\end{picture}
\be
+ \frac{1}{2}\ \cdot\overbrace{\left(v^\alpha+\frac{1}{2}v^\gamma(\p_\gamma v^\alpha) + \ldots\right)
\left(v^\beta+\frac{1}{2}v^\gamma(\p_\gamma v^\beta) + \ldots\right)\p_\alpha\p_\beta}^{
v^\alpha v^\beta \p_\alpha\p_\beta
+ \frac{1}{2}v^\gamma(\p_\gamma v^\alpha)v^\beta \p_\alpha\p_\beta
+\frac{1}{2} v^\alpha v^\gamma(\p_\gamma v^\beta) \p_\alpha\p_\beta
+ \frac{1}{4}v^\gamma(\p_\gamma v^\alpha)v^{\gamma'}(\p_{\gamma'} v^\alpha)\p_\alpha\p_\beta
+ \ldots
}
 + \nn
\ee
\begin{picture}(300,80)(-215,-20)
\put(-15,0){
\put(0,-2){\circle{6}}
\put(0,20){\vector(0,-1){20}}
\put(0,20){\circle*{4}}
}
\put(0,0){
\put(0,-2){\circle{6}}
\put(0,20){\vector(0,-1){20}}
\put(0,20){\circle*{4}}
}
\put(15,0){
\put(0,-2){\circle{6}}
\put(0,20){\vector(0,-1){20}}
\put(0,20){\circle*{4}}
}
\end{picture}
\be
+ \frac{1}{6}\overbrace{\Big(v^\alpha+ \ldots\Big)\Big(v^\beta + \ldots\Big)
\Big(v^\gamma + \ldots\Big) \p_\alpha\p_\beta\p_\gamma}^{
v^\alpha v^\beta v^\gamma \p_\alpha\p_\beta\p_\gamma + \ldots
} + \ldots = \nn
\ee

\be
= 1+\sum_{ {\cal F}} \frac{1}{{\rm Tree}({\cal F})!}
\prod_{ {\cal T}\in{\cal F}} \frac{\hat V_{{\cal T}}}{\sigma_{{\cal T}}{\cal T}!}
\ = \ : \exp\left(\
\boxed{\sum_{\cal T}\frac{\hat V_{{\cal T}}}{\sigma_{{\cal T}}{\cal T}!}}\
\right):
\label{BZfexpanded}
\ee

\noindent
In other words, at any vertex (except for the root) we put a vector field $\hat V$,
which acts on $v$ at exactly the next vertex towards the root.
The emerging combinatorial factors in the sums are of two kinds: the trivial ones,
associated with the forests (inverse factorials of the number of trees ${\rm Tree}({\cal F})$
in the forest), i.e. coefficients of the
Maclaurin expansion of the exponential, and the less trivial ones,
associated with the trees: they are described by recursively defined Connes-Moscovici
factorials \cite{CoMo}
\be
 {\cal T}! =  {\rm Vert}({\cal T})\cdot \prod_\tau {\cal T}_\tau!
\label{CoMoexp}
\ee
where ${\rm Vert}({\cal T})$ is the number of vertices in ${\cal T}$,
while the product goes over all sub-trees $\tau\subset{\cal T}$, in which ${\cal T}$
decays if the root (the bottom arrow) is cut away.
$\sigma_{\cal T}$ is just the tree symmetry factor
(in the above pictures, it is different from unity
only for the last tree in the first line, in this case, it is equal to 2).
The (non-trivial) fact that the forest dependence of combinatorial factors is so simple
allows one to rewrite the exponential of the vector field $\hat V$ as a {\it normal ordered}
exponential of another vector field (while one could expect that it would be some non-trivial
poly-vector). The normal ordering means that all the differential operators are put to the
right of all the coefficient functions,
\be
:e^{ \hat V}: \ =
 \sum_{n=0}^\infty
:v^{\alpha_1}\p_{\alpha_1}\ v^{\alpha_2}\p_{\alpha_2}\ \ldots\  v^{\alpha_n} \p_{\alpha_n} :
\ \equiv
\sum_{n=0}^\infty
v^{\alpha_1}v^{\alpha_2}\ldots v^{\alpha_n}\ \p_{\alpha_1}\p_{\alpha_2}\ldots \p_{\alpha_n}
\ee
Then this new vector field is just the sum over all trees in the box
in the second line of (\ref{BZfexpanded}): this is the statement of the last equation in this
formula.

\subsection{The Bogoliubov-Zimmermann tensor model}

The sum in (\ref{BZf}) goes over the rooted graphs with vertices of arbitrary valence.
In particular tensor models, one needs to restrict it to a particular valence, say $p+1$,
or, more precisely, $(p,1)$.
Each such vertex has one exiting link and $p$ incoming ones, and contributes
a factor of $\ v^{\otimes p}\left(\partial^{\otimes p} v\right)$.
Thus, contributions with a given $p$ will be the only surviving ones, if
$v$ are polynomials of exactly power $p$, then the answer for each tree
will consist of the product of $v(x)$'s at the end-points of the graph
times an $x$-independent number obtained by contraction of indices at all vertices.
Any vertex of higher valence will automatically drop out, the
vertices with lower valence will contain extra powers of $x$
and can be eliminated by putting $x=0$ in the expression for the graph with amputated
external vertices.

In other words, if
the vector field $\hat V$ should be a (rank-$p$ tensor-valued) vector field,
\be
\hat V\{x\} = v^J_{I_1\ldots I_p} x^{I_1}\ldots x^{I_p}\frac{\partial}{\partial x^J}
= \frac{\p v(x,\bar x)}{\p \bar x_J}\frac{\partial}{\partial x^J}
\ee
with $\bar x$-linear $v(x,\bar x) = v^J_{I_1\ldots I_p} x^{I_1}\ldots x^{I_p}\bar x_J$,
and one  considers its exponential, $e^{\hat V}$
applied to some function of $x$, say,
$e^{x}$, then
\be
{\cal Z}\{x\} = e^{{\cal F}\{x\}} = e^{\hat V\{x\}} e^x
\ee
will be expanded in graphs with valences $r+1$ or less.
It looks very much like a $W$-representation of the tree
(quasiclassical) approximation to a peculiar tensor model
\be
{Z}_{BZ} = \int d{\cal M}d\bar{\cal M}
\exp\left(-\mu \sum_I {\cal M}^I\bar{\cal M}_I +
t\underbrace{\sum_{I_1,\ldots,I_p,J}
v^J_{I_1\dots I_r}{\cal M}^{I_1}\ldots{\cal M}^{I_p}\bar{\cal M}_J}_
{v^J({\cal M})\, \bar{\cal M}_J}
\right)
\label{ZBZ}
\ee
which we naturally name {\it BZ model}.

In the context of the usual tensor models like the rainbow one,
the indices $I$ and $J$ play the role of multi-indices,
labeling the multicolored tubes/cables and the coefficients $v$ encode
their coupling via the keystone (e.g. tetrahedron) vertex,
in the rainbow model occasionally $p=r$.
Of course, all the indices can have different multi-coloring with one being distinguished,
associated with the contravariant index $J$, this is similar to
the colored (red) partition function in (\ref{Zpertr}).
The extended partition function contains all the tree operators,
which, in this case, all contain a single field ${\cal M}$ and arbitrarily large number
of ${\cal M}$:
\be
{\cal Z}_{BZ}\{t_{trees}\} =
 \int d{\cal M}d\bar{\cal M}
\exp\left(-\mu\cdot \sum_I {\cal M}^I\bar{\cal M}_I + \sum_{trees \ {\cal T}}
t_{\cal T} \cdot \underbrace{   T_{\cal T}({\cal M},\bar{\cal M})}_
{[{\cal M}^{\otimes \nu({\cal T})},\bar{\cal M}]_{\cal T}}\right)
\label{ZBZext}
\ee
Here $T_{\cal T}({\cal M},\bar{\cal M}) =
[{\cal M}^{\otimes \nu({\cal T})},\bar{\cal M}]_{\cal T} =
T^I_{\cal T}({\cal M})\cdot\bar{\cal M}_I$
is a polynomial in ${\cal M}$ of degree $\nu_{\cal T} = 1+ (p-1)\cdot {\rm Vert}({\cal T}$
and a linear function in $\bar{\cal M}$ with the coefficients from $v^{\otimes {\rm Vert}({\cal T})}$ and the conversion of indices
dictated by the tree ${\cal T}$.

Each $\bar{\cal M}$-linear function placed inside the Gaussian integral acts
on the ${\cal M}$-dependent objects as a vector field,
$V^J({\cal M})\bar{\cal M}_J \cong  \hat V \equiv v^J({\cal M})\frac{\p}{\p{\cal M}^J}$,
this is basically the meaning in which (\ref{BZfexpanded}) and (\ref{ZBZ}) are the same
(up to the factors $t$ and $\mu$).
In (\ref{ZBZext}) the vector field is a sum over all possible rooted trees,
taken with arbitrary coupling constants (time-variables) $t_{\cal T}$.

Composition of the vector fields induces
an associative non-commutative algebra structure on the "group algebra" of the rooted trees,
with multiplication
\be
\hat T_{{\cal T}_1} \circ \hat T_{{\cal T}_2} =
\left(T_1^I({\cal M})\,\frac{\partial T_2^J({\cal M})}{\partial {\cal M}^I}\right)
\frac{\partial}{\partial {\cal M}_J} =
T_{1\circ 2}({\cal M})^J \frac{\partial}{\partial {\cal M}_J} =
\hat T_{{{\cal T}_1}\circ {{\cal T}_2}}
\ee
which looks like just attaching a tree ${\cal T}_1$ to a vertex of ${\cal T}_2$,
summed over all vertices.
Of course, when the valence is restricted by $p$, the attachment is possible only
to the vertices which have free valencies.
This operation will play a crucial role in the structure of the universal
Ward identities in s.\ref{arcVir}.

\subsection{Archetypical/universal Virasoro constraint
\label{arcVir}}

The Ward identities in the BZ model form an archetypical set of the Virasoro constraints,
which is then inherited this or that way by all other QFT models.

There are two kinds of transformations generated by the "white" functions
(i.e. with no free/external indices):
$\bar{\cal M}$-independent $Q({\cal M})$ and $\bar{\cal M}$-linear
$S({\cal M},\bar{\cal M}) = S^J({\cal M})\bar{\cal M_J}$.
The two kinds of identities reflect the invariance with respect to the shifts
\be
\delta \bar{\cal M}_I = \frac{\partial Q}{\partial {\cal M}^I} \ \ \ \longrightarrow \ \ \
\mu \left<\left< \underbrace{{\cal M}^I  \frac{\partial Q}{\partial {\cal M}^I}}_
{{\rm deg}_Q \cdot Q({\cal M})}\right>\right> \ = \
\sum_{{\cal T}} t_{\cal T}\cdot\left<\left<
\underbrace{\frac{\partial T_{\cal T}}{\partial \bar{\cal M}_I}
\frac{\partial Q}{\partial {\cal M}^I}}_{\hat T_{\cal T} Q({\cal M})}\right>\right>
\label{arcVir0}
\ee
and
\be
\delta{\cal M}^I = \frac{\partial S}{\partial \bar{\cal M}_I} \ \ \ \longrightarrow \ \ \
\mu \left<\left< \underbrace{\bar{\cal M}_I  \frac{\partial S}{\partial \bar{\cal M}^I}}_{\hat S}
\right>\right> \ = \
\sum_{{\cal T}} t_{\cal T}\cdot
\left<\left<\underbrace{\frac{\partial S}{\partial \bar{\cal M}^I}
\frac{\partial T_{\cal T}}{\partial  {\cal M}_I}}_{\hat S \hat T_{\cal T}}\right>\right> +
\left<\left<\frac{\partial^2 S}{\partial \bar{\cal M}_I \partial {\cal M}^I}\right>\right>
\label{arcVir1}
\ee
The averages are in the model (\ref{ZBZext}), i.e. not Gaussian; moreover, they
should be understood as taken in the background ${\cal M}$-field
(or with an additional insertion of the source term like
$\exp\left({\cal J}^I\bar{\cal M}_I\right)$),
otherwise all averages are just zero.

An essential difference between (\ref{arcVir0}) and (\ref{arcVir1}) is that the
former does not contain a Jacobian contribution, while the latter does.
If we restrict to these two types of transformations,
at least the one in (\ref{arcVir1}),
then the $t$-linear terms
contain just the compositions of trees, i.e. are trees again, and can be
expressed as derivatives w.r.t. the variables $t_{T\circ Q}$ and $t_{S\circ T}$.
Jacobian contribution in (\ref{arcVir1}) can not, but instead it can be
treated within the context of (\ref{arcVir0}).

\subsection{Relation to Feynman diagrams}

At vanishing $t$, the Ward identities (\ref{arcVir0}) and (\ref{arcVir1}) and their
$t$-derivatives provide concrete recursion relations between particular
{\it Gaussian} correlators.
We can instead calculate these Gaussian correlators directly
by the Wick theorem.
Comparing these two types of calculations, one can note that the Ward identities
contain only action of the vector field $\hat S$ on the vector field $\hat T_{\cal T}$,
while the Wick theorem calculation {\it also} contains contributions with
$\hat T_{\cal T}$ acting on $\hat S$.

\begin{picture}(300,100)(-70,-50)
\put(-40,0){\line(1,3){6}}\put(-40,0){\line(1,-3){6}}
\put(-37,0){\line(1,3){6}}\put(-37,0){\line(1,-3){6}}
\put(37,0){\line(-1,3){6}}\put(37,0){\line(-1,-3){6}}
\put(40,0){\line(-1,3){6}}\put(40,0){\line(-1,-3){6}}
\put(-15,15){\vector(1,-1){10}}\put(-5,12){\mbox{$\ldots$}}\put(15,15){\vector(-1,-1){10}}
\put(0,0){\circle{12}}\put(-3.5,-2.5){\mbox{$S$}}
\put(0,-6){\vector(0,-1){10}}
\put(130,0){
\put(-40,0){\line(1,3){6}}\put(-40,0){\line(1,-3){6}}
\put(-37,0){\line(1,3){6}}\put(-37,0){\line(1,-3){6}}
\put(37,0){\line(-1,3){6}}\put(37,0){\line(-1,-3){6}}
\put(40,0){\line(-1,3){6}}\put(40,0){\line(-1,-3){6}}
\put(-15,15){\vector(1,-1){10}}\put(-5,12){\mbox{$\ldots$}}\put(15,15){\vector(-1,-1){10}}
\put(0,0){\circle{12}}\put(-3.5,-2.5){\mbox{$S$}}
\put(0,-6){\vector(0,-1){10}}
\qbezier(0,-16)(0,-26)(15,-26)
\qbezier(15,-26)(30,-26)(30,3)
\qbezier(15,15)(30,25)(30,3)
}
\put(62,-2){\mbox{$=$}}
\put(-57,-2){\mbox{$\mu\ \cdot$}}
\put(300,20){
\put(-50,-15){\line(1,3){10}}\put(-50,-15){\line(1,-3){10}}
\put(-47,-15){\line(1,3){10}}\put(-47,-15){\line(1,-3){10}}
\put(47,-15){\line(-1,3){10}}\put(47,-15){\line(-1,-3){10}}
\put(50,-15){\line(-1,3){10}}\put(50,-15){\line(-1,-3){10}}
\put(-15,15){\vector(1,-1){10}}\put(-5,12){\mbox{$\ldots$}}\put(15,15){\vector(-1,-1){10}}
\put(0,0){\circle{12}}\put(-3.5,-2.5){\mbox{$S$}}
\put(0,-6){\vector(0,-1){10}}
\put(0,-24){\circle{16}}\put(-5.5,-26){\mbox{$T_{\cal T}$}}
\put(-20,-6){\vector(1,-1){12}}\put(20,-6){\vector(-1,-1){12}}
\put(0,-32){\vector(0,-1){10}}
}
\put(188,-2){\mbox{$+ \ \ \sum_{\cal T} \ t_{\cal T}\ \cdot $}}
\end{picture}

\noindent
The resolution of this "paradox" is that the Wick theorem application actually provides
a combination of {\it two} Ward identities.

\subsection{BZ resolvents}

Of course, expansions like (\ref{BZf}) exist for all other functions of
vector fields, not obligatory exponentials, the only difference is
in combinatorial coefficients.
The tree dependent coefficient is provided by a recursive analogue
of the Connes-Moscovici (recursive Maclaurin) formula (\ref{CoMoexp}).
In particular, one can define a BZ-resolvent
as a Laplace transform of the BZ exponential (\ref{BZf}),
\be
\frac{1}{z-\hat V} = \int_0^\infty e^{-zt} e^{z\hat V}
\ee

\subsection{The message}

Generalization from the vector to exponentials of poly-vector fields, i.e. to actions
non-linear in $\bar{\cal M}$, do not possess any enhanced reparametrization symmetry:
closed algebra is formed only for the scalar and vector transforms.
Indeed, the three types of terms in the Virasoro constraints have $\bar{\cal M}$-powers
\be
\bar n \ \oplus \ (\bar n + \bar k -1) \oplus (\bar n-1)
\ee
and potentially closed are only the two cases: either $\bar k=1$ or $\bar n=1$.
A possible way out is to make an infinite tower of powers $\bar k$, tying them to the
powers of ${\cal M}$, as it is done in the complex matrix model,
where the operators $\Tr(MM^\dagger)^k$ have $\bar k = k$.
This, however, leads to problems with the Jacobians.

A hope can be that the loop operators arising from the Jacobians are {\it algebraically}
dependent on the tree operators, like it happens in the one-matrix models.
This can happen in particular models, specifically, in the rainbow model.
However, this option deserves further investigation.

An alternative remark is that the action of trees is defined on {\it all} poly-vectors,
and the Jacobians are needed only from the {\it acting} tree, thus they are always the same.
In case of poly-vectors, this does not provide the necessary recursion (because no
poly-vectors arise in this way from the vector fields), still this provides a set of
relations describing the extended poly-vector generating functions as {\it representations}
of the tree algebra.

What definitely exists are {\it recursions} like (\ref{basicrecursion})
between the Gaussian correlators, which allow one to evaluate {\it all} averages
recursively in the power of fields: first all correlators with two fields, then
with four, then with six, and so on.
These recursions are obtained from the generic Ward identities when they are expanded
in powers of time-variables around the Gaussian point.
Such evaluation of the Gaussian correlators is the necessary stage of development
in the theory of tensor models, which can hardly be avoided,
and we presented some examples in this paper.
Lifting to the true Ward identities is important for non-perturbative calculations,
i.e. for the study of expansions around non-Gaussian points,
and for development of related more sophisticated techniques:
character expansions, integrability (KP/Toda and Hurwitz),
quasiclassical integrability, spectral curves, AMM/EO topological recursion,
$W$ and Kontsevich representations etc.
This is also a long work for the future.

\section{RG-closed tensor generalization of the complex matrix model
\label{redmod}}

\subsection{Partition function}

Substitute now the rectangular matrix $M_i^j$ by a tensor $A_i^{j_1,\ldots,j_{r-1}}$
of rank $r$
with one covariant and $r-1$ contravariant indices.
Adding a conjugate tensor $\bar A^{i}_{j_1,\ldots,j_{r-1}}$, one can make a kinetic term
and consider the following model:
\be
Z_{TC} = \int d^2A \exp\left(\sum_{i,j_1,\ldots,j_{r-1}}
A_i^{j_1 \ldots j_{r-1}} \bar A^{i}_{j_1\ldots j_{r-1}}\right)
\label{ZTC1}
\ee
where the measure is induced by the norm
$||\delta A||^2 = \delta A_i^{j_1 \ldots j_{r-1}} \delta \bar A^{i}_{j_1\ldots j_{r-1}}$.
Each index can be rotated by its own unitary group of its own size,
so that the model has the symmetry $\otimes_{a=1}^r U(N_a)$.
In fact, the symmetry in (\ref{ZTC1}) is much higher:
the integral is just the same as for the rectangular {\it matrix} model
of size  $N_1\times \Big(N_2\cdot\ldots\cdot N_r\Big)$, with the symmetry
$U(N_1)\otimes U(N_2\cdot\ldots\cdot N_r)$.
What distinguishes the tensor model from such an {\it enveloping} matrix model is
the choice of allowed operators.
If they have lower (tensor-model) symmetry than that of the matrix model,
their correlators are {\it not} among the matrix model ones and should
be calculated from the dedicated Ward identities, which need to be
separately derived.
For this, it is important to know the what we call
"RG (renormalization-group)-closed" sets of operators,
for which the Ward identities are self-sufficient and self-consistent,
at least, in the certain large-$N$ limits.
In this section, we provide some primary examples of such considerations.
For the sake of simplicity, we draw the pictures and write the formulas for the model with $r=3$, which we call
RGB (red-green-blue) or Aristotelian (since Aristotle distinguished 3 colors in the rainbow \cite{Arist2}, \cite[p.107]{Arist}, only Newton raised the number to the canonical seven).
In most cases, generalization to arbitrary $r$  is obvious
just like the further resolution of colors in the spectrum.

\subsection{Notation: two types of diagrams}

If one thinks about the tensor models, the main problem is to find a workable
description of indices and their contractions.
Algebraically, there are no notions like matrix product and trace and even a small number
of tensors can be contracted in many different ways.
Drawing pictures can help, but this interferes with already existing
technique of the Feynman diagrams in QFT.
In fact, this problem already exists with matrices, but there a simple way
out was invented:
the Feynman propagators in Yang-Mills theory are depicted as double lines,
and gauge invariance requires the thin lines to be
trivially rearranged (cyclically connected) at the vertices.
For rank $r$ tensor fields, the Feynman propagators are
thick  lines, tubes or cables consisting of $r$ thin lines.
The real problem are interaction vertices, where these thin lines
can be interconnected in many complicated ways.
Thus, there is a separate task of drawing the vertices, i.e. of
drawing the gauge-invariant local operators,
and most of pictures in this and other tensor model papers are trying
to depict them.
Things are greatly simplified in the rainbow models of \cite{rainbow},
where thin lines have as many different colorings as only possible,
and this both decreases the number of invariant operators
and simplifies pictures for them.

Coming back to the simplest possible Gaussian rainbow model (\ref{ZTC1}),
we use it to introduce the convenient notation,
which allow one to separately treat the drawings (diagrams) for the
local operators and for the Feynman diagrams for their correlators and interactions.

The vertices ("local" operators) are represented by "thin" diagrams,
where the vertices are fields (tensors) and they connected by thin colored lines,
which describe the contraction of indices.

The Feynman diagrams ("thick diagrams") describe averages (correlators of "local" operators):
they are also diagrams where "local" operators
shrink to thick points of different kinds (with different internal structures)
and different external valencies associated with the fields, which were the vertices
in the "thin" diagrams.
The Feynman propagators are depicted as thick ($r$-colored) lines (tubes/cables).

This {\it double-level diagram technique}, where the Feynman vertices and propagators
have their own non-trivial internal structure,
is getting more and more important in modern theory:
for example, something very similar appears under the name of double-fat diagrams
in the effective theory of arborescent knots in \cite{arbor}.


\subsubsection*{Kinetic term and Feynman propagator}

As an operator, $\Tr A\bar A$ in (\ref{ZTC1}) can be depicted by three thin lines
of different colors, connecting two vertices $A$ and $\bar A$,
which we will usually depict as a circle of "unit length".
Directions of arrows depend on the choice of covariant and contravariant indices,
which is not essential for the models in this paper.
However, we choose them in accord with the tetrahedron model, despite it
is beyond the scope of the present text.

\begin{picture}(300,150)(-250,-120)

\put(-30,0){
\put(-200,-2){\mbox{$\Tr A\bar A \ \ \ \ \ =
A_{i_r}^{j_g \alpha_b}
\bar A^{i_r}_{j_g\alpha_b}  =
A_{{\cre i_r}}^{{\cg j_g}{\cb \alpha_b}}
\bar A^{{\cre i_r}}_{{\cg j_g}{\cb \alpha_b}} =$}}
\put(0,3){{\cre \vector(1,0){70}}}
\put(70,0){{\cb \vector(-1,0){70}}}
\put(70,-3){{\cg \vector(-1,0){70}}}
\put(-15,-2){\mbox{$A$}}
\put(77,-2){\mbox{$\bar A$}}
\put(95,-2){\mbox{$=$}}
\put(230,0){\circle*{5}}

\put(130,0){
\put(0,0){\cre\qbezier(-15,0)(-15,15)(0,15)\qbezier(15,0)(15,15)(0,15)\put(-1,15){\vector(1,0){2}}}
\put(0,0){\cg\qbezier(-15,0)(-15,-15)(0,-15)\qbezier(15,0)(15,-15)(0,-15)
\put(1,-15){\vector(-1,0){2}}}
\put(0,0){\cb\qbezier(-15,0)(0,-20)(15,0)\put(1,-10){\vector(-1,0){2}}}
}
\put(155,-2){\mbox{$=$}}
\put(185,0){
\put(0,0){\cre\qbezier(-15,0)(-15,15)(0,15)\qbezier(15,0)(15,15)(0,15)\put(-1,15){\vector(1,0){2}}}
\put(0,0){\cg\qbezier(-15,0)(-15,-15)(0,-15)\qbezier(15,0)(15,-15)(0,-15)
\put(1,-15){\vector(-1,0){2}}}
\put(0,0){\cb\qbezier(-15,0)(0,20)(15,0)\put(1,10){\vector(-1,0){2}}}
}
\put(210,-2){\mbox{$=$}}
}
\put(-215,-40){
\put(50,-2){\mbox{$\Big< A \ \bar A\ \Big>$}}
\put(103,-2){\mbox{$=$}}
\put(125,-2){\mbox{$A$}}
\put(217,-2){\mbox{$\bar A$}}
\put(190,0){\line(-2,1){10}}\put(190,0){\line(-2,-1){10}}
\linethickness{0.8mm}
\put(140,0){\line(1,0){70}}
\put(250,-2){\mbox{$=\ \ \ \delta_{{\cre i}}^{{\cre i'}}\cdot
\delta_{{\cg j'}}^{{\cg j}}\cdot \delta_{{\cb \alpha'}}^{{\cb\alpha}} $}}
}

\put(-65,-90){
\put(0,0){\cre\qbezier(-15,0)(-15,15)(0,15)\qbezier(15,0)(15,15)(0,15)\put(-1,15){\vector(1,0){2}}}
\put(0,0){\cg\qbezier(-15,0)(-15,-15)(0,-15)\qbezier(15,0)(15,-15)(0,-15)
\put(1,-15){\vector(-1,0){2}}}
\put(0,0){\cb\qbezier(-15,0)(0,-20)(15,0)\put(1,-10){\vector(-1,0){2}}}
\linethickness{0.8mm}
\qbezier(-15,0)(-25,-0)(-25,10)\qbezier(-25,10)(-25,20)(0,20)
\qbezier(15,0)(25,-0)(25,10)\qbezier(25,10)(25,20)(0,20)
\put(35,-2){\mbox{$=$}}
}

\put(10,-90){
\put(0,0){\cre\qbezier(-15,0)(-15,15)(0,15)\qbezier(15,0)(15,15)(0,15)\put(-1,15){\vector(1,0){2}}}
\put(0,0){\cg\qbezier(-15,0)(-15,-15)(0,-15)\qbezier(15,0)(15,-15)(0,-15)
\put(1,-15){\vector(-1,0){2}}}
\put(0,0){\cb\qbezier(-15,0)(0,-20)(15,0)\put(1,-10){\vector(-1,0){2}}}
\linethickness{0.8mm}
\put(-15,0){\line(1,0){30}}
}

\put(-320,-90){
\put(60,-2){\mbox{$\Big< \Tr A\bar A\ \Big>$}}
\put(113,-2){\mbox{$=$}}
\linethickness{0.8mm}
\put(160,0){\qbezier(-15,0)(-15,15)(0,15) \qbezier(-15,0)(-15,-15)(0,-15)
\qbezier(15,0)(15,15)(0,15) \qbezier(15,0)(15,-15)(0,-15)}
\put(160,-15){\circle*{5}}
\put(200,-2){\mbox{$=$}}
}

\put(-160,-90){
\put(207,-2){\mbox{$=$}}
\put(260,0){\cre \circle{30}} \put(260,0){\cb \circle{26}} \put(260,0){\cg \circle{22}}
\put(300,-2){\mbox{$= \ \ \ {\cre N_r}\cdot{\cb N_b}\cdot{\cg N_g}$}}
}

\end{picture}

From the point of view of Feynman diagrams, this operator is just a vertex of valence $2$,
to be denoted by a fat point with just two vacancies for possible attachment of the
Feynman propagators.
The propagator depicted by the thick black line is itself defined by the same kinetic
term $\Tr A\bar A$ or, if one prefers, as a correlator of $A$ and $\bar A$.
In this sense, the thick black line is a tube or a cable consisting of three
thing colored lines.
The average $\Big<\Tr A\bar A\Big>$ is a closed circle made from the Feynman propagator.

\subsubsection*{"Single-trace" non-chiral operators}

The matrix model single-trace operators $\Tr (AA^\dagger)^k = \Tr (A\bar A)^k$
are now substituted by
$K_k={\cre K_k}$ and $\tilde K_k={\cg K_k}$

\begin{picture}(300,110)(-100,-60)

{\cre
\qbezier(0,0)(10,17)(20,34)\put(10,17){\vector(1,2){2}}
\qbezier(80,0)(70,17)(60,34)\put(70,17){\vector(1,-2){2}}
\qbezier(20,-34)(40,-34)(60,-34)\put(40,-34){\vector(-1,0){2}}
}
\put(0,0){{\cg
\qbezier(0,0)(10,-17)(20,-34)\put(10,-17){\vector(-1,2){2}}
\qbezier(80,0)(70,-17)(60,-34)\put(70,-17){\vector(-1,-2){2}}
\qbezier(20,34)(40,34)(60,34)\put(40,34){\vector(1,0){2}}
}}
{\cb
\qbezier(0,0)(25,-10)(20,-34)\put(17,-13){\vector(-1,2){2}}
\qbezier(80,0)(55,-10)(60,-34)\put(63,-13){\vector(-1,-2){2}}
\qbezier(20,34)(40,16)(60,34)\put(40,25){\vector(1,0){2}}
}
\put(22,-55){\mbox{$K_3={\cre K_3}$}}
\put(-12,-3){\mbox{$A$}}
\put(8,32){\mbox{$\bar A$}}
\put(65,32){\mbox{$A$}}
\put(85,-3){\mbox{$\bar A$}}
\put(65,-40){\mbox{$A$}}
\put(8,-40){\mbox{$\bar A$}}

\put(150,0){
{\cre
\qbezier(0,0)(10,17)(20,34)\put(10,17){\vector(1,2){2}}
\qbezier(80,0)(70,17)(60,34)\put(70,17){\vector(1,-2){2}}
\qbezier(20,-34)(40,-34)(60,-34)\put(40,-34){\vector(-1,0){2}}
}
\put(0,0){{\cg
\qbezier(0,0)(10,-17)(20,-34)\put(10,-17){\vector(-1,2){2}}
\qbezier(80,0)(70,-17)(60,-34)\put(70,-17){\vector(-1,-2){2}}
\qbezier(20,34)(40,34)(60,34)\put(40,34){\vector(1,0){2}}
}}
{\cb
\qbezier(0,0)(25,10)(20,34)\put(17,13){\vector(-1,-2){2}}
\qbezier(80,0)(55,10)(60,34)\put(63,13){\vector(-1,2){2}}
\qbezier(20,-34)(40,-16)(60,-34)\put(40,-25){\vector(1,0){2}}
}
\put(22,-55){\mbox{$\tilde K_3={\cg K_3}$}}
\put(-12,-3){\mbox{$A$}}
\put(8,32){\mbox{$\bar A$}}
\put(65,32){\mbox{$A$}}
\put(85,-3){\mbox{$\bar A$}}
\put(65,-40){\mbox{$A$}}
\put(8,-40){\mbox{$\bar A$}}
}

\end{picture}

\noindent
which we often call respectively red and green circles
(or {\it benzene rings}) of length $k$ referring to the color and number
of single-line sides.
The circle of unit length can be considered as either "red" or "green",
${\cal K}_{\cre 1} = {\cal K}_{\cg 1}$.
In fact, there are also a blue cousins ${\cal K}_{\cb m}$ of the green operators,
but we begin from just red, then add green, and blue then automatically emerges.

\subsection{Ordinary Virasoro constraints for an oversimplified tensor model}

The simplest possibility for an extension of the partition function (\ref{ZTC1})
is to include only the operators $K_k$, i.e.
only the red circles, this makes one of the colors
distinguished:
\be
{\cre {\cal Z}}\{{\cre t}\} = \int d^2\!A \exp\Big(-\mu\,\Tr A\bar A + \sum_k { \cre t_kK_k}\Big)
\label{Zpertr}
\ee
and we call this "red" tensor model.
Considering a deformation
\be
\delta A = \frac{\partial K_{n+1}}{\partial \bar A} = {\bf \nabla}(K_{n+1})
\ee
of the integration variable in this integral,
we deduce that ${\cre {\cal Z}}$
satisfies the nearly conventional Virasoro constraints (with $n\geq 0$):
\be
\left(-\mu\frac{\partial}{\partial {\cre t_{n+1}}} +
\sum_k k{\cre t_k} \frac{\partial}{\partial {\cre t_{k+n}}} +
\sum_{a=1}^{n-1} \frac{\partial^2}{\partial {\cre t_a}\partial {\cre t_{n-a}}}
+ \underbrace{({\cre N_r} + {\cg N_g}{\cb N_b})}_{{\cre \alpha}}
(1-\delta_{n,0})\frac{\partial}{\partial {\cre t_{n}}}
+ \underbrace{{\cre N_r}{\cg N_g}{\cb N_b}}_\beta\cdot \delta_{n,0} \right)
{\cre Z\{t\}} = 0
\label{Virpertr}
\ee
or
\be
\left(-\mu+\frac{{\cre \alpha}}{z}\right){\cre \rho} + {\cre\rho}^2 + \frac{\beta}{z^2}
+ \nabla_z {\cre\rho} = 0
\ee
for the resolvent
\be
{\cre \rho}(z) = \nabla_z \log {\cre Z} = \frac{1}{{\cre Z}}\sum_{n=1}^\infty \frac{1}{z^{n+1}}
\frac{\partial}{\partial {\cre t}_n}\log {\cre Z}
\ee
at ${\cre t}=0$.

The Virasoro constrains (\ref{Virpertr})
provide a rigorous identification of the minimally-extended partition function
(\ref{Zpertr}) with that of rectangular matrix model:
they differ only in {\it interpretation} of the parameters $\alpha$ and $\beta$.
As usual, we consider this identification for the partition functions
analytically continued in $N$.

\subsection{Spectral curve as the leading term of the genus expansion}

Neglecting $\nabla_z {\cre\rho}$, we get the spectral curve
\be
{\cre\rho}_0 = \frac{1}{2}\left(\mu-\frac{{\cre\alpha}}{z} - \sqrt{\left(\mu-\frac{{\cre\alpha}}{z}\right)^2
- \frac{4\beta}{z^2}}\right) = \nn \\
= \frac{\beta}{\mu z^2} + \frac{{\cre\alpha}\beta}{\mu^2z^3} +
\frac{({\cre\alpha}^2+\beta)\beta}{\mu^3 z^4} + \frac{({\cre\alpha}^2+3\beta){\cre\alpha}\beta}{\mu^4z^5}
+  \frac{({\cre\alpha}^4+6{\cre\alpha}^2\beta+2\beta^2)\beta}{\mu^5 z^6}
+ \ldots
\label{rho0r}
\ee
or
\be
y^2 =  \left(\mu-\frac{\alpha}{z}\right)^2
- \frac{4\beta}{z^2}
\ee
which describes the single-trace averages in the limit of large $\alpha$ and $\beta$.

\subsection{Examples of averages
\label{exavRCM}}

The simplest correlators are recursively deduced from (\ref{Virpertr}),
and they are basically the same as in s.\ref{averc}, if expressed through $\alpha$ and $\beta$:

\be
\begin{array}{c|cccc}
{\cal O}_{[1]}=\Big<{\cal K}_{\cre 1}\Big> = \underline{\frac{\beta}{\mu}}  \\ \\
{\cal O}_{[2]} =\Big<{\cal K}_{\cre 2}\Big>=  \underline{\frac{{\cre \alpha}\beta}{\mu^2}}
& {\cal O}_{[1,1]}=\Big<{\cal K}_{\cre 1}{\cal K}_{\cre 1}\Big>
=  \frac{\beta(\beta+1)}{\mu^2} \\ \\
{\cal O}_{[3]} =\Big<{\cal K}_{\cre 3}\Big>=
\frac{\beta(\underline{{\cre \alpha}^2+\beta}+1)}{\mu^3}
& {\cal O}_{[2,1]}=\Big<{\cal K}_{\cre 2}{\cal K}_{\cre 1}\Big>
=   \frac{{\cre \alpha}\beta(\beta+2)}{\mu^3}
& {\cal O}_{[1,1,1]} = \Big<{\cal K}_{\cre 1}{\cal K}_{\cre 1}{\cal K}_{\cre 1}\Big> =
\frac{\beta(\beta+1)(\beta+2)}{\mu^3} \\ \\
{\cal O}_{[4]} = \frac{{\cre \alpha}\beta\Big(\underline{{\cre \alpha}^2+3\beta}+5\Big)}{\mu^4}
& {\cal O}_{[3,1]} = \frac{\beta(\beta+3)({\cre \alpha}^2+\beta+1)}{\mu^4}
&\boxed{ {\cal O}_{[2,2]}
= \frac{\beta({\cre \alpha}^2\beta+4{\cre \alpha}^2+2\beta+2)}{\mu^4} }\\ \\
& {\cal O}_{[2,1,1]} = \frac{{\cre \alpha}\beta(\beta+2)(\beta+3)}{\mu^4}
& {\cal O}_{[1,1,1,1]} = \frac{\beta(\beta+1)(\beta+2)(\beta+3)}{\mu^4}  \\ \\
{\cal O}_{[5]} = \frac{\beta\Big(\underline{{\cre \alpha}^4+6{\cre \alpha}^2\beta+2\beta^2}
+15{\cre \alpha}^2+10\beta+8\Big)}{\mu^5}
& {\cal O}_{[4,1]} = \frac{{\cre \alpha}\beta(\beta+4)({\cre \alpha}^2+3\beta+5)}{\mu^5}
& \boxed{{\cal O}_{[3,2]} = \frac{{\cre \alpha}\beta
({\cre \alpha}^2\beta+6{\cre \alpha}^2+\beta^2+13\beta+18)}{\mu^5}} \\ \\
& {\cal O}_{[3,1,1]} = \frac{({\cre \alpha}^2+\beta+1)\beta(\beta+3)(\beta+4)}{\mu^5}
& {\cal O}_{[2,2,1]} =
\frac{\beta(\beta+4)({\cre \alpha}^2\beta+4{\cre \alpha}^2+2\beta+2)}{\mu^5}\\ \\
 & {\cal O}_{[2,1,1,1]}=\frac{{\cre \alpha}\beta(\beta+2)(\beta+3)(\beta+4)}{\mu^5}
 & {\cal O}_{[1,1,1,1,1]} = \frac{ \beta(\beta+1)(\beta+2)(\beta+3)(\beta+4)}{\mu^5} \\ \\
{\cal O}_{[6]} = \frac{{\cre\alpha}\beta\Big(
\underline{{\cre\alpha}^4+10{\cre\alpha}^2\beta+10\beta^2} +
35{\cre\alpha}^2+70\beta+84\Big)}{\mu^6}     & \\ \\
\ldots \\
\end{array}
\label{corrZpertr}
\ee
As usual, every step of recursion produces an extra power of $\mu^{-1}$.
The underlined terms in the single-trace averages
(for the single-line Young diagrams in the first column)
are described by the spectral curve formula (\ref{rho0r}).
From these formulas, it is clear what the genus zero approximation
actually means in this case:
one picks up the highest possible powers
in ${\alpha}$ and $\beta$, irrespective of actual relation between $\alpha$ and $\beta$.
Of course, other interesting large-$N$ limits are also possible in this case.

\bigskip

These correlators satisfy an analogue of the sum rules (\ref{suruN1N2}), e.g.
\be
{\cal O}_{[2]}+{\cal O}_{[1,1]} = \frac{({\cre\alpha}+\beta+1)\beta}{\mu^2}
= \frac{{\cre N_r}({\cre N_r}+1){\cg N_g}{\cb N_b}({\cg N_g}{\cb N_b}+1)}{\mu^2}
\nn \\
\boxed{
\left.\frac{1}{d_{[k]}{\cre{\cal Z}}}\chi_{[k]}\left\{\frac{\p}{\p {\cre t_n}}\right\}{\cre{\cal Z}}\right|_{t=0} =
\sum_{\Lambda\vdash k} \varphi_{[k]}(\Lambda)\cdot {\cal O}_{[\Lambda]}^{N_1\times N_2}
= \frac{\Gamma({\cre N_r}+k)}{\Gamma({\cre N_r})}\frac{\Gamma({\cg N_g}{\cb N_b}+k)}
{\Gamma({\cg N_g}{\cb N_b})}
}
\ee
and so on,
but they are not expressible through the FT functions (which are now triple-graded)
as simply as their matrix model predecessors in s.\ref{averc}. Moreover, one can obtain formulas similar to those in s.2.3 for arbitrary correlators and the partition function
\be\boxed{
{\cal O}_{\Lambda} = \frac{1}{\mu^{|\Lambda|}}\sum_{ R\,\vdash |\Lambda|}
\frac{ D_R({\cre N_r})D_R({\cg N_g}{\cb N_b})}{d_R}\cdot \psi_R(\Lambda)
}
\ee
\be\label{partab}
\boxed{
{\cre Z}\{{\cre t}\} = \sum_{R} \frac{1}{\mu^{|R|}}\frac{D_R({\cre N_r})\,D_R({\cg N_g}{\cb N_b})}{d_R} \cdot \chi_R\{{\cre t}\}
}
\ee
in terms of sizes of tensors, or with the replace $({\cre N_r},{\cg N_g}{\cb N_b})\to {\cre\alpha}/2\pm\sqrt{{\cre\alpha}^2/4-\beta}$ in terms of ${\cre\alpha}$ and $\beta$.

\bigskip

Factorization (\ref{factHe}) in the case, when the Young diagram has a single-line tale,
is now:
\be
{\cal O}_{[\Lambda,1^n]}
= {\mu^{\beta}} \left.\left(\frac{\partial}{\partial t_1}\right)^n
\left(\frac{1}{\mu^{\beta}} {\cal O}_{\Lambda}\right) \right|_{t=0} =
 {\mu^{\beta}} \left(-\frac{\partial}{\partial \mu}\right)^n\left(\frac{1}{\mu^\beta}
{\cal O}_{\Lambda}\right)
= \frac{1}{\mu^{ n}} {\cal O}_{\Lambda} \cdot
\prod_{i=0}^{n-1} \left(\beta+|\Lambda|+i\right)
\label{factab}
\ee
where   ${\cal O}_\Lambda \sim \mu^{-|\Lambda|}$, so that
\be
\boxed{
{\cal O}_{[\Lambda,1^{n}]} = \frac{1}{\mu^{n}}\,
{\cal O}_{[\Lambda]}  \prod_{i=|\Lambda|}^{|\Lambda|+n-1} (\beta+i)
}
\label{factRCM}
\ee
for arbitrary Young diagram $\Lambda$.
This property resembles a similar structure of symmetric group characters,
see \cite{MMN1}.
Irreducible in the above table are just the averages in the first column
and the averages in boxes.

\subsection{Recursions
\label{recuRCM}}

With the above formulas, it is easy to check the simplest recursions following from
the Virasoro constraints (\ref{Virpertr}) and their $t$-derivatives at $t=0$.
In fact, they can also be considered as examples of the basic Ward identity
\be
\boxed{
\mu \left<A\,\frac{\partial {\cal K}}{\partial A}\right> =
\left<\frac{\p^2{\cal K}}{\p A \, \p \bar A} \right>
}
\label{basrec}
\ee
reflecting the invariance under the shift $\delta \bar A = \frac{\p {\cal K}}{\p A}$
of the integration variable.
Here ${\cal K}$ can be any product
of the operators ${\cal K}_{\cre m}$, and, for homogeneous ${\cal K}$, the l.h.s. is
proportional to ${\cal K}$ itself.
The r.h.s. contains contributions from gluing (underlined) and cutting,
cutting of the circle of unit length provides a factor of ${\cre\alpha}$,
gluing the two vertices of ${\cal K}_{\cre 1}$
(if it is present at the l.h.s.) provides a factor of $\beta$ (double underlined):

\begin{landscape}
\thispagestyle{empty}
{\tiny
\be
\mu\cdot\Big<{\cal K}_{\cre 1}\Big> = \Big<1\Big> = \beta & \ \ \ &\mu\cdot\frac{\beta}{\mu}=\beta
\nn
\ee
\be
\begin{array}{lll}
\mu\cdot \Big<{\cal K}_{\cre 2}\Big> = {\cre\alpha}\cdot\Big<{\cal K}_{\cre 1}\Big> & &
\mu\cdot \frac{{\cre \alpha}\beta}{\mu^2} = {\cre \alpha}\cdot \frac{\beta}{\mu}
\nn \\
\mu\cdot \Big<{\cal K}_{\cre 1}{\cal K}_{\cre 1}\Big> = \underline{\Big<{\cal K}_{\cre 1}\Big>}
+ \underline{\underline{\beta\cdot\Big<{\cal K}_{\cre 1}\Big>}}  & &
\mu\cdot \frac{\beta(\beta+1)}{\mu^2}
= (1+\beta)\cdot \frac{\beta}{\mu}
\nn \\
\nn \\
\mu\cdot\Big<{\cal K}_{\cre 3}\Big> = {\cre\alpha}\cdot\Big<{\cal K}_{\cre 2}\Big>
+ \Big<{\cal K}_{\cre 1}{\cal K}_{\cre 1}\Big>
& & \mu\cdot\frac{\beta({\cre\alpha}^2+\beta+1)}{\mu^3}
= {\cre\alpha}\cdot\frac{{\cre\alpha}\beta}{\mu^2} + \frac{\beta(\beta+1)}{\mu^2}
\nn \\
3\mu\cdot\Big<{\cal K}_{\cre 2}{\cal K}_{\cre 1}\Big> =
\underline{2\cdot2\cdot\Big<{\cal K}_{\cre 2}\Big>}
+ 2{\cre \alpha}\cdot\Big<{\cal K}_{\cre 1} {\cal K}_{\cre 1}\Big>
+  \underline{\underline{\beta\cdot\Big<{\cal K}_{\cre 2}\Big>}}  & &
3\mu\cdot\frac{{\cre\alpha}\beta(\beta+2)}{\mu^3} =
(4+\beta)\cdot\frac{{\cre\alpha}\beta}{\mu^2} + 2{\cre\alpha}\cdot\frac{\beta(\beta+1)}{\mu^2}
\nn \\
\mu\cdot\Big<{\cal K}_{\cre 1}{\cal K}_{\cre 1}{\cal K}_{\cre 1}\Big> =
\underline{ 2\cdot \Big<{\cal K}_{\cre 1}{\cal K}_{\cre 1}\Big>}
+ \underline{\underline{\beta\cdot\Big<{\cal K}_{\cre 1}{\cal K}_{\cre 1}\Big>}}  & &
\mu\cdot\frac{\beta(\beta+1)(\beta+2)}{\mu^3} =
 (2+\beta)\cdot\frac{\beta(\beta+1)}{\mu^2}
\end{array}
\nn
\ee
\be
\!\!\!\!\!\!\!\!\!
\begin{array}{lll}
\mu\cdot\Big<{\cal K}_{\cre 4}\Big> =  {\cre\alpha}\cdot \Big<{\cal K}_{\cre 3}\Big>
+ 2\cdot \Big<{\cal K}_{\cre 2}{\cal K}_{\cre 1}\Big>
& & \mu\cdot \frac{{\cre\alpha}\beta({\cre\alpha}^2+3\beta+5) }{\mu^4}
= {\cre\alpha}\cdot \frac{\beta({\cre\alpha}^2+\beta+1)}{\mu^3}
+2\cdot \frac{{\cre\alpha}\beta(\beta+2)}{\mu^3}
\nn \\ \nn \\
4\mu\cdot\Big<{\cal K}_{\cre 3}{\cal K}_{\cre 1}\Big> =
\underline{3\cdot2\cdot\Big<{\cal K}_{\cre 3}\Big>}
+ 3{\cre \alpha}\cdot\Big<{\cal K}_{\cre 2} {\cal K}_{\cre 1}\Big> +
&& 4\mu\cdot \frac{\beta(\beta+3)({\cre\alpha}^2+\beta+1)}{\mu^4} =
(6+\beta)\cdot \frac{\beta({\cre\alpha}^2+\beta+1)}{\mu^3}
+ 3{\cre\alpha}\cdot \frac{{\cre\alpha}\beta(\beta+2)}{\mu^3}
+ 3\cdot\frac{\beta(\beta+1)(\beta+2)}{\mu^3}
 + 3\Big<{\cal K}_{\cre 1} {\cal K}_{\cre 1}{\cal K}_{\cre 1}\Big>
+ \underline{\underline{\beta\cdot \Big<{\cal K}_{\cre 3}\Big>}}
\nn \\ \nn \\
\mu\cdot\Big<{\cal K}_{\cre 2}{\cal K}_{\cre 2}\Big> = \underline{2\cdot\Big<{\cal K}_{\cre 3}\Big>}
+ {\cre\alpha}\cdot\Big<{\cal K}_{\cre 2}{\cal K}_{\cre 1}\Big>
& & \mu\cdot \frac{\beta({\cre\alpha}^2\beta+4{\cre\alpha}^2+2\beta+2)}{\mu^4}
= 2\cdot \frac{\beta({\cre\alpha}^2+\beta+1)}{\mu^3} +
{\cre\alpha}\cdot  \frac{{\cre\alpha}\beta(\beta+2)}{\mu^3}
\nn \\ \nn \\
4\mu\cdot\Big<{\cal K}_{\cre 2}{\cal K}_{\cre 1}{\cal K}_{\cre 1} \Big> =
 \underline{(2\cdot 2\cdot 2+2)\cdot\Big<{\cal K}_{\cre 2}{\cal K}_{\cre 1}\Big>} +
 & & 4\mu\cdot\frac{{\cre \alpha}\beta(\beta+2)(\beta+3)}{\mu^4}
 = 2(5+\beta)\cdot  \frac{{\cre\alpha}\beta(\beta+2)}{\mu^3}
 + 2{\cre\alpha}\cdot\frac{\beta(\beta+1)(\beta+2)}{\mu^3}
 + 2{\cre\alpha}\cdot\Big<{\cal K}_{\cre 1}{\cal K}_{\cre 1}{\cal K}_{\cre 1} \Big>
 + \underline{\underline{2\beta\cdot \Big<{\cal K}_{\cre 2}{\cal K}_{\cre 1}  \Big>}}
 \nn \\ \nn \\
\mu\Big<{\cal K}_{\cre 1}{\cal K}_{\cre 1}{\cal K}_{\cre 1}{\cal K}_{\cre 1}\Big> =
\underline{3\cdot \Big<{\cal K}_{\cre 1}{\cal K}_{\cre 1}{\cal K}_{\cre 1} \Big>}+
\underline{\underline{\beta\cdot \Big<{\cal K}_{\cre 1}{\cal K}_{\cre 1}{\cal K}_{\cre 1}\Big> }}
& &
\mu\cdot\frac{\beta(\beta+1)(\beta+2)(\beta+3)}{\mu^4} =
 (3+\beta)\cdot\frac{\beta(\beta+1)(\beta+2)}{\mu^3}
\end{array}
\nn
\ee
\be
\!\!\!\!\!\!\!\!\!\!\!\! \!\!\!\!\!\!\!
\begin{array}{lll}
\mu\cdot\Big<{\cal K}_{\cre 5}\Big> =  {\cre \alpha}\cdot\Big<{\cal K}_{\cre 4}\Big>
+ 2\cdot \Big<{\cal K}_{\cre 3}{\cal K}_{\cre 1}\Big>
+ \Big<{\cal K}_{\cre 2}{\cal K}_{\cre 2}\Big>
& & \mu\cdot\frac{\beta({\cre\alpha}^4+6{\cre\alpha^2}\beta+15{\cre\alpha}^2+
2\beta^2+10\beta+8)}{\mu^5}
= {\cre \alpha}\cdot  \frac{{\cre\alpha}\beta({\cre\alpha}^2+3\beta+5) }{\mu^4}
 + 2\cdot  \frac{\beta(\beta+3)({\cre\alpha}^2+\beta+1)}{\mu^4}
+ \frac{\beta({\cre\alpha}^2\beta+4{\cre\alpha}^2+2\beta+2)}{\mu^4}
\nn \\ \nn \\ \nn \\
5\mu\cdot\Big<{\cal K}_{\cre 4}{\cal K}_{\cre 1}\Big> =
\underline{8\cdot\Big<{\cal K}_{\cre 4}\Big>}
+ 4{\cre\alpha}\cdot\Big<{\cal K}_{\cre 3}{\cal K}_{\cre 1}\Big>
+ 8\cdot \Big<{\cal K}_{\cre 2}{\cal K}_{\cre 1}{\cal K}_{\cre 1}\Big>
+ \underline{\underline{\beta\cdot \Big<{\cal K}_{\cre 4} \Big>}}
&& 5\mu\cdot\frac{{\cre\alpha}\beta(\beta+4)({\cre\alpha}^2+3\beta+5)}{\mu^5}
=(8+\beta)\cdot \frac{{\cre\alpha}\beta({\cre\alpha}^2+3\beta+5) }{\mu^4}
+ 4{\cre\alpha}\cdot  \frac{\beta(\beta+3)({\cre\alpha}^2+\beta+1)}{\mu^4}
+ 8\cdot \frac{{\cre \alpha}\beta(\beta+2)(\beta+3)}{\mu^4}
\nn \\ \nn \\ \nn \\
5\mu\cdot\Big<{\cal K}_{\cre 3}{\cal K}_{\cre 2}\Big> =
\underline{6\cdot 2\cdot\Big<{\cal K}_{\cre 4}\Big>}
+ 3{\cre\alpha}\cdot \Big<{\cal K}_{\cre 2}{\cal K}_{\cre 2}\Big>
+2{\cre\alpha}\cdot \Big<{\cal K}_{\cre 3}{\cal K}_{\cre 1}\Big>
+ 3\cdot\Big<{\cal K}_{\cre 2}{\cal K}_{\cre 1}{\cal K}_{\cre 1}\Big>
& &  5\mu\cdot\frac{{\cre\alpha}\beta({\cre\alpha}^2\beta+6{\cre\alpha}^2+\beta^2
 +13\beta+18)}{\mu^5}
= 12\cdot \frac{{\cre\alpha}\beta({\cre\alpha}^2+3\beta+5) }{\mu^4}
+ 3{\cre\alpha}\cdot \frac{\beta({\cre\alpha}^2\beta+4{\cre\alpha}^2+2\beta+2)}{\mu^4}
+2{\cre \alpha}\cdot \frac{\beta(\beta+3)({\cre\alpha}^2+\beta+1)}{\mu^4}
+3\cdot \frac{{\cre \alpha}\beta(\beta+2)(\beta+3)}{\mu^4}
\nn \\ \nn \\ \nn \\
5\mu\cdot\Big<{\cal K}_{\cre 3}{\cal K}_{\cre 1}{\cal K}_{\cre 1} \Big> =
\underline{(6\cdot 2+2)\cdot\Big<{\cal K}_{\cre 3}{\cal K}_{\cre 1}\Big>}
+ 3{\cre\alpha}\cdot \Big<{\cal K}_{\cre 2}{\cal K}_{\cre 1}{\cal K}_{\cre 1} \Big>
+ 3\cdot\Big<{\cal K}_{\cre 1}{\cal K}_{\cre 1}{\cal K}_{\cre 1}{\cal K}_{\cre 1} \Big>
+ \underline{\underline{2\beta\cdot\Big<{\cal K}_{\cre 3}{\cal K}_{\cre 1}\Big>}}
 & &  5\mu\cdot\frac{\beta(\beta+3)(\beta+4)({\cre\alpha}^2+\beta+1)}{\mu^5}
= (14+2\beta)\cdot  \frac{\beta(\beta+3)({\cre\alpha}^2+\beta+1)}{\mu^4}
 + 3{\cre\alpha}\cdot \frac{{\cre \alpha}\beta(\beta+2)(\beta+3)}{\mu^4}
 +3\cdot \frac{\beta(\beta+1)(\beta+2)(\beta+3)}{\mu^4}
 \nn\\ \nn \\ \nn \\
5\mu\cdot\Big<{\cal K}_{\cre 2}{\cal K}_{\cre 2}{\cal K}_{\cre 1} \Big> =
\underline{4\cdot 2\cdot\Big<{\cal K}_{\cre 3} {\cal K}_{\cre 1} \Big>
+2\cdot 4\cdot \Big<{\cal K}_{\cre 2}{\cal K}_{\cre 2}  \Big>}
+ 4{\cre\alpha}\cdot \Big<{\cal K}_{\cre 2}{\cal K}_{\cre 1}{\cal K}_{\cre 1} \Big>
+\underline{\underline{\beta\cdot \Big<{\cal K}_{\cre 2}{\cal K}_{\cre 2}  \Big>}}
& &  5\mu\cdot\frac{\beta(\beta+4)({\cre \alpha}^2\beta+4{\cre \alpha}^2+2\beta+2)}{\mu^5}
= 8 \cdot \frac{\beta(\beta+3)({\cre\alpha}^2+\beta+1)}{\mu^4}
+ (8+\beta)\cdot \frac{\beta({\cre\alpha}^2\beta+4{\cre\alpha}^2+2\beta+2)}{\mu^4}
+4{\cre \alpha} \cdot \frac{{\cre \alpha}\beta(\beta+2)(\beta+3)}{\mu^4}
\nn \\ \nn \\ \nn \\
5\mu\cdot\Big<{\cal K}_{\cre 2}{\cal K}_{\cre 1}{\cal K}_{\cre 1}{\cal K}_{\cre 1}\Big> =
\underline{(3\cdot 4+3\cdot 2)\cdot \Big<{\cal K}_{\cre 2}{\cal K}_{\cre 1}{\cal K}_{\cre 1}\Big>}
+2{\cre \alpha}\cdot \Big<{\cal K}_{\cre 1}{\cal K}_{\cre 1}{\cal K}_{\cre 1}{\cal K}_{\cre 1}\Big>
+\underline{\underline{3\beta\cdot \Big<{\cal K}_{\cre 2}{\cal K}_{\cre 1}{\cal K}_{\cre 1} \Big>}}
& &   5\mu\cdot\frac{{\cre \alpha}\beta(\beta+2)(\beta+3)(\beta+4)}{\mu^5}
= (18+3\beta) \cdot \frac{{\cre \alpha}\beta(\beta+2)(\beta+3)}{\mu^4}
+ 2{\cre\alpha}\cdot \frac{\beta(\beta+1)(\beta+2)(\beta+3)}{\mu^4}
\nn \\ \nn \\ \nn \\
\mu\cdot\Big<{\cal K}_{\cre 1}{\cal K}_{\cre 1}
{\cal K}_{\cre 1}{\cal K}_{\cre 1}{\cal K}_{\cre 1}\Big> =
\underline{4\cdot \Big<{\cal K}_{\cre 1}{\cal K}_{\cre 1}
{\cal K}_{\cre 1}{\cal K}_{\cre 1}\Big>}
+ \underline{\underline{\beta\cdot \Big<{\cal K}_{\cre 1}{\cal K}_{\cre 1}
{\cal K}_{\cre 1}{\cal K}_{\cre 1}\Big>}}
& &  \mu\cdot\frac{\beta(\beta+1)(\beta+2)(\beta+3)(\beta+4)}{\mu^5} = (4+\beta)\cdot\frac{\beta(\beta+1)(\beta+2)(\beta+3)}{\mu^4}
\nn \\ \nn \\
&\!\!\!\!\!\!\!\!\!\!\!\!\!\!\!\!\!\!\!\!\!\!\!\!\!\!\!\!\!\!\!\!\!\!\!\! \ldots
\end{array}
\ee
}
\end{landscape}
Such corollaries of the universal recursion (\ref{basrec})
is the kind of relations that one can first look for in more complicated
tensor models, where their generating functions like (\ref{Virpertr}) are not
immediately available.

\subsection{Integrability properties: does Virasoro imply integrability?}

Since the model (\ref{Zpertr}) is equivalent to the rectangular complex matrix model,
which, according to \cite{CMM,AMM3} is integrable, at least, when the matrix is square,
the partition function  ${\cre Z}\{{\cre t}\}$ at $\beta={\cre\alpha}^2/4$
is actually  again a $\tau$-function of the (forced) Toda chain hierarchy, however, it is a different solution to the hierarchy: $C_k$ in the determinant representation (\ref{detrep}) is now of the form
\be
C_k=\int_0^\infty dx\ x^k\exp\left(-\mu x+\sum_k t_kx^k\right)
\ee
In fact, since the model (\ref{Zpertr}) has the representation (\ref{partab}), similarly to the rectangular complex model (\ref{CZotau}), it can be associated \cite{MMi} for arbitrary ${\cre\alpha}$ and $\beta$ with $Z_{(1,2)}$
\be
{\cre Z}\{{\cre t}\} =Z_{(1,2)}\Big\{\mu,{\cre N_r},{\cg N_g}{\cb N_b}\Big|{\cre t_k}\Big\}
\ee
which is a hypergeometric KP $\tau$-function \cite{GKM2,Charint,AMMN,AMMN2}. Moreover, by switching on another set of times, one makes of it $Z_{(2,2)}$ which is a Toda lattice $\tau$-function, however, different from the Toda chain.

\subsubsection{$W$-representation}

The $W$-representation for this model also can be read off from formulas of \cite{AMMN2} upon its identification with $Z_{(1,2)}$:
\be
{\cre Z}\{{\cre t}\}=\exp\left\{{1\over\mu}\Big(\beta t_1+{\cre \alpha}\hat L_{1}\{{\cre t}\} + \hat W_{1}\{{\cre t}\}\Big)\right\}\cdot 1
\ee
with
\be
\hat L_1\{t\} =\sum_m  (m+1)t_{m+1}\frac{\p}{\p t_m},  \\
\hat W_1\{t\} =\sum_{a,b} abt_at_b \frac{\p}{\p t_{a+b-1}} + (a+b+1) t_{a+b+1}\frac{\p^2}{\p t_a\p t_b}
\ee

\subsection{The message}

Thus, the "red" tensor model (\ref{Zpertr}) is not really tensor: it is equivalent
to a rectangular matrix model, all its averages are deducible from a single set
of Virasoro constraints and satisfy the linear sum rules,
i.e. the model is {\it solvable} in extreme sense.

The only thing we lose are the advantages of the Fourier transform in $N$.
It worked perfectly for the single-trace correlators ${\cal O}_{[k]}$
in the Hermitian model, i.e. in the "vertical" direction in the table of averages.
However, in the "horizontal" direction, a "composite" variable   $N_1N_2$ appears
through the factorization formula,
and there is no nice way to express the Fourier transform in such variable
through the two separate transformations in $N_1$
and $N_2$ (unless both $N$'s are large and the Fourier sums can be substituted
by integrals).
Still, the double Fourier transform allowed one to find simple {\it combinations}
of averages, where the complexities drop out, and this led to the powerful sum rules.
At the tensor level, however, we have a triple (for three colorings) Fourier transform
and the triply-composed variable $N_rN_gN_b$, with no efficient way to work with it.
Of course, the correlators in the red model can still be found,
but with the reference to the matrix models,
not by an adequate triple-Fourier technique.
Thus, if one wants to look at tensor peculiarities in the simplest possible place,
the issue of the Fourier transform is the right choice.
In the next section, we see the need for a solution to this problem
even better.

Still, to really penetrate in the field of tensor models and see the deviations from
the matrix model intuition, we need more than just the "red" model.
Thus, we add blue to red, and then green emerges as promised in \cite{Arist2}.
Actually, we begin from adding green.

\section{A non-trivial RG-closed extension of the Aristotelian tensor model
\label{redgreenmod}}

\subsection{The red-green partition function}

Of course, in considerations of the previous section one could use
operators ${\cg {\cal K}_{k}}$ instead of ${\cre {\cal K}_k}$.
The only difference between them is the switch of some contravariant
indices to covariant ones, and it plays no role at this stage.
The Gaussian averages of ${\cg {\cal K}_k}$ are given by the same formulas
(\ref{corrZpertr}).

However, we can also consider mixed correlators like
$\Big<{\cre{\cal K}_k}\ {\cg{\cal K}_m}\Big>$,
i.e. study a less trivial ``mixed'' partition function
\be
{Z}\{{\cre t},{\cg t}
\} = \int d^2\!A \exp\left(-\mu\,\Tr A\bar A +
\sum_k {\cre t_kK_k}+\sum_k {\cg t_kK_k}
\right)
\label{Zpertrg1}
\ee
For this purpose the standard Virasoro constraints are no longer sufficient.
Moreover, to get the Ward identities, we now need to deform, say,
${\cre K_k}$ by ${\bf \nabla}({\cg K_{m+1}})$, and this produces
{\it new} operators like

\begin{picture}(300,80)(-180,-40)
\put(-60,-2){\mbox{${\cal K}_{{\cre 3},{\cg 3}} \ \ = $}}
{\cre
\put(0,0){\vector(0,1){20}}\put(15,35){\vector(1,0){20}}\put(50,20){\vector(0,-1){20}}
\put(50,-20){\vector(-1,-1){15}}\put(15,-35){\vector(-1,1){15}}
}
\put(-3,0){
{\cg  \put(0,20){\vector(1,1){15}} \put(35,35){\vector(1,-1){15}}
\put(50,0){\vector(0,-1){20}}\put(35,-35){\vector(-1,0){20}}\put(0,-20){\vector(0,1){20}}
}
}
{\cb \put(50,0){\vector(-1,0){50}}
\qbezier(0,20)(15,20)(15,35)\put(12,25){\vector(1,1){2}}
\qbezier(50,20)(35,20)(35,35)\put(38,25){\vector(1,-1){2}}
\qbezier(0,-20)(15,-20)(15,-35)\put(12,-25){\vector(1,-1){2}}
\qbezier(50,-20)(35,-20)(35,-35)\put(38,-25){\vector(1,1){2}}
}
\put(125,0){
\put(-50,-2){\mbox{$=$}}
\put(0,0){\circle{40}}
{\cb \put(20,0){\vector(-1,0){40}}}
}
\end{picture}

\noindent
This means that (\ref{Zpertrg1}) is not RG-complete, and more operators
and time-variables should be added to its extended action.
Clearly, just ${\cal K}_{{\cre k},{\cg m}}$ are not enough:
further variations of {\it these} operators can produce even more.
Typical examples of diagrams contributing to the full set of Ward identities are

\begin{picture}(300,80)(-250,-50)

\put(-100,0){
\put(0,0){\circle{40}}
{\cb \put(10,17){\line(-1,0){20}}\put(10,-17){\line(-1,0){20}}\put(20,0){\line(-1,0){40}}}
\put(-10,-40){\mbox{$kt\frac{\partial}{\partial t}$}}
}

\put(0,0){\circle{40}}
\put(50,0){\circle{40}}
{\cb \put(20,0){\line(-1,0){40}}
\put(67,10){\line(0,-1){20}}
\put(60,-17){\line(-1,0){20}}}
{\cb\qbezier(0,20)(25,0)(50,-20)\qbezier(0,-20)(25,0)(50,20)}
\put(15,-40){\mbox{$\frac{\partial^2}{\partial t \partial t }$}}

\end{picture}

\noindent
The question is what the {\it minimal} RG-completion is.

We can look for it by examining the tree and loop descendants of the
keystone operators, but a more practical approach is to do this directly
at the Ward identities.
Namely, there is a universal recursion for the Gaussian averages,
\be
\mu \left<A_{{\cre i_r}{\cb \alpha_b}}^{{\cg j_g}}
\frac{\partial {\cal K}}{\partial A_{{\cre i_r}{\cb \alpha_b}}^{{\cg j_g}}}\right> =
\left<\frac{\partial^2{\cal K}}{\partial A_{{\cre i_r}{\cb \alpha_b}}^{{\cg j_g}}
\partial \bar A^{{\cre i_r}{\cb \alpha_b}}_{{\cg j_g}}  }\right>
\label{basicrecursion}
\ee
applicable to arbitrary operator ${\cal K}$.
It can be considered either as the Ward identity for the shift
$\delta\bar A = \p {\cal K}/\p A$ evaluated at the point where all $t_k=0$,
or just as a simple consequence of the Wick theorem
(Wick recursion, result of just one propagator insertion).
The point is that the operators arising at the r.h.s. can be of more general
nature than those at the l.h.s.
That is, if one takes for ${\cal K}$ arbitrary functions of keystone operator(s),
their $A,\bar A$-derivatives contain both the tree and loop descendants of keystones.
Thus, looking at (\ref{basicrecursion}) and nothing else we iteratively reconstruct
the entire RG-closed set of operators generated by the given keystone ones.

\subsection{Hierarachy/tower of Virasoro-like constraints}

Moreover, the same procedure can be promoted to the level of Virasoro-like
identities, i.e. for {\it the generating functions} of recursion relations.

\paragraph{Step 1:} To this end, we can start from a single keystone operator ${\cal K}_{\cre 2}$,
discover that the recursion generates from it all ${\cal K}_{\cre k}$
with arbitrary $k$ as the tree operators,
while all loop operators appear algebraic function of those.
Thus, at this stage, one needs to add all these ${\cal K}_{\cre k}$
to extend the action with independent coefficients (time-variables) $t_{\cre k}$,
i.e. to obtain the extended red model (\ref{Zpertr}).
The above reasoning is now expressed in form of the basic level Virasoro constraint
(\ref{Virpertr}), which we naturally call "red-Virasoro".

\paragraph{Step 2:} We could instead begin from another keystone operator
${\cal K}_{\cg 2}$ and arrive at the green model satisfying the green-Virasoro constraints.

\paragraph{Step 3:} What we need, is the model where both ${\cal K}_{\cre 2}$
and ${\cal K}_{\cg 2}$ are included as keystones, and the natural starting point
for the extended action is the one in (\ref{Zpertrg1}).
However, the red-Virasoro constraints hold for this model only at the point where all
green times $t_{\cg k}=0$, while the green-Virasoro constraints hold at the point where all $t_{\cre k}=0$.
When both types of time-variables are non-vanishing, we should include the descendants
coming from the recursion relation (\ref{basicrecursion}) with
${\cal K} = {\cal K}_{\cre m}{\cal K}_{\cg n}$, i.e. the operators ${\cal K}_{\cre m,\cg n}$.
Since they are algebraically independent of ${\cal K}_{\cre k}$ and ${\cal K}_{\cg k}$,
we should add them to extended the action with their own new couplings (time-variables)
$t_{{\cre m},{\cg n}}$.
Then, in the Aristotelian model (\ref{Zpertrg1}), we immediately obtain
the new Virasoro constraint of the next level
\be
\!\!\!\!\!\!\!\!\!\!\!\!\!\!\!\!\!\!\!\!\!\!\!\!\!\!\!\!\!\!\!\!\!
\left(-\mu\, \frac{\partial}{\partial t_{{\cre m+1}}} +
\sum_k kt_{{\cre k}} \frac{\partial}{\partial t_{{\cre k+m}}} +
\underline{   \sum_k kt_{{\cg k}} \frac{\partial}{\partial t_{{\cre m+1},{\cg k}}}} +
\right. \nn \\
\left. +
\sum_{a=1}^{m-1} \frac{\partial^2}{\partial t_{{\cre a}}\partial t_{{\cre m-a}}}
+ \underbrace{({\cre N_r} + {\cg N_g}{\cb N_b})}_{{\cre \alpha}}
(1-\delta_{{\cre m},0})\frac{\partial}{\partial t_{{\cre m}}}
+ \underbrace{{\cre N_r}{\cg N_g}{\cb N_b}}_\beta\cdot \delta_{{\cre m},0} \right)
 {\cal Z}\{{\cre t},{\cg t}\} = 0
\label{Virpertrg}
\ee
valid for arbitrary values of $t_{\cre k}$ and $t_{\cg n}$
(new as compared to (\ref{Virpertr}) is just the underlined term).
Eq.(\ref{Virpertrg}) is the Ward identity for the shift
$\delta \bar A = \p{\cal K}_{{\cre m}}/\p A$.
Of course, (\ref{Virpertrg}) has a "green" counterpart associated with the shift
$\delta \bar A = \p{\cal K}_{{\cg n}}/\p A$:
\be\label{Virpertrg2}
\!\!\!\!\!\!\!\!\!\!\!\!\!\!\!\!\!\!\!\!\!\!\!\!\!\!\!\!\!\!\!\!\!
\left(-\mu\, \frac{\partial}{\partial t_{{\cg n+1}}} +
\sum_k kt_{{\cg k}} \frac{\partial}{\partial t_{{\cg k+n}}} +
\underline{   \sum_k kt_{{\cre k}} \frac{\partial}{\partial t_{{\cre k},{\cg n+1}}}} +
\right. \nn \\
\left. +
\sum_{a=1}^{n-1} \frac{\partial^2}{\partial t_{{\cg a}}\partial t_{{\cg n-a}}}
+ \underbrace{({\cg N_r} + {\cre N_g}{\cb N_b})}_{{\cg \alpha}}
(1-\delta_{{\cg n},0})\frac{\partial}{\partial t_{{\cg n}}}
+ \underbrace{{\cre N_r}{\cg N_g}{\cb N_b}}_\beta\cdot \delta_{{\cg n},0} \right)
 {\cal Z}\{{\cre t},{\cg t}\} = 0
\ee
These two sets of constraints taken at all zero times $t_k$'s lead to the relations for the correlators that we already know from the previous section:
\be
-\mu\Big<{\cal K}_{\cre m+1}\Big>+\sum_{a=1}^{m-1}\Big<{\cal K}_{\cre a}{\cal K}_{\cre m-a}\Big>
+ {\cre \alpha}
(1-\delta_{{\cre m},0})\Big<{\cal K}_{\cre m}\Big>
+ \beta\cdot \delta_{{\cre m},0}= 0\\
-\mu\Big<{\cal K}_{\cg n+1}\Big>+\sum_{a=1}^{n-1}\Big<{\cal K}_{\cg a}{\cal K}_{\cg n-a}\Big>
+ {\cg \alpha}
(1-\delta_{{\cg n},0})\Big<{\cal K}_{\cg n}\Big>
+ \beta\cdot \delta_{{\cg n},0}= 0\nn
\ee

However, one can go further and take the first derivative of (\ref{Virpertrg}) w.r.t. $t_{{\cg p}}$ at all zero times $t_k$'s which gives
\be
-\mu\Big<{\cal K}_{\cre m}{\cal K}_{\cg p}\Big>+p\Big<{\cal K}_{{\cre m},{\cg p}}\Big>+\sum_{a=1}^{m-2}\Big<{\cal K}_{\cre a}{\cal K}_{\cre m-a-1}{\cal K}_{\cg p}\Big>
+{\cre \alpha}
(1-\delta_{{\cre m},1})\Big<{\cal K}_{\cre m-1}{\cal K}_{\cg p}\Big>
+ \beta\cdot\Big<{\cal K}_{\cg p}\Big>\cdot \delta_{{\cre m},1}= 0
\ee
Similarly, the derivative of (\ref{Virpertrg2}) w.r.t. $t_{{\cre q}}$ gives
\be
-\mu\Big<{\cal K}_{\cg n}{\cal K}_{\cre q}\Big>+q\Big<{\cal K}_{{\cg n},{\cre q}}\Big>+\sum_{a=1}^{n-2}\Big<{\cal K}_{\cg a}{\cal K}_{\cg n-a-1}{\cal K}_{\cre q}\Big>
+{\cg \alpha}
(1-\delta_{{\cg n},1})\Big<{\cal K}_{\cg n-1}{\cal K}_{\cre q}\Big>
+ \beta\cdot\Big<{\cal K}_{\cre q}\Big>\cdot \delta_{{\cg n},1}= 0
\ee
Choosing in these expressions ${\cre q}={\cre m}$ and ${\cg p}={\cg n}$ and adding them with the coefficients $m$ and $n$ respectively, one arrives at the recurrent relation (we take into account that ${\cal K}_{{\cg n},{\cre q}}={\cal K}_{{\cre q},{\cg n}}$)
\be
-({\cre m}+{\cg n})\,\mu\cdot \Big<{\cal K}_{\cre m}{\cal K}_{\cg n}\Big> +
2{\cre m}{\cg n}\cdot \Big<{\cal K}_{{\cre m} ,{\cg n}}\Big>
+{\cre m}\, {\cre \alpha}(1-\delta_{{\cre m},1})\cdot \Big<{\cal K}_{\cre m-1}{\cal K}_{\cg n}\Big>
+ {\cre m}\cdot \!\!\!\!\!\!\!\!\!\sum_{\stackrel{m_1,m_2\geq 1}{m_1+m_2=m-1}}\!\!\!
\Big<{\cal K}_{\cre m_1}{\cal K}_{\cre m_2}{\cal K}_{\cg n}\Big>+\nn\\
+{\cg n}\,{\cg \alpha}(1-\delta_{{\cg n},1})\cdot \Big<{\cal K}_{\cre m}{\cal K}_{\cg n-1}\Big>
+ {\cg n}\cdot\!\!\!\!\!\!\!\!\!\sum_{\stackrel{n_1,n_2\geq 1}{n_1+n_2=n-1}}\!\!\!
\Big<{\cal K}_{\cre m}{\cal K}_{\cg n_1}{\cal K}_{\cg n_2}\Big>+\beta \cdot\Big<{\cal K}_{\cg n}\Big>\cdot \delta_{{\cre m},1}+\beta \cdot\Big<{\cal K}_{\cre m}\Big>\cdot \delta_{{\cg n},1}=0
\label{recmn}
\ee

In fact, using (\ref{Virpertrg}) or (\ref{Virpertrg2}), one can easily calculate the correlators $\Big<{\cal K}_{\cre m,\cg n}\Big>$ in the planar limit. The simplest way to do this is to get the equation for the generating functions. This is done immediately, however, requires first evaluating the two-resolvents in the planar limit. This calculation will be presented elsewhere, and, in the next section, we perform just direct calculation through the Gaussian integration.

\paragraph{Next steps:}
(\ref{Virpertrg}) holds, however, only at vanishing values of the newly-added times:
$t_{{\cre m},{\cg n}}=0$.
If we want to release {\it these} time-variables, we will need new operators to add,
include them with new couplings and get the Virasoro-like constraints of the next level,
and so on.
In this way, we arrive at the clearly ordered tower of embedded constraints,
which is generically infinite, though can sometimes terminate,
if the newly emerging operators get algebraically dependent of the previous ones.
This unavoidably happens, for example, if we keep the values of $N_r$, $N_g$ and $N_b$ fixed and integer.
This option was not considered interesting in the simplest matrix models,
where much more can be achieved: a single generating function for all Ward identities
at all $N$.
However, for the tensor models this option can turn out to be much more interesting.

\paragraph{Possible culmination} will appear when one manages to understand this
entire well-structured tower of constraints well enough and find a top level
generating function(al) unifying them all.
There is, however, a long way to go before we reach this point,
and one of the ways is a reformulation of particular models in the BZ terms
summarized in above s.\ref{BZrem}.

\paragraph{Advantages} of reformulation in terms of Virasoro-like constraints
are not exhausted by their beauty: important is an ability to go beyond the
Gaussian phases and to true non-perturbative considerations.
We briefly mentioned in s.\ref{mamorem} possible techniques involved in this part of
the story, and they should be also looked at and for in the study of
tensor models.

\bigskip

In the remaining part of this section, we come back down to earth and start
doing the first of above steps for the Aristotelian model, the first
non-trivial one among the tensor models.
Our main concern is developing some technique for the Gaussian calculations.
The first results are tested with
the basic recursions (\ref{basicrecursion}) and (\ref{recmn}).
The main goals, however, are the lifting from the recursions to their
generating functions and functionals and
the search for general formulas and relations like (\ref{surN1N2}).

\subsection{Gaussian averages in the Aristotelian model by direct computation
\label{aveAr}}

\subsubsection{Red rings and green rings
\label{rorgrings}}

Like in the red model from s.\ref{redmod},
when only operators ${\cal K}_{\cre m}$ are present in the correlators,
they can be considered as those in the rectangular matrix model of size
${\cre N_b}\times {\cg N_g}{\cb N_b}$, i.e. taken just from the table in
s.\ref{averc} with ${\cre\alpha} = {\cre N_r} + {\cg N_g}{\cb N_b}$
and $\beta={\cre N_r}{\cg N_g}{\cb N_b}$:
\be
\Big< \prod_{i} {\cal K}_{\cre m_i} \Big> =
{\cal O}_{[\Lambda]}^{{\cre N_b}\times {\cg N_g}{\cb N_b}}
\ee
for $\Lambda=[m_1\geq m_2\geq \ldots \geq 0]$.

The same is true when only green operators are present:
\be
\Big< \prod_{j} {\cal K}_{\cg n_j} \Big> =
{\cal O}_{[\Lambda]}^{{\cg N_g}\times {\cre N_r}{\cb N_b}}
\ee
only this time ${\cg \alpha} = {\cg N_g} + {\cre N_r}{\cb N_b}$
and $\Lambda=[n_1\geq n_2\geq \ldots \geq 0]$, while
$\beta={\cre N_r}{\cg N_g}{\cb N_b}$ remains the same.

The operators ${\cal K}_{\cre m}$ and ${\cal K}_{\cg n}$ are naturally
depicted as red and green circles of lengths $m$ and $n$ respectively,
as will be more accurately explained in s.\ref{rgcomplrg} below.
Since ${\cal K}_{\cre 1} = {\cal K}_{\cg 1}$,
the circles of unit length can be considered either  red or green.
In the rectangular matrix model, the insertion of such unit circles changes
averages in a very simple way described by the factorization formula (\ref{factRCM}).

\subsubsection{Reductions at $N=1$}

In more complicated cases, the averages in the Aristotelian model (\ref{Zpertrg1})
are not reduced to those for the rectangular matrices.
However, they do so, when any of the three colorings disappear, i.e. when
any one of the three numbers
${\cre N_r}$, ${\cg N_g}$ or ${\cb N_b}$ becomes unity.
This provides a convenient check for formulas
and also helps to build them by lifting from the three different rectangular model limits:
such calculations can provide the answers modulo $({\cre N_r}^2-1)({\cg N_g}^2-1)({\cb N_b}^2-1)$.
We illustrate this approach in examples below in this section.

\subsubsection{Red and green rings together}

The next correlators to look at are the collections of rings of different colors,
beginning from $\Big<{\cal K}_{\cre m}\,{\cal K}_{\cg n}\Big>$.

First of all, if we put, say, ${\cg N_g}=1$ then the operators ${\cal K}_{\cre m}$ and
${\cal K}_{\cg n}$
turn respectively into  ${\cal K}_{{\cre m}}$,
$ {\cal K}_{1}^{\cg n}$ of the rectangular ${\cre N_r}\times {\cb N_b}$ matrix model,
i.e. will be easily calculated with the help of (\ref{factRCM}):
\be
\Big<{\cal K}_{\cre m}\,{\cal K}_{\cg n}\Big> \ \stackrel{{\cg N_g}=1}{=} \
{\cal O}^{{\cre N_r}\times {\cb N_b}}_{[{\cre m},1^{\cg n}]}
= \frac{1}{\mu^n}\,{\cal O}^{{\cre N_r}\times {\cb N_b}}_{[{\cre m}]}
\prod_{j=0}^{{\cg n}-1} \Big({\cre N_r}{\cb N_b} + {\cre m}+j\Big)
\label{KmKng1}
\ee
Likewise,
\be
\Big<{\cal K}_{\cre m}\,{\cal K}_{\cg n}\Big> \ \stackrel{{\cre N_r}=1}{=} \
{\cal O}^{{\cg N_g}\times {\cb N_b}}_{[{\cg n },1^{\cre m}]}
= \frac{1}{\mu^n}\,{\cal O}^{{\cg N_g}\times {\cb N_b}}_{[{\cg n}]}
\prod_{i=0}^{{\cre m}-1} \Big({\cg N_g}{\cb N_b} + {\cg n}+i\Big)
\label{KmKnr1}
\ee
If ${\cb N_b}=1$, then the both types of operators turn into the circles in the
rectangular model ${\cre N_r}\times{\cg N_g}$ and
\be
\Big<{\cal K}_{\cre m}\,{\cal K}_{\cg n}\Big> \ \stackrel{{\cb N_b}=1}{=} \
{\cal O}^{{\cre N_r}\times {\cg N_g}}_{[{\cre m},{\cg n } ]}
\label{KmKnb1}
\ee
where this time we should substitute $\alpha={\cb \alpha} \equiv {\cre N_r}+{\cg N_g}$.
As already mentioned, these formulas provide a nice starting point for
evaluating the correlator at generic values of ${\cre N_r}$, ${\cg N_g}$ and ${\cb N_b}$.

For example,
\be
\Big<{\cal K}_{\cre 2}\,{\cal K}_{\cg 2}\Big>
= \frac{N_rN_gN_b}{\mu^4}\Big( {\cre N_r}\!^2{\cg N_g}\!^2{\cb N_b}\!^3 +
({\cre N_r}\!^2 +{\cg N_g}\!^2+4){\cre N_r}{\cg N_g}{\cb N_b}\!^2
+ ({\cre N_r}\!^2{\cg N_g}\!^2+4{\cre N_r}\!^2+4{\cg N_g}\!^2+2){\cb N_b}
+6{\cre N_r}{\cg N_g}\Big)
\label{K2K2}
\ee
and
\be
\Big<{\cal K}_{\cre 3}\,{\cal K}_{\cg 2}\Big>
= \frac{N_rN_gN_b}{\mu^5}\Big(
{\cre N_r}\!^2 {\cg N_g}\!^3 {\cb N_b}\!^4
+(3 {\cre N_r}\!^2+{\cg N_g}\!^2+6) {\cre N_r} {\cg N_g}\!^2 {\cb N_b}\!^3 +\nn \\
+({\cre N_r}\!^4+ 3 {\cre N_r}\!^2 {\cg N_g}\!^2 +19 {\cre N_r}\!^2+6 {\cg N_g}\!^2+6)
{\cg N_g} {\cb N_b}\!^2
+( {\cre N_r}\!^2 {\cg N_g}\!^2+6{\cre N_r}\!^2+25{\cg N_g}\!^2+18){\cre N_r} {\cb N_b}
+12({\cre N_r}\!^2+1){\cg N_g}
\Big)
\label{K3K2}
\ee
are fully defined by the three reductions (\ref{KmKng1})-(\ref{KmKnb1}).
The matrix model calculation becomes insufficient beginning from
\be
\Big<{\cal K}_{\cre 4}\,{\cal K}_{\cg 2}\Big>
= \frac{N_rN_gN_b}{\mu^6}\Big(
{\cre N_r}\!^2 {\cg N_g}\!^4 {\cb N_b}\!^5
+ (6{\cre N_r}\!^2+{\cg N_g}\!^2+8){\cre N_r}{\cg N_g}\!^3{\cb N_b}\!^4 +
\nn \\
+ (6{\cre N_r}\!^4+6{\cre N_r}\!^2{\cg N_g}\!^2+53{\cre N_r}\!^2+8{\cg N_g}\!^2+12)
{\cg N_g}\!^2{\cb N_b}\!^3
+({\cre N_r}\!^4+6{\cre N_r}\!^2{\cg N_g}\!^2+53{\cre N_r}\!^2+65{\cg N_g}\!^2+100)
{\cre N_r} {\cg N_g} {\cb N_b}\!^2 + \nn \\
+ ({\cre N_r}\!^4{\cg N_g}\!^2+8{\cre N_r}\!^4+85{\cre N_r}\!^2{\cg N_g}\!^2
+80{\cre N_r}\!^2+68{\cg N_g}\!^2+32){\cb N_b}
+ 20({\cre N_r}\!^2+5){\cre N_r} {\cg N_g}
\Big)
\label{K4K2}
\ee
and
\be
\Big<{\cal K}_{\cre 3}\,{\cal K}_{\cg 3}\Big>
= \frac{N_rN_gN_b}{\mu^6}\Big(
{\cre N_r}\!^3 {\cg N_g}\!^3 {\cb N_b}\!^5+3 ({\cre N_r}\!^2+{\cg N_g}\!^2+3)
 {\cre N_r}\!^2 {\cg N_g}\!^2 {\cb N_b}\!^4 +
 \nn \\
+({\cre N_r}\!^4+9{\cre N_r}\!^2 {\cg N_g}\!^2 +{\cg N_g}\!^4+28 {\cre N_r}\!^2
+28 {\cg N_g}\!^2  +18) {\cre N_r}  {\cg N_g}  {\cb N_b}\!^3 +
\nn \\
+3({\cre N_r}\!^4{\cg N_g}\!^2+{\cre N_r}\!^2{\cg N_g}\!^4
+3 {\cre N_r}\!^4 +35 {\cre N_r}\!^2{\cg N_g}\!^2 +3{\cg N_g}\!^4
+15 {\cre N_r}\!^2+15 {\cg N_g}\!^2+2) {\cb N_b}\!^2 +
\nn \\
+( {\cre N_r}\!^2{\cg N_g}\!^2 +46 {\cre N_r}\!^2 +46 {\cg N_g}\!^2+181)
 {\cre N_r} {\cg N_g} {\cb N_b}
+30 ({\cre N_r}\!^2+1) ({\cg N_g}\!^2+1)
\Big)
\label{K3K3}
\ee
From these examples, supplemented by other corollaries of (\ref{KmKng1})-(\ref{KmKnb1}),
one can observe an emerging structure of the answer:
\be
\Big<{\cal K}_{\cre m}\,{\cal K}_{\cg 2}\Big>
= \frac{N_rN_gN_b}{\mu^{m+1}}\left\{
{\cre N_r}\!^2 {\cg N_g}\!^m {\cb N_b}\!^{m+1}
+ \left(\frac{m(m-1)}{2}{\cre N_r}\!^2+{\cg N_g}\!^2+2m\right)
{\cre N_r}{\cg N_g}\!^{m-1}{\cb N_b}\!^m +
\right.
\nn \\
\!\!\!\!\!\!\!\!\!\!
+ \left(\frac{{\cre m}({\cre m}-1)^2({\cre m}-2)}{12}\cdot {\cre N_r}\!^4
+\frac{m(m-1)}{2}{\cre N_r}\!^2{\cg N_g}\!^2
+\frac{m(m-1)(m^2+23m-2)}{24} \cdot {\cre N_r}\!^2+2m{\cg N_g}\!^2+m(m-1)\right)
{\cg N_g}\!^{m-2}{\cb N_b}\!^{m-1} +
\nn
\ee
\vspace{-0.6
cm}
\be
\left. \phantom{\frac{m(m-1)}{2}}     + O\Big({\cb N_b}\!^{m-2}\Big)\right\}
\ee
and, further,
\be
\Big<{\cal K}_{\cre m}\,{\cal K}_{\cg n}\Big>
= \frac{N_rN_gN_b}{\mu^{m+1}}\left\{
{\cre N_r}\!^{\cg n} {\cg N_g}\!^{\cre m} {\cb N_b}\!^{{\cre m}+{\cg n}-1}
+ \left(\frac{{\cre m}({\cre m}-1)}{2}{\cre N_r}\!^2+\frac{{\cg n}({\cg n}-1)}{2}{\cg N_g}\!^2
+{\cre m}{\cg n}\right)
{\cre N_r}\!^{{\cg n}-1}{\cg N_g}\!^{{\cre m}-1}{\cb N_b}\!^{{\cre m}+{\cg n}-2} +
\right.
\nn \\
+ \left(\frac{{\cre m}({\cre m}-1)^2({\cre m}-2)}{12}\cdot {\cre N_r}\!^4
+\frac{{\cre m}({\cre m}-1)}{2}\frac{{\cg n}({\cg n}-1)}{2}\cdot{\cre N_r}\!^2{\cg N_g}\!^2
+ \frac{{\cg n}({\cg n}-1)^2({\cg n}-2)}{12}\cdot {\cg N_g}\!^4
+
\right.
\nn \\
+\frac{{\cre m}({\cre m}-1)\cdot\big(({\cre m}+1)({\cre m}-2) +12{\cre m}{\cg n} \big)}{24}
{\cre N_r}\!^2
+ \frac{{\cg n}({\cg n}-1)\cdot\big(({\cg n}+1)({\cg n}-2)+12{\cre m}{\cg n}\big)}{24}
 {\cg N_g}\!^2 +
\nn
\ee
\vspace{-0.6cm}
\be
\left.\left.
\phantom{\frac{m(m-1)}{2}}
+ \frac{{\cre m}({\cre m}-1)\,{\cg n}({\cg n}-1)}{2} \right)
{\cre N_r}\!^{{\cg n}-2}{\cg N_g}\!^{{\cre m}-2}{\cb N_b}\!^{{\cre m}+{\cg n}-3}
\ \ + \ \ O\Big({\cb N_b}\!^{{\cre m}+{\cg n}-4}\Big)\right\}
\label{KmKnexpan}
\ee
but more data is needed to fully reconstruct it even for this simplest kind of
Gaussian correlators in the simplest of tensor models.

As usual, many terms in the expansion (\ref{KmKnexpan}) can be restored from the reduction to
${\cg N_g}=1$, when it should match the large-${\cb N_b}$ expansion
\be
{\cal O}^{{\cre N_r}\times {\cb N_b}}_{[{\cre m},1^{\cg n}]} =
{\cal O}^{{\cre N_r}\times {\cb N_b}}_{[{\cre m}]}
\prod_{i={\cre m}}^{{\cre m}+{\cg n}-1}({\cre N_r}{\cb N_b} +i) = \nn \\
= {\cre N_r}\!^{\cg n}{\cb N_b}\!^{{\cre m}+{\cg n}-1}
\left(1 + \frac{{\cre m}({\cre m}-1)}{2}\frac{{\cre N_r}}{{\cb N_b}}
+ \frac{{\cre m}({\cre m}-1)^2({\cre m}-2)}{12}\frac{{\cre N_r}\!^2}{{\cb N_b}\!^2}
+ \frac{({\cre m}+1){\cre m}({\cre m}-1)({\cre m}-2)}{{\cb N_b}\!^2} + \ldots\right)\cdot\nn\\
\cdot\left(1 + \frac{(2{\cre m}+{\cg n}-1){\cg n}}{2{\cre N_r}{\cb N_b}}
+ \frac{{\cg n}({\cg n}-1)(12{\cre m}^2+12{\cre m}{\cg n}+3{\cg n}^2-12{\cre m}-7{\cg n}+2)}
{24{\cre N_r}\!^2{\cb N_b}\!^2}
+\ldots\right)
\ee
Efficient handling of such formulas requires
an adequate multi-color generalization of the FT calculus,
which will be developed elsewhere.

\subsubsection{Direct evaluation of $\Big<{\cal K}_{{\cre 2},{\cg 2}}\Big>$}

In our notation, the operators ${\cal K}_{{\cre m},{\cg 1}} = {\cal K}_{{\cre m}}$,
thus, their Gaussian averages
are just the same as in (\ref{corrZpertr}).
The averages of ${\cal K}_{{\cre 1},{\cg n}} = {\cal K}_{{\cg n}}$ are obtained by the
substitution of ${\cre \alpha} = {\cre N_r}+{\cg N_g}{\cb N_b}$ by
${\cg \alpha} = {\cg N_g}+{\cre N_r}{\cb N_b}$.

\bigskip

Thus, the first non-trivial example is the average of ${\cal K}_{{\cre 2},{\cg 2}}$:

\begin{picture}(300,80)(-120,-40)
\put(-60,-2){\mbox{${\cal K}_{{\cre 2},{\cg 2}} \ \ = $}}
{\cre
\put(0,0){\vector(0,1){20}}
\put(50,20){\vector(0,-1){20}}
\put(50,-20){\line(-1,0){50}}\put(6,-20){\vector(-1,0){2}}
}
\put(-3,0){
{\cg
\put(50,0){\vector(0,-1){20}}
\put(0,20){\line(1,0){50}}\put(44,20){\vector(1,0){2}}
\put(0,-20){\vector(0,1){20}}
}
}
{\cb
\put(50,0){\vector(-1,0){50}}
\qbezier(0,20)(25,35)(50,20)\put(44,23){\vector(3,-1){2}}
\qbezier(50,-20)(25,-35)(0,-20)\put(44,-23){\vector(3,1){2}}
}
\put(-10,-2){\mbox{$0$}}\put(-10,18){\mbox{$\bar 1$}}
\put(55,18){\mbox{$1$}} \put(55,-2){\mbox{$\bar 0$}}
\put(-10,-22){\mbox{$\bar 2$}} \put(55,-22){\mbox{$2$}}
\put(125,0){
\put(-50,-2){\mbox{$=$}}
\put(0,0){\circle{40}}
{\cb \put(20,0){\vector(-1,0){40}}}
\put(-20,0){\circle*{3}}\put(20,0){\circle*{3}}
\put(-10,17){\circle*{3}}\put(10,17){\circle*{3}}
\put(-10,-17){\circle*{3}}\put(10,-17){\circle*{3}}
}
\put(215,0){
\put(-50,-2){\mbox{$=$}}
{\cb \put(20,0){\vector(-1,0){40}}}
\put(-20,20){\mbox{${\cre 2}$}} \put(-20,-25){\mbox{${\cg 2}$}}
\linethickness{0.8mm}
\put(0,0){{\cre\qbezier(-20,0)(-20,20)(0,20)\qbezier(20,0)(20,20)(0,20)} }
\put(0,0){{\cg\qbezier(-20,0)(-20,-20)(0,-20)\qbezier(20,0)(20,-20)(0,-20)} }
}
%
\end{picture}

\noindent
By the Wick theorem, this average
is a sum of six different pairings
\be
\Big<{\cal K}_{{\cre 2},{\cg 2}}\Big> \ = \ \frac{N_rN_gN_b}{\mu^3}\Big(
\underbrace{N_rN_gN_b^2}_{{\cb <0\bar 0>}<1\bar 1><2\bar 2>}
+  \!\!\! \!\!\!\overbrace{N_r^2N_b}^{{\cre <0\bar 1><1\bar 0>}<2\bar 2>} \!\!\! \!\!\!
+  \!\!\! \!\!\!\underbrace{N_g^2N_b}_{{\cg <0\bar 2>}<1\bar 1>{\cg <2\bar 0}>}  \!\!\! \!\!\!
+  \!\!\! \!\!\!\overbrace{N_b}^{{\cb <0\bar 0>}<1\bar 2><2\bar 1>}  \!\!\! \!\!\!
+ \!\!\!\!\!\!\!\! \!\!\! \!\!\!
\underbrace{2 N_rN_g}_{{\cre <0\bar 1>}<1\bar 2>{\cg <2\bar 0>}
+{\cg <0\bar 2>}{\cre <1\bar 0>}<2\bar 1>}
 \!\!\!  \Big)
\label{avK22_01}
\ee

An alternative calculation uses the recursion relation
i.e. the Virasoro constraint at $t=0$.
Namely, eq.(\ref{basicrecursion}) in this case says
\be
3\mu\Big<{\cal K}_{{\cre 2},{\cg 2}}\Big> =
\mu \left<A_{{\cre i_r}{\cb \alpha_b}}^{{\cg j_g}}
\frac{\partial {\cal K}_{{\cre 2},{\cg 2}}}{\partial A_{{\cre i_r}{\cb \alpha_b}}^{{\cg j_g}}}\right> =
\left<\frac{\partial^2{\cal K}_{{\cre 2},{\cg 2}}}{\partial A_{{\cre i_r}{\cb \alpha_b}}^{{\cg j_g}}
\partial \bar A^{{\cre i_r}{\cb \alpha_b}}_{{\cg j_g}}  }\right>
= \sum_{a,b=0}^2 \left<
\frac{\partial^2{\cal K}_{{\cre 2},{\cg 2}}}{\partial A(a)\partial \bar A(\bar b) }  \right>
\ee
where the sum goes over pairs of vertices labeled by
$0,1,2$ and $\bar 0,\bar 1,\bar 2$ in the picture.
Explicitly contracting the indices in these 9 terms, one obtains
\be
3\mu\Big<{\cal K}_{{\cre 2},{\cg 2}}\Big> =
\underbrace{(2{\cre N_r}+{\cg N_g}{\cb N_b})\Big<{\cal K}_{\cg 2}\Big>}
_{<0\bar 1>,<1\bar 0>,<1\bar 1>}
+ \overbrace{(2{\cg N_g}+{\cre N_r}{\cb N_b}) \Big<{\cal K}_{\cre 2}\Big>}
^{<0\bar 2>,<2\bar 0>,<2\bar 2>} +
\underbrace{2 \Big<{\cal K}_{\cb 2}\Big>}_{<1\bar 2>,<2\bar 1>} +
\overbrace{{\cb N_b}\Big<{\cal K}_{\cre 1}{\cal K}_{\cg 1}\Big>}^{<0\bar 0>} =
\nn
\ee
$$
\ \stackrel{(\ref{corrZpertr})}{=} \
\frac{N_rN_gN_b}{\mu^2}\Big( (2{\cre N_r}+{\cg N_g}{\cb N_b})({\cre N_r}{\cb N_b}+{\cg N_g})
+(2{\cg N_g}+{\cre N_r}{\cb N_b})({\cg N_g}{\cb N_b}+{\cre N_r})
+2({\cre N_r}{\cg N_g}+{\cb N_b}) + {\cb N_b}({\cre N_r}{\cg N_g}{\cb N_b}+1)\Big) =
$$
\be
= \frac{3N_rN_gN_b}{\mu^2}\Big({\cre N_r}{\cg N_g}{\cb N_b}\!\!^2
+ {\cre N_r}\!\!^2{\cb N_b} + {\cg N_g}\!\!^2{\cb N_b} + 2{\cre N_r}{\cg N_g}   + {\cb N_b}\Big)
\ee
in full accordance with (\ref{avK22_01}).
We remind that there is a symmetry between $N_r$ and $N_g$, but not $N_b$.

A specifics (simplicity) of this example is that the next-level operator with more blue lines
(two in this case)
arising in the recursion appeared to be equivalent to what we denoted as
${\cal K}_{\cb 2}$, for which we actually now the answer:
its average is the same $\alpha\beta/\mu^2$ as that of ${\cal K}_{\cre 2}$,
with ${\cre \alpha}={\cre N_r} + {\cg N_g}{\cb N_b}$ substituted by
${\cb \alpha}={\cb N_b} + {\cre N_r}{\cg N_g}$.



The FT generating function is remarkably simple in this case
${\cal K}_{{\cre 2},{\cg 2}}$:
\be
\boxed{
\sum_{{\cre N_r},{\cg N_g},{\cb N_b}}
{\cre \lambda_r^{ N_r}}{\cg \lambda_g^{ N_g}}{\cb \lambda_b^{ N_b}}\cdot
\Big<{\cal K}_{{\cre 2},{\cg 2}}^{{\cre N_r}\times{\cg N_g}\times{\cb N_b}}\Big>
= \frac{6{\cre \lambda_r}{\cg \lambda_g}{\cb \lambda_b}}{(1-{\cre \lambda_r})^4
(1-{\cg \lambda_g})^4(1-{\cb \lambda_b})^4}\Big((1-{\cre \lambda_r}{\cg \lambda_g})^2
- ({\cre \lambda_r}-{\cg \lambda_g})^2{\cb\lambda_b}^2\Big)
}
\ee
but it becomes sophisticated for averages with higher ${\cre m}$ and ${\cg n}$.

\subsubsection{Some other Gaussian averages $\Big<{\cal K}_{{\cre m},{\cg n}}\Big>$
and their reductions}

At ${\cg N_g}=1$ the average of ${\cal K}_{{\cre m},{\cg n}}$ becomes equivalent
to that of $\Tr (M\bar M)^m \Big(\Tr M\bar M\Big)^{n-1}$  in the rectangular
matrix model with the matrix size ${\cre N_r}\times {\cb N_b}$.
For ${\cre N_r}=1$, the same is true for the ${\cg N_g}\times {\cb N_b}$ model.
Finally, at ${\cb N_b}=1$ we get the equivalence to just $\Tr (M\bar M)^{m+n-1}$
in the  ${\cre N_r}\times {\cg N_g}$ model.
Thus,
we get the following expressions through the {\it single-hook} averages:
\be
\Big<{\cal K}_{{\cre m},{\cg n}}\Big> \ \stackrel{{\cg N_g}=1}{ =}\
{\cal O}_{[{\cre m},1^{{\cg n}-1}]}^{{\cre N_r}\times {\cb N_b}} \nn \\
\nn \\
\Big<{\cal K}_{{\cre m},{\cg n}}\Big> \ \stackrel{{\cre N_r}=1}{ =}\
{\cal O}_{[{\cg n},1^{{\cre m}-1}]}^{{\cg N_g}\times {\cb N_b}}  \nn \\
\nn \\
\Big<{\cal K}_{{\cre m},{\cg n}}\Big> \ \stackrel{{\cb N_b}=1}{ =}\
{\cal O}_{[{\cre m}+{\cg n}-1 ]}^{{\cre N_r}\times {\cg N_g}}
\label{redto2}
\ee
When only one of the three $N$'s is different from unity, the
rectangular model reduces to the vector one and we get a universal answer:
\be
\Big<{\cal K}_{{\cre m},{\cg n}}\Big> \ \stackrel{{\cre N_r}={\cg N_g}=1}{ =}\
{\cb N_b}({\cb N_b}+1)({\cb N_b}+2)\ldots ({\cb N_b}+{\cre m}+{\cg n}-1)
= \frac{({\cb N_b}+{\cre m}+{\cg n}-1)!}{({\cb N_b}-1)!} \nn \\
\Big<{\cal K}_{{\cre m},{\cg n}}\Big> \ \stackrel{{\cre N_r}={\cb N_b}=1}{ =}\
{\cg N_g}({\cg N_g}+1)({\cg N_g}+2)\ldots ({\cg N_g}+{\cre m}+{\cg n}-1)
= \frac{({\cg N_g}+{\cre m}+{\cg n}-1)!}{({\cg N_g}-1)!}
\nn \\
\Big<{\cal K}_{{\cre m},{\cg n}}\Big> \ \stackrel{{\cg N_g}={\cb N_b}=1}{ =}\
{\cre N_r}({\cre N_r}+1)({\cre N_r}+2)\ldots ({\cre N_r}+{\cre m}+{\cg n}-1)
= \frac{({\cre N_r}+{\cre m}+{\cg n}-1)!}{({\cre N_r}-1)!}
\ee
With the help of   (\ref{corrZpertr}), one can check that (\ref{redto2}) is indeed true:

\be
\Big<{\cal K}_{{\cre 2},{\cg 2}}\Big> \ = \
\frac{N_rN_gN_b}{\mu^3}\Big({\cre N_r}{\cg N_g}{\cb N_b}\!\!^2
+ ({\cre N_r}\!\!^2  + {\cg N_g}\!\!^2+1){\cb N_b}
+ 2{\cre N_r}{\cg N_g}    \Big)\nn\\
\begin{array}{cccc}
 N_g=1 && N_rN_b(N_r+N_b)(N_rN_b+2)
 &=  {\cal O}_{[2,1]}^{{\cre N_r}\times {\cb N_b}} \\ \\
N_b=1 && N_rN_g(N_r^2+N_g^2+3N_rN_g+1)
&=  {\cal O}_{[3]}^{{\cre N_r}\times {\cg N_g}}
\end{array}
\label{K22}
\ee

$\bullet$ Similarly for ${\cal K}_{{\cre 3},{\cg 2}}$ we get:

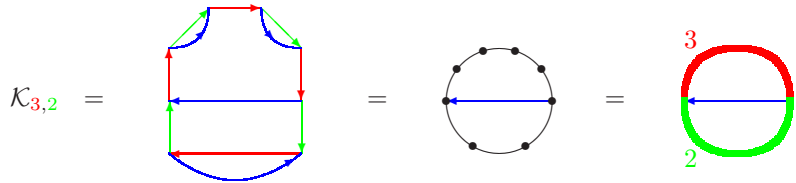
\begin{picture}(300,80)(-120,-40)
\put(-60,-2){\mbox{${\cal K}_{{\cre 3},{\cg 2}} \ \ = $}}
{\cre
\put(0,0){\vector(0,1){20}}\put(15,35){\vector(1,0){20}}\put(50,20){\vector(0,-1){20}}
\put(50,-20){\vector(-1,0){50}}
}
\put(-3,0){
{\cg  \put(0,20){\vector(1,1){15}} \put(35,35){\vector(1,-1){15}}
\put(50,0){\vector(0,-1){20}} \put(0,-20){\vector(0,1){20}}
}
}
{\cb \put(50,0){\vector(-1,0){50}}
\qbezier(0,20)(15,20)(15,35)\put(12,25){\vector(1,1){2}}
\qbezier(50,20)(35,20)(35,35)\put(38,25){\vector(1,-1){2}}
\qbezier(0,-20)(25,-40)(50,-20)\put(46,-23){\vector(2,1){2}}

}
\put(125,0){
\put(-50,-2){\mbox{$=$}}
\put(0,0){\circle{40}}
{\cb \put(20,0){\vector(-1,0){40}}}
\put(-10,-17){\circle*{3}}\put(10,-17){\circle*{3}}
\put(-16,12){\circle*{3}}\put(16,12){\circle*{3}}
\put(-6,19){\circle*{3}}\put(6,19){\circle*{3}}
\put(-20,0){\circle*{3}}\put(20,0){\circle*{3}}
}
\put(215,0){
\put(-50,-2){\mbox{$=$}}
{\cb \put(20,0){\vector(-1,0){40}}}
\put(-20,20){\mbox{${\cre 3}$}} \put(-20,-25){\mbox{${\cg 2}$}}
\linethickness{0.8mm}
\put(0,0){{\cre\qbezier(-20,0)(-20,20)(0,20)\qbezier(20,0)(20,20)(0,20)} }
\put(0,0){{\cg\qbezier(-20,0)(-20,-20)(0,-20)\qbezier(20,0)(20,-20)(0,-20)} }
}
\end{picture}

\be
\Big<{\cal K}_{{\cre 3},{\cg 2}}\Big> \ = \ \frac{N_rN_gN_b}{\mu^4}\Big(
{\cre N_r}{\cg N_g}\!^2{\cb N_b}\!^3 + (3{\cre N_r}\!^2+{\cg N_g}\!^2+2){\cg N_g}{\cb N_b}\!^2
+ ({\cre N_r}\!^2+5{\cg N_g}\!^2+5){\cre N_r}{\cb N_b} + 3({\cre N_r}\!^2+1){\cg N_g}
\Big)
\nn \\
\begin{array}{cccc}
N_g=1 && N_rN_b(N_rN_b+3)(N_r^2+N_b^2+3N_rN_b+1)
&=   {\cal O}_{[3,1]}^{{\cre N_r}\times {\cb N_b}}\\ \\
N_r=1 && N_gN_b(N_g+N_b)(N_gN_b+2)(N_gN_b+3)
&=   {\cal O}_{[2,1,1]}^{{\cg N_g}\times {\cb N_b}}\\ \\
N_b=1 &&  N_rN_g(N_r+N_g)\Big( N_r^2+N_g^2+5N_rN_g+5\Big)
&=   {\cal O}_{[4]}^{{\cre N_r}\times {\cg N_g}}
\end{array}
\ee
Note that for $N_b\neq 1$ there is no symmetry between $N_r$ and $N_g$ in this case:
by exchanging $N_r$ and $N_g$, one gets the expression for $\Big<{\cal K}_{{\cre 2},{\cg 3}}\Big>$.
In particular, in this case, we would get
\be
\Big<{\cal K}_{{\cre 2},{\cg 3}}\Big> \ = \ \frac{N_rN_gN_b}{\mu^4}\Big(
{\cre N_r}\!^2{\cg N_g} {\cb N_b}\!^3 + ({\cre N_r}\!^2+3{\cg N_g}\!^2+2){\cre N_r}{\cb N_b}\!^2
+ (5{\cre N_r}\!^2+{\cg N_g}\!^2+5){\cg N_g}{\cb N_b} + 3({\cg N_g}\!^2+1){\cre N_r}
\Big)
\nn\\
\begin{array}{cccc}
N_g=1 && N_rN_b(N_r+N_b)(N_rN_b+2)(N_rN_b+3)
&=  {\cal O}_{[2,1,1]}^{{\cre N_r}\times {\cb N_b}} \\ \\
N_r=1 && N_gN_b(N_gN_b+3)(N_g^2+N_b^2+3N_gN_b+1)
&=  {\cal O}_{[3,1]}^{{\cg N_g}\times {\cb N_b}}\\ \\
\end{array}
\ee

\bigskip

$\bullet$ For ${\cal K}_{{\cre 3},{\cg 3}}$ we obtain:

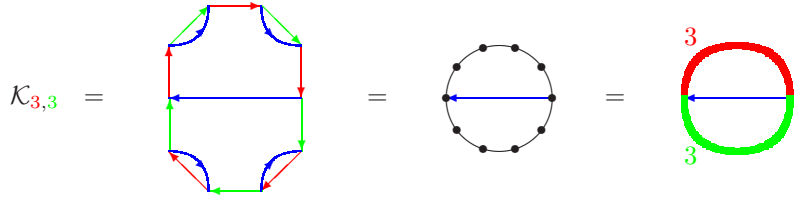
\begin{picture}(300,80)(-120,-40)
\put(-60,-2){\mbox{${\cal K}_{{\cre 3},{\cg 3}} \ \ = $}}
{\cre
\put(0,0){\vector(0,1){20}}\put(15,35){\vector(1,0){20}}\put(50,20){\vector(0,-1){20}}
\put(50,-20){\vector(-1,-1){15}}\put(15,-35){\vector(-1,1){15}}
}
\put(-3,0){
{\cg  \put(0,20){\vector(1,1){15}} \put(35,35){\vector(1,-1){15}}
\put(50,0){\vector(0,-1){20}}\put(35,-35){\vector(-1,0){20}}\put(0,-20){\vector(0,1){20}}
}
}
{\cb \put(50,0){\vector(-1,0){50}}
\qbezier(0,20)(15,20)(15,35)\put(12,25){\vector(1,1){2}}
\qbezier(50,20)(35,20)(35,35)\put(38,25){\vector(1,-1){2}}
\qbezier(0,-20)(15,-20)(15,-35)\put(12,-25){\vector(1,-1){2}}
\qbezier(50,-20)(35,-20)(35,-35)\put(38,-25){\vector(1,1){2}}
}
\put(125,0){
\put(-50,-2){\mbox{$=$}}
\put(0,0){\circle{40}}
{\cb \put(20,0){\vector(-1,0){40}}}
\put(-16,12){\circle*{3}}\put(16,12){\circle*{3}}
\put(-6,19){\circle*{3}}\put(6,19){\circle*{3}}
\put(-20,0){\circle*{3}}\put(20,0){\circle*{3}}
\put(-16,-12){\circle*{3}}\put(16,-12){\circle*{3}}
\put(-6,-19){\circle*{3}}\put(6,-19){\circle*{3}}
}
\put(215,0){
\put(-50,-2){\mbox{$=$}}
{\cb \put(20,0){\vector(-1,0){40}}}
\put(-20,20){\mbox{${\cre 3}$}} \put(-20,-25){\mbox{${\cg 3}$}}
\linethickness{0.8mm}
\put(0,0){{\cre\qbezier(-20,0)(-20,20)(0,20)\qbezier(20,0)(20,20)(0,20)} }
\put(0,0){{\cg\qbezier(-20,0)(-20,-20)(0,-20)\qbezier(20,0)(20,-20)(0,-20)} }
}
\end{picture}

\be
\Big<{\cal K}_{{\cre 3},{\cg 3}}\Big> \ = \ \frac{N_rN_gN_b}{\mu^5}\Big(
{\cre N_r}\!^2{\cg N_g}\!^2{\cb N_b}\!^4
+ \big(3{\cre N_r}\!^2+3{\cg N_g}\!^2+4\big){\cre N_r}{\cg N_g}{\cb N_b}\!^3 +
\nn
\ee
\be
+\big({\cre N_r}\!^4+{\cg N_g}\!^4+13{\cre N_r}\!^2{\cg N_g}\!^2
+  9{\cre N_r}\!^2+9{\cg N_g}\!^2 + 2\big){\cb N_b}\!^2
+\big(7{\cre N_r}\!^2   +  7{\cg N_g}\!^2 +36\big){\cre N_r}{\cg N_g}{\cb N_b}
+   6({\cre N_r}\!^2+1)({\cg N_g}\!^2+1)
\Big)
\nn
\ee
\vspace{-0.5cm}
\be
\begin{array}{cccc}
N_g=1 && N_rN_b(N_rN_b+3)(N_rN_b+4)(N_r^2+N_b^2+3N_bN_r+1)
&=  {\cal O}_{[3,1,1]}^{{\cre N_r}\times {\cb N_b}}\\ \\
N_b=1 && N_rN_g\Big(N_r^4+10N_r^3N_g+20N_r^2N_g^2+10N_rN_g^3+N_g^4
 + 15N_r^2+40N_rN_g+15N_g^2+8 \Big)
& =  {\cal O}_{[5]}^{{\cre N_r}\times {\cg N_g}}
\end{array}
\ee

\bigskip

\be
\Big<{\cal K}_{{\cre 4},{\cg 2}}\Big> \ = \ \frac{N_rN_gN_b}{\mu^5}\Big(
{\cre N_r}{\cg N_g}\!^3{\cb N_b}\!^4 +(6{\cre N_r}\!^2 +{\cg N_g}\!^2 +3 )
{\cg N_g}\!^2 {\cb N_b}\!^3
+(6{\cre N_r}\!^2+9{\cg N_g}\!^2+20){\cre N_r}{\cg N_g} {\cb N_b}\!^2 + \nn \\
+ ({\cre N_r}\!^4+14{\cre N_r}\!^2{\cg N_g}\!^2+15{\cre N_r}\!^2 + 12{\cg N_g}\!^2+8)
 {\cb N_b}
+4{\cre N_r}{\cg N_g}({\cre N_r}\!^2+5)
\Big)
 \nn
\ee

\vspace{-0.5cm}

\be
\begin{array}{cccc}
N_g=1 && N_rN_b(N_rN_b+4)(N_r^2+N_b^2+5N_rN_b+5)
&=   {\cal O}_{[4,1]}^{{\cre N_r}\times {\cb N_b}}\\ \\
N_r=1 && N_gN_b(N_g+N_b)(N_gN_b+2)(N_gN_b+3)(N_gN_b+4)
&=   {\cal O}_{[2,1,1,1]}^{{\cg N_g}\times {\cb N_b}}\\ \\
N_b=1 &&  N_rN_g\Big(N_r^4+10N_r^3N_g+20N_r^2N_g^2+10N_rN_g^3+N_g^4
 + 15N_r^2+40N_rN_g+15N_g^2+8 \Big)
&=   {\cal O}_{[5]}^{{\cre N_r}\times {\cg N_g}}
\end{array}
\ee
Again, for $N_b\neq 1$ there is no symmetry between $N_r$ and $N_g$ in this case:
by exchanging $N_r$ and $N_g$, one gets the expression for $\Big<{\cal K}_{{\cre 2},{\cg 4}}\Big>$.

\bigskip

$$
\Big<{\cal K}_{{\cre 4},{\cg 3}}\Big> \ = \ \frac{N_rN_gN_b}{\mu^6}\Big(
{\cre N_r}\!^2{\cg N_g}\!^3{\cb N_b}\!^5
+3(2{\cre N_r}\!^2+{\cg N_g}\!^2+2){\cre N_r}{\cg N_g}\!^2{\cb N_b}\!^4 +
$$
$$
+(6{\cre N_r}\!^4+{\cg N_g}\!^4+24{\cre N_r}\!^2{\cg N_g}\!^2+35{\cre N_r}\!^2+13{\cg N_g}\!^2+6)
{\cg N_g}{\cb N_b}\!^3
+({\cre N_r}\!^4+12{\cg N_g}\!^4+34{\cre N_r}\!^2{\cg N_g}\!^2+25{\cre N_r}\!^2+119{\cg N_g}\!^2+34)
{\cre N_r}{\cb N_b}\!^2  +
$$
$$
+(9{\cre N_r}\!^4+25{\cre N_r}\!^2{\cg N_g}\!^2+140{\cre N_r}\!^2+22{\cg N_g}\!^2+78)
{\cg N_g}{\cb N_b}
+10({\cre N_r}\!^2+5)({\cg N_g}\!^2+1){\cre N_r}\Big)
\nn
$$
\vspace{-0.5cm}
\be
\begin{array}{cccc}
N_g=1 && N_rN_b(N_r+N_b)(N_rN_b+4)(N_rN_b+5)(N_r^2+5N_bN_r+N_b^2+5)
&=  {\cal O}_{[4,1,1]}^{{\cre N_r}\times {\cb N_b}}\\ \\
N_r=1 && N_gN_b (N_gN_b+3)(N_gN_b+4)(N_gN_b+5)(N_g^2+3N_gN_b+N_b^2+1)
&=  {\cal O}_{[3,1,1,1]}^{{\cg N_g}\times {\cb N_b}}\\ \\
N_b=1 &&  \!\!\!\! N_rN_g(N_r+N_g)\Big( N_r^4+14N_r^3N_g+36N_r^2N_g^2+14N_rN_g^3+N_g^4
+35N_r^2+140N_rN_g+35N_g^2+84\Big)
&=   {\cal O}_{[6]}^{{\cre N_r}\times {\cg N_g}}
\end{array}
\ee

\subsubsection{Wheel operators}

We consider also the wheel operators:

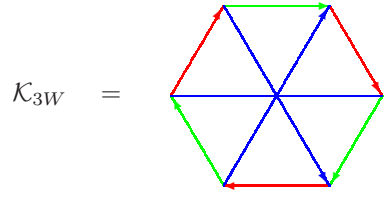
\begin{picture}(300,100)(-190,-50)

\put(-60,-2){\mbox{${\cal K}_{3W}\ \ \ = $}}

\put(0,0){\cre
\qbezier(0,0)(10,17)(20,34) \qbezier(80,0)(70,17)(60,34) \put(60,-34){\vector(-1,0){40}}
\put(17,29){\vector(1,2){2}}\put(77,5){\vector(1,-2){2}}
}
\put(0,0){\cg
\qbezier(0,0)(10,-17)(20,-34) \qbezier(80,0)(70,-17)(60,-34) \put(20,34){\vector(1,0){40}}
\put(63,-29){\vector(-1,-2){2}}\put(3,-5){\vector(-1,2){2}}
}
\put(0,0){\cb
\put(80,0){\line(-1,0){80}}
\qbezier(20,34)(40,0)(60,-34)\qbezier(20,-34)(40,0)(60,34)
\put(57,-29){\vector(1,-2){2}} \put(57,29){\vector(1,2){2}}
}

\end{picture}

They are totally symmetric in the three colors and they can have only odd lengths $m$.
The first non-trivial example after ${\cal K}_{1W} = {\cal K}_1$ with the average
$\Big< {\cal K}_{1W} \Big> = N_rN_gN_b$ and the FT function
\be
\sum_{{\cre N_r},{\cg N_g},{\cb N_b}} {\cre \lambda_r^{N_r}}{\cg \lambda_g^{N_g}}
{\cb \lambda_b^{N_b}}\cdot
\Big< {\cal K}_{1W}^{{\cre N_r}\times{\cg N_g}\times{\cb N_b}} \Big>
= \frac{{\cre \lambda_r}{\cg \lambda_g}{\cb \lambda_b}}
{(1-{\cre \lambda_r})^2(1-{\cg \lambda_g})^2(1-{\cb \lambda_b})^2}
\ee
is
\be
\Big< {\cal K}_{3W} \Big> = \frac{N_rN_gN_b}{\mu^3}\Big(3{\cre N_r}{\cg N_g}{\cb N_b} +
{\cre N_r}\!^2+{\cg N_g}\!^2 + {\cb N_b}\!^2\Big)
\label{K3W}
\ee
Unfortunately, the FT formula is somewhat long in this case.

When any of the color numbers is one, this operators turns into an ordinary circle in the rectangular model with the remaining two colorings:
\be
{\cg N_g=1} && \Big< {\cal K}_{mW} \Big> = {\cal O}^{{\cre N_r}\times {\cb N_b}}_{[m]}\nn\\
{\cre N_r=1} && \Big< {\cal K}_{mW} \Big> = {\cal O}^{{\cg N_g}\times {\cb N_b}}_{[m]}\nn\\
{\cb N_b=1} && \Big< {\cal K}_{mW} \Big> = {\cal O}^{{\cre N_r}\times {\cg N_g}}_{[m]}
\ee
Thus, the average of the wheel operators is the three-coloring continuation of the basic
two-color averages in the first column of the table in s.\ref{averc}.

\subsubsection{Examples of recursion checks}

We have now enough explicit formulas to check the non-trivial example (\ref{recmn})
of the recursion in the Aristotelian model.
For example, for ${\cre m}=2$ and ${\cg n}=2$ we get:
\be
4\mu\cdot \Big< {\cal K}_{\cre 2}{\cal K}_{\cg 2}\Big>
= 8\cdot\Big<{\cal K}_{{\cre 2},{\cg 2}}\Big>
+ 2{\cre \alpha}\cdot
\overbrace{\Big< {\cal K}_{\cg 2}{\cal K}_{\cg 1}\Big>}^{
\Big< {\cal K}_{\cre 1}{\cal K}_{\cg 2}\Big>} +
2{\cg \alpha}\cdot
\overbrace{\Big< {\cal K}_{\cre 2}{\cal K}_{\cre 1}\Big>}^{
\Big< {\cal K}_{\cre 2}{\cal K}_{\cg 1}\Big>}
\ee
Note that the color of the unit circle operator ${\cal K}_1$ in the last two terms should
be adjusted to that of ${\cal K}_2$, if one wants to apply the  prescription from the
end of s.\ref{rorgrings}.
Substituting the averages from (\ref{K2K2}), (\ref{K22}) and (\ref{corrZpertr})
we obtain the identity
(the overall factor $4\beta\mu^{-3}$ is omitted):
\be
N_r^2N_g^2N_b^3 + (N_r^2+N_g^2+4)N_rN_gN_b^2 + (N_r^2N_g^2+4N_r^2+4N_g^2+2)N_b+6N_rN_g =\nn \\
= 2\Big(N_rN_gN_b^2+(N_r^2+N_g^2+1)N_b+2N_rN_g\Big)
+ (N_r+N_gN_b)(N_g+N_rN_b)(N_rN_gN_b+2)
\ee
Likewise, for ${\cre m}=3$ and ${\cg n}=2$
\be
5\mu\cdot \Big< {\cal K}_{\cre 3}{\cal K}_{\cg 2}\Big> =
12\cdot \Big<{\cal K}_{{\cre 3},{\cg 2}}\Big>
+ 3{\cre \alpha}\cdot \Big< {\cal K}_{\cre 2}{\cal K}_{\cg 2}\Big>
+ 3 \cdot \overbrace{\Big< {\cal K}_{\cg 2}{\cal K}_{\cg 1}{\cal K}_{\cg 1}\Big>}^{
\Big< {\cal K}_{\cre 1}{\cal K}_{\cre 1}{\cal K}_{\cg 2}\Big>}
+2{\cg \alpha}\cdot \overbrace{\Big< {\cal K}_{\cre 3}{\cal K}_{\cre 1}\Big>}^{
\Big< {\cal K}_{\cre 3}{\cal K}_{\cg 1}\Big>}
\ee
what is indeed true:
\be
5\cdot\Big({\cre N_r}\!^2 {\cg N_g}\!^3 {\cb N_b}\!^4
+(3 {\cre N_r}\!^2+{\cg N_g}\!^2+6) {\cre N_r} {\cg N_g}\!^2 {\cb N_b}\!^3 +\nn \\
+({\cre N_r}\!^4+ 3 {\cre N_r}\!^2 {\cg N_g}\!^2 +19 {\cre N_r}\!^2+6 {\cg N_g}\!^2+6)
{\cg N_g} {\cb N_b}\!^2
+( {\cre N_r}\!^2 {\cg N_g}\!^2+6{\cre N_r}\!^2+25{\cg N_g}\!^2+18){\cre N_r} {\cb N_b}
+12({\cre N_r}\!^2+1){\cg N_g}\Big) = \nn \\
= 12\cdot \Big({\cre N_r}{\cg N_g}\!^2{\cb N_b}\!^3 + (3{\cre N_r}\!^2+{\cg N_g}\!^2+2)
{\cg N_g}{\cb N_b}\!^2
+ ({\cre N_r}\!^2+5{\cg N_g}\!^2+5){\cre N_r}{\cb N_b} + 3({\cre N_r}\!^2+1){\cg N_g}
\Big) + \nn \\
+ 3\cdot ({\cre N_r}+{\cg N_g}{\cb N_b})
\Big({\cre N_r}\!^2{\cg N_g}\!^2{\cb N_b}\!^3
+ ({\cre N_r}\!^2+{\cg N_g}\!^2+4){\cre N_r}{\cg N_g}{\cb N_b}\!^2
+ ({\cre N_r}\!^2{\cg N_g}\!^2+4{\cre N_r}\!^2+4{\cg N_g}\!^2+2){\cb N_b}\!
+6{\cre N_r}{\cg N_g}\Big)
+ \nn \\
+ ({\cg N_g}+{\cre N_r}{\cb N_b})(N_rN_gN_b+3)
\Big(3\cdot  (N_rN_gN_b+2)
+ 2\cdot \left(({\cre N_r}+{\cg N_g}{\cb N_b})^2+N_rN_gN_b+1\right)\Big)
\ee

\bigskip

When ${\cg N_g}=1$, the operators ${\cal K}_{\cre m}$, ${\cal K}_{\cg n}$ and
${\cal K}_{{\cre m} ,{\cg n}}$
turn respectively into  ${\cal K}_{{\cre m}}$,
$ {\cal K}_{1}^{\cg n}$ and ${\cal K}_{{\cre m}}{\cal K}_{{\cre 1}}^{\cg n-1}$
of the ${\cre N_r}\times{\cb N_b}$
rectangular matrix model, and the recursion relation (\ref{recmn}) reduces to
\be
({\cre m}+{\cg n})\,\mu\cdot {\cal O}^{{\cre N_r}\times{\cb N_b}}_{[{\cre m},1^{\cg n}]} =
2{\cre m}{\cg n}\cdot  {\cal O}^{{\cre N_r}\times{\cb N_b}}_{[{\cre m},1^{\cg n-1}]} + \nn \\
+{\cre m} \,{\cre \alpha}\cdot {\cal O}^{{\cre N_r}\times{\cb N_b}}_{[{\cre m-1},1^{\cg n}]}
+ {\cre m}\cdot \!\!\!\!\!\!\!\!\!\sum_{\stackrel{m_1,m_2\geq 1}{m_1+m_2=m-1}}\!\!\!
 {\cal O}^{{\cre N_r}\times{\cb N_b}}_{[{\cre m_1,m_2},1^{\cg n}]}
+{\cg n}\,{\cg \alpha}\cdot{\cal O}^{{\cre N_r}\times{\cb N_b}}_{[{\cre m},1^{\cg n-1}]}
+ {\cg n}({\cg n}-2)\cdot
{\cal O}^{{\cre N_r}\times{\cb N_b}}_{[{\cre m},1^{\cg n-1}]}
\label{recmnG1}
\ee
For example, at ${\cg n}=1$ we get:
\be
(m+1)\,\mu\cdot {\cal O}_{[m,1]}=(2m+\beta)\cdot {\cal O}_{[m]} + m\alpha\cdot {\cal O}_{[m-1,1]}
+ m\cdot \!\!\!\!\!\!\sum_{\stackrel{m_1,m_2\geq 1}{m_1+m_2=m-1}}\!\!\! {\cal O}_{[m_1,m_2,1]}
\ee
(note that for ${\cg n}=1$ the sum ${\cg \alpha} + {\cg n}-2 = {\cre N_r}{\cb N_b} = \beta$).
The factorization property (\ref{factRCM}) reduces this to
$m(m+\beta-1)$ times the basic recursion of the rectangular
complex model,
\be
\mu\cdot{\cal O}_{m} = \alpha\cdot {\cal O}_{m-1}+
\!\!\!\!\!\!\!\!\!\sum_{\stackrel{m_1,m_2\geq 1}{m_1+m_2=m-1}}\!\!\! {\cal O}_{[m_1,m_2]}
\ee
see  s.\ref{recuRCM}.
The data from s.\ref{exavRCM} can be used to check other particular cases  (\ref{recmnG1}).

Similarly, at ${\cre N_r}=1$ we have
\be
({\cre m}+{\cg n})\,\mu\cdot {\cal O}^{{\cg N_g}\times{\cb N_b}}_{[{\cg n},1^{\cre m}]} =
2{\cre m}{\cg n}\cdot  {\cal O}^{{\cg N_g}\times{\cb N_b}}_{[{\cg n},1^{\cre m-1}]} + \nn \\
+{\cre m} \,{\cre \alpha}\cdot {\cal O}^{{\cg N_g}\times{\cb N_b}}_{[{\cg n},1^{\cre m-1}]}
+ {\cre m}({\cre m}-2)\cdot
 {\cal O}^{{\cg N_g}\times{\cb N_b}}_{[{\cg n},1^{\cre m-1}]}
+{\cg n}\,{\cg \alpha}\cdot{\cal O}^{{\cg N_g}\times{\cb N_b}}_{[{\cg n},1^{\cre m-1}]}
+  {\cg n}\cdot\!\!\!\!\!\!\!\!\!\sum_{\stackrel{n_1,n_2\geq 1}{n_1+n_2=n-1}}\!\!\!
{\cal O}^{{\cg N_g}\times{\cb N_b}}_{[{\cg n_1,n_2},1^{\cre m-1}]}
\ee

For ${\cb N_b}=1$ the operators ${\cal K}_{\cre m}$, ${\cal K}_{\cg n}$ and
${\cal K}_{{\cre m} ,{\cg n}}$
turn into  ${\cal K}_{{\cre m}}$,
$ {\cal K}_{\cg n}$ and ${\cal K}_{{\cre m}+{\cg n}-1} $
of the ${\cre N_r}\times{\cg N_g}$ model, and  (\ref{recmn}) becomes
\be
({\cre m}+{\cg n})\,\mu\cdot {\cal O}^{{\cre N_r}\times{\cg N_g}}_{[{\cg n},{\cre m}]} =
2{\cre m}{\cg n}\cdot  {\cal O}^{{\cre N_r}\times{\cg N_g}}_{[{\cre m}+{\cg n}-1]} + \nn \\
+{\cre m} \,{\cre \alpha}\cdot {\cal O}^{{\cre N_r}\times{\cg N_g}}_{[{\cre m}-1,{\cg n}]}
+ {\cre m}\cdot \!\!\!\!\!\!\!\!\!\sum_{\stackrel{m_1,m_2\geq 1}{m_1+m_2=m-1}}\!\!\!
 {\cal O}^{{\cre N_r}\times{\cg N_g}}_{[{\cre m_1,m_2},{\cg n}]}
+{\cg n}\,{\cg \alpha}\cdot{\cal O}^{{\cre N_r}\times{\cg N_g}}_{[{\cre m},{\cg n}-1]}
+  {\cg n}\cdot\!\!\!\!\!\!\!\!\!\sum_{\stackrel{n_1,n_2\geq 1}{n_1+n_2=n-1}}\!\!\!
{\cal O}^{{\cre N_r}\times{\cg N_g}}_{[{\cre m},{\cg n_1,n_2}]}
\ee

\subsection{On RG completion of the Aristotelian model
\label{rgcomplrg}}

To study the problem of RG completion in this model, we introduce an
additional notation.

The red and green propagators are

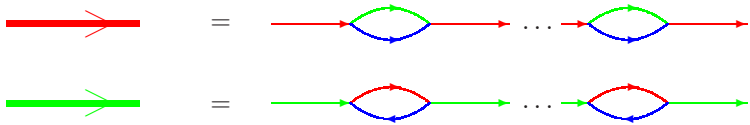
\begin{picture}(300,70)(-50,-50)

\put(100,0){
\put(0,0){\cre \put(0,0){\vector(1,0){30}} \put(60,0){\vector(1,0){30}}  \put(150,0){\vector(1,0){30}}
\put(110,0){\vector(1,0){10}}  }
\put(0,0){\cg  \qbezier(30,0)(45,12)(60,0) \qbezier(120,0)(135,12)(150,0)
\put(43,6){\vector(1,0){6}}\put(133,6){\vector(1,0){6}} }
\put(0,0){\cb  \qbezier(30,0)(45,-12)(60,0) \qbezier(120,0)(135,-12)(150,0)
\put(43,-6){\vector(1,0){6}}\put(133,-6){\vector(1,0){6}} }
 \put(95,-2){\mbox{$\ldots$}}
}
\put(77,-2){\mbox{$=$}}
\put(100,-30){
\put(0,0){\cg \put(0,0){\vector(1,0){30}} \put(60,0){\vector(1,0){30}}  \put(150,0){\vector(1,0){30}}
\put(110,0){\vector(1,0){10}} }
\put(0,0){\cre  \qbezier(30,0)(45,12)(60,0) \qbezier(120,0)(135,12)(150,0)
\put(43,6){\vector(1,0){6}}\put(133,6){\vector(1,0){6}}}
\put(0,0){\cb  \qbezier(30,0)(45,-12)(60,0) \qbezier(120,0)(135,-12)(150,0)
\put(49,-6){\vector(-1,0){6}}\put(139,-6){\vector(-1,0){6}}}
 \put(95,-2){\mbox{$\ldots$}}
}
\put(77,-33){\mbox{$=$}}

\put(40,0){\cre
\put(0,0){\line(-2,1){10}}\put(0,0){\line(-2,-1){10}}
}

\put(40,-30){\cg
\put(0,0){\line(-2,1){10}}\put(0,0){\line(-2,-1){10}}
}

\linethickness{0.8mm}
\put(0,0){\cre  \put(0,0){\line(1,0){50}} }
\put(0,0){\cg  \put(0,-30){\line(1,0){50}} }

\end{picture}

\noindent
We denote them by the thick lines, but these are not propagators in Feynman diagrams,
where the thick lines are multi-colored tubes/cables,
here these are just the "dressed" thin-line trees.

The main operation in taking averages with the help of the Wick theorem is
connecting two vertices by a Feynman propagator, i.e. just eliminating the two
vertices and connecting the lines inside them.

If we consider a diagram which is a set of thick red and green lines
meeting at the vertices and further connected by thin blue lines,
then we should distinguish between five cases:

\bigskip

(a) merged are two vertices inside a thick propagator

(b) merged are two vertices in two thick propagators of the same color

(c) merged are two vertices in two thick propagators of two different colors

(d) merged are two inter-propagator vertices

(e) merged are the inter-propagator vertex and that inside a thick propagator.

\bigskip

\noindent
It is easy to see that

\bigskip

(a) leads to decoupling of a closed piece of the thick line,
i.e. of the average $\Big<{\cal K}_{{\cre m_2}}\Big>$

(b) leads to overcrossing  of the two propagators

(c) leads to emerging of two new inter-propagator vertices connected by a blue line

(d) connects the two remote vertices and releases two thick propagators

(e) exchanges a piece of the thick propagator

\bigskip

\noindent
Pictorially:

\begin{picture}(300,600)(-100,-560)

\put(-100,-2){\mbox{$(a)$}}
\put(100,-2){\mbox{$\longrightarrow$}}

\put(-100,-122){\mbox{$(b)$}}
\put(100,-122){\mbox{$\longrightarrow$}}

\put(-100,-242){\mbox{$(c)$}}
\put(100,-242){\mbox{$\longrightarrow$}}

\put(-100,-362){\mbox{$(d)$}}
\put(100,-362){\mbox{$\longrightarrow$}}

\put(-100,-482){\mbox{$(e)$}}
\put(100,-482){\mbox{$\longrightarrow$}}

\put(5,-15){\mbox{$m_1$}} \put(30,-15){\mbox{$m_2$}} \put(60,-15){\mbox{$m_3$}}
\put(135,25){\mbox{$m_1+m_3-1$}} \put(130,-17){\mbox{$m_2$}}

\put(0,-120){
\put(60,15){\cre
\put(0,0){\line(-2,1){10}}\put(0,0){\line(-2,-1){10}}
}
\put(20,-15){\cre
\put(0,0){\line(2,-1){10}}\put(0,0){\line(2,1){10}}
}
\put(191,15){\cre
\put(0,0){\line(-2,1){10}}\put(0,0){\line(-2,-1){10}}
}
\put(143,-15){\cre
\put(0,0){\line(2,-1){10}}\put(0,0){\line(2,1){10}}
}
\put(15,20){\mbox{$m_1$}}\put(60,20){\mbox{$m_2$}}
\put(15,-25){\mbox{$m_3$}}\put(60,-25){\mbox{$m_4$}}
\put(115,20){\mbox{$m_1+m_3$}}\put(190,20){\mbox{$m_2+m_4$}}
}

\put(0,-240){
\put(0,0){{\cb \put(160,0){\vector(1,0){15}}}}
\put(60,15){\cre
\put(0,0){\line(-2,1){10}}\put(0,0){\line(-2,-1){10}}
}
\put(20,-15){\cg
\put(0,0){\line(2,-1){10}}\put(0,0){\line(2,1){10}}
}
\put(195,15){\cre
\put(0,0){\line(-2,1){10}}\put(0,0){\line(-2,-1){10}}
}
\put(143,-15){\cg
\put(0,0){\line(2,-1){10}}\put(0,0){\line(2,1){10}}
}
\put(15,20){\mbox{$m_1$}}\put(60,20){\mbox{$m_2$}}
\put(15,-25){\mbox{$n_2$}}\put(60,-25){\mbox{$n_1$}}
\put(115,20){\mbox{$m_1+n_2$}}\put(190,20){\mbox{$n_1+m_2$}}
}

\put(0,-360){
\put(0,0){{\cb \put(-10,0){\vector(1,0){20}}}}
\put(0,0){{\cb \put(40,0){\vector(1,0){20}}}}
\put(0,0){{\cb \put(170,0){\vector(1,0){20}}}}
\put(-20,20){\mbox{$m_1$}}\put(5,20){\mbox{$m_2$}}
\put(35,20){\mbox{$m_3$}}\put(65,20){\mbox{$m_4$}}
\put(145,20){\mbox{$m_1$}}\put(165,30){\mbox{$m_2+m_3$}}\put(210,20){\mbox{$m_4$}}
\put(-20,-25){\mbox{$n_1$}}\put(5,-25){\mbox{$n_2$}}
\put(35,-25){\mbox{$n_3$}}\put(65,-25){\mbox{$n_4$}}
\put(145,-25){\mbox{$n_1$}}\put(165,-35){\mbox{$n_2+n_3$}}\put(210,-25){\mbox{$n_4$}}
\put(11.5,15){\cre
\put(0,0){\line(-1,-2){5}}\put(0,0){\line(1,-2){5}}
}
\put(41.5,5){\cre
\put(0,0){\line(-1,2){5}}\put(0,0){\line(1,2){5}}
}
\put(-15,7){\cre
\put(0,0){\line(-1,0){8}}\put(0,0){\line(0,1){8}}
}
\put(74,12){\cre
\put(0,0){\line(-1,0){8}}\put(0,0){\line(0,-1){8}}
}
\put(165,7){\cre
\put(0,0){\line(-1,0){8}}\put(0,0){\line(0,1){8}}
}
\put(204,12){\cre
\put(0,0){\line(-1,0){8}}\put(0,0){\line(0,-1){8}}
}
\put(177,15.5){\cre
\put(0,0){\line(2,-1){10}}\put(0,0){\line(2,1){10}}
}
}

\put(0,-480){
\put(0,0){{\cb \put(-10,0){\vector(1,0){20}}}}
\put(0,0){{\cb \put(155,0){\vector(1,0){20}}}}
\put(-20,20){\mbox{$m_1$}}\put(5,20){\mbox{$m_2$}}
\put(45,20){\mbox{$m_3$}}
\put(145,20){\mbox{$m_1$}}\put(175,20){\mbox{$m_3$}}\put(210,20){\mbox{$m_2+m_4$}}
\put(-20,-25){\mbox{$n_1$}}\put(5,-25){\mbox{$n_2$}}
\put(45,-25){\mbox{$m_4$}}
\put(145,-25){\mbox{$n_1$}}\put(175,-25){\mbox{$n_2 $}}
\put(11.5,15){\cre
\put(0,0){\line(-1,-2){5}}\put(0,0){\line(1,-2){5}}
}
\put(51.5,15){\cre
\put(0,0){\line(-1,-2){5}}\put(0,0){\line(1,-2){5}}
}
\put(-15,7){\cre
\put(0,0){\line(-1,0){8}}\put(0,0){\line(0,1){8}}
}
\put(163.5,0){
\put(11.5,15){\cre
\put(0,0){\line(-1,-2){5}}\put(0,0){\line(1,-2){5}}
}
\put(51.5,15){\cre
\put(0,0){\line(-1,-2){5}}\put(0,0){\line(1,-2){5}}
}
\put(-15,7){\cre
\put(0,0){\line(-1,0){8}}\put(0,0){\line(0,1){8}}
}
}
}

\put(-30,-60){\mbox{$
\acontraction{\Big(\Tr}{{\partial\over\partial \bar A}}{{\partial\over\partial A}\Big)\Big[
\ldots A_{{\cre i}}^{{\cg p}{\cb a}}}{
\bar {A}}
\acontraction[2ex]{\Big(\Tr{\partial\over\partial \bar A}}{{\partial\over\partial A}}{\Big)\Big[
\ldots A_{{\cre i}}^{{\cg p}{\cb a}}
\bar {A}^{{\cre k}}_{{\cg p}{\cb a}}\hat O^{{\cre j}}_{{\cre k}}}
{A}
\Big(\Tr{\partial\over\partial \bar A}{\partial\over\partial A}\Big)\Big[
\ldots A_{{\cre i}}^{{\cg p}{\cb a}}
\bar {A}^{{\cre k}}_{{\cg p}{\cb a}}\hat O^{{\cre j}}_{{\cre k}}
{A}_{{\cre j}}^{{\cg q}{\cb b}} \bar A^{{\cre l}}_{{\cg q}{\cb b}}\ldots\Big]
=\Big[\ldots A_{{\cre i}}^{{\cg p}{\cb a}}\hat O^{{\cre k}}_{{\cre k}} \bar A^{{\cre l}}_{{\cg p}{\cb a}}\ldots\Big]
$}}

\put(-100,-180){\mbox{$
\acontraction{\Big(\Tr}{{\partial\over\partial \bar A}}{{\partial\over\partial A}\Big)\Big[
\ldots A_{{\cre i}}^{{\cg p}{\cb a}}}{
\bar {A}}
\acontraction[2ex]{\Big(\Tr{\partial\over\partial \bar A}}{{\partial\over\partial A}}{\Big)\Big[
\ldots A_{{\cre i}}^{{\cg p}{\cb a}}
\bar {A}^{{\cre k}}_{{\cg p}{\cb a}}\hat O_{{\cre k}}^{{\cre j}}\Big]\Big[\hat O'^{{\cre m}}_{{\cre n}}}
{A}
\Big(\Tr{\partial\over\partial \bar A}{\partial\over\partial A}\Big)\Big[
\ldots A_{{\cre i}}^{{\cg p}{\cb a}}
\bar {A}^{{\cre k}}_{{\cg p}{\cb a}}\hat O_{{\cre k}}^{{\cre j}}\Big]\Big[\hat O'^{{\cre m}}_{{\cre n}}
{A}_{{\cre m}}^{{\cg q}{\cb b}} \bar A^{{\cre l}}_{{\cg q}{\cb b}}\ldots\Big]
=\Big[\ldots A_{{\cre i}}^{{\cg p}{\cb a}} \hat O_{{\cre k}}^{{\cre j}}\Big]\Big[\hat O'^{{\cre k}}_{{\cre n}}\bar A^{{\cre l}}_{{\cg p}{\cb a}}\ldots\Big]=\Big[\ldots A_{{\cre i}}^{{\cg p}{\cb a}}\bar A^{{\cre l}}_{{\cg p}{\cb a}}\ldots \Big]\Big[(\hat O\hat O')^{{\cre j}}_{{\cre n}}\Big]
$}}

\put(-90,-300){\mbox{$
\acontraction{\Big(\Tr}{{\partial\over\partial \bar A}}{{\partial\over\partial A}\Big)\Big[
\ldots A_{{\cre i}}^{{\cg p}{\cb a}}}{
\bar {A}}
\acontraction[2ex]{\Big(\Tr{\partial\over\partial \bar A}}{{\partial\over\partial A}}{\Big)\Big[
\ldots A_{{\cre i}}^{{\cg p}{\cb a}}
\bar {A}^{{\cre i}}_{{\cg q}{\cb a}}\hat O_{{\cg r}}^{{\cg q}}\Big]\Big[\hat O'^{{\cre j}}_{{\cre k}}}
{A}
\Big(\Tr{\partial\over\partial \bar A}{\partial\over\partial A}\Big)\Big[
\ldots A_{{\cre i}}^{{\cg p}{\cb a}}
\bar {A}^{{\cre i}}_{{\cg q}{\cb a}}\hat O_{{\cg r}}^{{\cg q}}\Big]\Big[\hat O'^{{\cre j}}_{{\cre k}}
{A}_{{\cre j}}^{{\cg s}{\cb b}} \bar A^{{\cre l}}_{{\cg s}{\cb b}}\ldots\Big]
=\Big[\ldots A_{{\cre i}}^{{\cg p}{\cb a}} \hat O_{{\cg r}}^{{\cg q}}\Big]\Big[\hat O'^{{\cre i}}_{{\cre k}}\bar A^{{\cre l}}_{{\cg q}{\cb a}}\ldots\Big]=\Big[\ldots A_{{\cre i}}^{{\cg p}{\cb a}} \hat O'^{{\cre i}}_{{\cre k}}\Big]\Big[\hat O_{{\cg r}}^{{\cg q}}\bar A^{{\cre l}}_{{\cg q}{\cb a}}\ldots\Big]
$}}

\put(-110,-420){\mbox{$
\acontraction{\Big(\Tr}{{\partial\over\partial \bar A}}{{\partial\over\partial A}\Big)\Big[\ldots A_{{\cre i}}^{{\cg p}{\cb a}} \hat O^{{\cre i}}_{{\cre k}}\Big]\Big[\hat O'^{{\cg q}}_{{\cg r}}}{\bar A}
\acontraction[2ex]{\Big(\Tr{\partial\over\partial \bar A}}{{\partial\over\partial A}}{\Big)\Big[\ldots A_{{\cre i}}^{{\cg p}{\cb a}} \hat O^{{\cre i}}_{{\cre k}}\Big]\Big[\hat O'^{{\cg q}}_{{\cg r}}\bar A^{{\cre l}}_{{\cg q}{\cb a}}\ldots\Big]\Big[\ldots}{ A}
\Big(\Tr{\partial\over\partial \bar A}{\partial\over\partial A}\Big)\Big[\ldots A_{{\cre i}}^{{\cg p}{\cb a}} \hat O^{{\cre i}}_{{\cre k}}\Big]\Big[\hat O'^{{\cg q}}_{{\cg r}}\bar A^{{\cre l}}_{{\cg q}{\cb a}}\ldots\Big]\Big[\ldots A_{{\cre j}}^{{\cg s}{\cb b}} \hat {\bar O}^{{\cre j}}_{{\cre m}}\Big]\Big[\hat {\bar O}'^{{\cg u}}_{{\cg t}}\bar A^{{\cre n}}_{{\cg u}{\cb b}}\ldots\Big]
=\Big[\ldots A_{{\cre i}}^{{\cg p}{\cb a}} \hat O^{{\cre i}}_{{\cre k}}\Big]\Big[\hat {\bar O}'^{{\cg q}}_{{\cg t}}\bar A^{{\cre n}}_{{\cg q}{\cb a}}\ldots\Big]\Big[\hat O'^{{\cg s}}_{{\cg r}}\ldots\Big]\Big[\ldots\hat {\bar O}^{{\cre l}}_{{\cre m}}\Big]
$}}

\put(-80,-540){\mbox{$
\acontraction{\Big(\Tr}{{\partial\over\partial \bar A}}{{\partial\over\partial A}\Big)\Big[\ldots A_{{\cre i}}^{{\cg p}{\cb a}} \hat O^{{\cre i}}_{{\cre k}}\Big]\Big[\ldots}{\bar A}
\acontraction[2ex]{\Big(\Tr{\partial\over\partial \bar A}}{{\partial\over\partial A}}{\Big)\Big[\ldots A_{{\cre i}}^{{\cg p}{\cb a}} \hat O^{{\cre i}}_{{\cre k}}\Big]\Big[\ldots\bar A^{{\cre o}}_{{\cg q}{\cb a}}\hat O'^{{\cre l}}_{{\cre o}}\Big]\Big[
\ldots}{ A}
\Big(\Tr{\partial\over\partial \bar A}{\partial\over\partial A}\Big)\Big[\ldots A_{{\cre i}}^{{\cg p}{\cb a}} \hat O^{{\cre i}}_{{\cre k}}\Big]\Big[\ldots\bar A^{{\cre o}}_{{\cg q}{\cb a}}\hat O'^{{\cre l}}_{{\cre o}}\Big]\Big[
\ldots A_{{\cre m}}^{{\cg s}{\cb b}}
\bar {A}^{{\cre j}}_{{\cg s}{\cb b}}\hat O''^{{\cre n}}_{{\cre j}}\Big]
=\Big[\ldots A_{{\cre i}}^{{\cg p}{\cb a}} \hat O^{{\cre i}}_{{\cre k}}\Big]\Big[\ldots\hat O'^{{\cre l}}_{{\cre m}}\Big]\Big[
\ldots \bar {A}^{{\cre j}}_{{\cg q}{\cb a}}\hat O''^{{\cre n}}_{{\cre j}}\Big]
$}}

\linethickness{0.8mm}

{\cre \put(0,0){\line(1,0){70}}}
\qbezier(20,0)(35,25)(50,0)
\put(30,0){
{\cre
\put(100,20){\line(1,0){50}}
\put(125,-5){\qbezier(-14,0)(-14,14)(0,14) \qbezier(14,0)(14,14)(0,14)
\qbezier(-14,0)(-14,-14)(0,-14)\qbezier(14,0)(14,-14)(0,-14) }
 }
}

\put(0,-120){
{\cre \put(10,15){\line(1,0){50}}  \put(60,-15){\line(-1,0){50}}  }
\put(35,15){\line(0,-1){30}}
{\cre \qbezier(140,15)(155,15)(155,0) \qbezier(155,0)(155,-15)(140,-15)
\put(5,0){\qbezier(175,15)(160,15)(160,0) \qbezier(175,-15)(160,-15)(160,0)}
\put(140,15){\line(-1,0){5}}   \put(180,15){\line(1,0){5}}
\put(140,-15){\line(-1,0){5}}  \put(180,-15){\line(1,0){5}}  }
}

\put(0,-240){
{\cre \put(10,15){\line(1,0){50}}}  \put(0,0){{\cg  \put(60,-15){\line(-1,0){50}}  }}
\put(35,15){\line(0,-1){30}}
{\cre \qbezier(140,15)(155,15)(155,0)  \put(140,15){\line(-1,0){5}} }
\put(0,0){{\cg \qbezier(155,0)(155,-15)(140,-15)  \put(140,-15){\line(-1,0){5}}  }}
 \put(5,0){ {\cre\qbezier(175,15)(160,15)(160,0)  \put(175,15){\line(1,0){5}}  } }
 \put(5,0){ {\cg \qbezier(175,-15)(160,-15)(160,0)  \put(175,-15){\line(1,0){5}}   }}
}

\put(0,-360){
\put(-40,0){
{\cre \qbezier(10,15)(17,8)(25,0) \put(45,0){\line(0,1){15}}  }
\put(0,0){{\cg \qbezier(10,-15)(17,-8)(25,0) \put(45,0){\line(0,-1){15}}     }}
}
\put(10,0){
{\cre \put(25,0){\line(0,1){15}} \qbezier(45,0)(50,5)(60,15)   }
\put(0,0){{\cg  \put(25,0){\line(0,-1){15}} \qbezier(45,0)(50,-5)(60,-15)   }}
}
{\cre   \qbezier(195,25)(180,5)(165,25) }
\put(0,0){{\cg  \qbezier(195,-25)(180,-5)(165,-25)  }}
\put(8,0){\line(1,0){30}}

\put(140,0){
{\cre \qbezier(10,15)(17,8)(25,0)   }
\put(0,0){{\cg \qbezier(10,-15)(17,-8)(25,0)       }}
{\cre   \qbezier(45,0)(50,5)(60,15)   }
\put(0,0){{\cg    \qbezier(45,0)(50,-5)(60,-15)   }}
}
}

\put(-40,-480){
{\cre \qbezier(10,15)(17,8)(25,0) \put(45,0){\line(0,1){15}} \put(85,-15){\line(0,1){30}} }
\put(0,0){{\cg \qbezier(10,-15)(17,-8)(25,0) \put(45,0){\line(0,-1){15}}     }}
\put(45,0){\line(1,0){40}}
\put(160,0){
{\cre \qbezier(10,15)(17,8)(25,0) \put(45,0){\line(0,1){15}} \put(85,-15){\line(0,1){30}} }
\put(0,0){{\cg \qbezier(10,-15)(17,-8)(25,0) \put(45,0){\line(0,-1){15}}     }}
}
}

\end{picture}

\bigskip

\noindent
In result, we obtain that a closed set of operators is formed by tri-valent vertices
connected by thin blue lines and thick red and green lines, which carry additional
"length" labels.
It is now clear that the Laplace operator at the r.h.s. of the recursion relation
 (\ref{basicrecursion})
does not take us away from this restricted set of operators, i.e. we get
a closed set of Ward identities.
This makes such
models potentially solvable, though there is still
a long way to go before these solutions are made as clear and explicit as
in \cite{UFN23} for the ordinary matrix models.



\subsection{The structure of RG-complete set of operators}

The previous subsection provides a description of the theory (\ref{Zpertrg1}) in terms
of the decorated tri-valent graphs.

The keystone operators ${\cal K}_{\cre 2}$ and ${\cal K}_{\cg 2}$ are now depicted
as red and green circles of length 2.

The tree operators are chains of these operators connected by thick black lines, which can
be then eliminated by the rules (a)-(e). The same is true about the loop operators.

The trees made from ${\cal K}_{\cre 2}$ alone are fully handled by rule (b),
and they are just red circles of arbitrary length $m$.
The loops made from ${\cal K}_{\cre 2}$ are also handled by rule (a) and they are
disconnected collections of red circles of arbitrary lengths.
This simple structure of the RG-completion of ${\cal K}_{\cre 2}$ was actually underlying
the solution of the model in sec.\ref{redmod}:

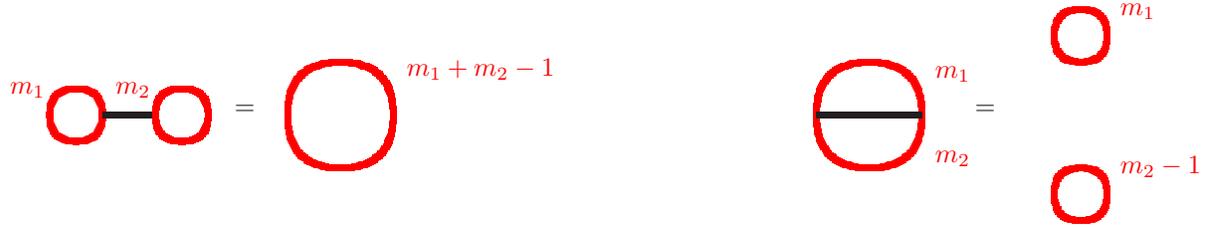
\begin{picture}(400,100)(-35,-50)
\linethickness{0.8mm}
\put(0,0){\cre\qbezier(-10,0)(-10,10)(0,10)}
\put(0,0){\cre\qbezier(10,0)(10,10)(0,10)}
\put(0,0){\cre\qbezier(-10,0)(-10,-10)(0,-10)}
\put(0,0){\cre\qbezier(10,0)(10,-10)(0,-10)}
\put(0,0){\cre\put(-25,8){\mbox{$m_1$}}}

\put(10,0){\line(1,0){20}}

\put(40,0){
\put(0,0){\cre\qbezier(-10,0)(-10,10)(0,10)}
\put(0,0){\cre\qbezier(10,0)(10,10)(0,10)}
\put(0,0){\cre\qbezier(-10,0)(-10,-10)(0,-10)}
\put(0,0){\cre\qbezier(10,0)(10,-10)(0,-10)}
\put(0,0){\cre\put(-25,8){\mbox{$m_2$}}}
}

\put(60,0){\mbox{$=$}}

\put(100,0){
\put(25,15){\cre\mbox{$m_1+m_2-1$}}
\put(0,0){\cre\qbezier(-20,0)(-20,20)(0,20)}
\put(0,0){\cre\qbezier(20,0)(20,20)(0,20)}
\put(0,0){\cre\qbezier(-20,0)(-20,-20)(0,-20)}
\put(0,0){\cre\qbezier(20,0)(20,-20)(0,-20)}
}

\put(300,0){
\put(25,14){\cre\mbox{$m_1$}}\put(25,-18){\cre\mbox{$m_2$}}
\put(0,0){\cre\qbezier(-20,0)(-20,20)(0,20)}
\put(0,0){\cre\qbezier(20,0)(20,20)(0,20)}
\put(0,0){\cre\qbezier(-20,0)(-20,-20)(0,-20)}
\put(0,0){\cre\qbezier(20,0)(20,-20)(0,-20)}
\put(-20,0){\line(1,0){40}}
}

\put(340,0){\mbox{$=$}}

\put(380,30){
\put(0,0){\cre\qbezier(-10,0)(-10,10)(0,10)}
\put(0,0){\cre\qbezier(10,0)(10,10)(0,10)}
\put(0,0){\cre\qbezier(-10,0)(-10,-10)(0,-10)}
\put(0,0){\cre\qbezier(10,0)(10,-10)(0,-10)}
\put(0,0){\cre\put(15,8){\mbox{$m_1$}}}
}

\put(380,-30){
\put(0,0){\cre\qbezier(-10,0)(-10,10)(0,10)}
\put(0,0){\cre\qbezier(10,0)(10,10)(0,10)}
\put(0,0){\cre\qbezier(-10,0)(-10,-10)(0,-10)}
\put(0,0){\cre\qbezier(10,0)(10,-10)(0,-10)}
\put(0,0){\cre\put(15,8){\mbox{$m_2-1$}}}
}

\end{picture}

The same is true for the RG-completion of ${\cal K}_{\cg 2}$:

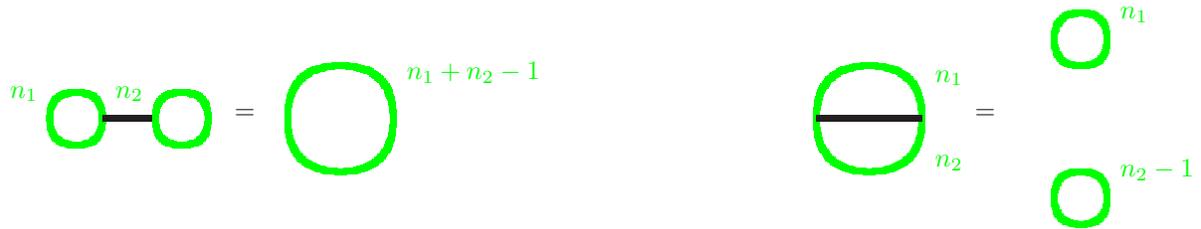
\begin{picture}(400,100)(-35,-50)
\linethickness{0.8mm}
\put(0,0){\cg\qbezier(-10,0)(-10,10)(0,10)}
\put(0,0){\cg\qbezier(10,0)(10,10)(0,10)}
\put(0,0){\cg\qbezier(-10,0)(-10,-10)(0,-10)}
\put(0,0){\cg\qbezier(10,0)(10,-10)(0,-10)}
\put(0,0){\cg\put(-25,8){\mbox{$n_1$}}}

\put(10,0){\line(1,0){20}}

\put(40,0){
\put(0,0){\cg\qbezier(-10,0)(-10,10)(0,10)}
\put(0,0){\cg\qbezier(10,0)(10,10)(0,10)}
\put(0,0){\cg\qbezier(-10,0)(-10,-10)(0,-10)}
\put(0,0){\cg\qbezier(10,0)(10,-10)(0,-10)}
\put(0,0){\cg\put(-25,8){\mbox{$n_2$}}}
}

\put(60,0){\mbox{$=$}}

\put(100,0){
\put(25,15){\cg\mbox{$n_1+n_2-1$}}
\put(0,0){\cg\qbezier(-20,0)(-20,20)(0,20)}
\put(0,0){\cg\qbezier(20,0)(20,20)(0,20)}
\put(0,0){\cg\qbezier(-20,0)(-20,-20)(0,-20)}
\put(0,0){\cg\qbezier(20,0)(20,-20)(0,-20)}
}

\put(300,0){
\put(25,14){\cg\mbox{$n_1$}}\put(25,-18){\cg\mbox{$n_2$}}
\put(0,0){\cg\qbezier(-20,0)(-20,20)(0,20)}
\put(0,0){\cg\qbezier(20,0)(20,20)(0,20)}
\put(0,0){\cg\qbezier(-20,0)(-20,-20)(0,-20)}
\put(0,0){\cg\qbezier(20,0)(20,-20)(0,-20)}
\put(-20,0){\line(1,0){40}}
}

\put(340,0){\mbox{$=$}}

\put(380,30){
\put(0,0){\cg\qbezier(-10,0)(-10,10)(0,10)}
\put(0,0){\cg\qbezier(10,0)(10,10)(0,10)}
\put(0,0){\cg\qbezier(-10,0)(-10,-10)(0,-10)}
\put(0,0){\cg\qbezier(10,0)(10,-10)(0,-10)}
\put(0,0){\cg\put(15,8){\mbox{$n_1$}}}
}

\put(380,-30){
\put(0,0){\cg\qbezier(-10,0)(-10,10)(0,10)}
\put(0,0){\cg\qbezier(10,0)(10,10)(0,10)}
\put(0,0){\cg\qbezier(-10,0)(-10,-10)(0,-10)}
\put(0,0){\cg\qbezier(10,0)(10,-10)(0,-10)}
\put(0,0){\cg\put(15,8){\mbox{$n_2-1$}}}
}

\end{picture}

Something new arises when the tree operators involve chains with both types of
the Feynman diagram vertices ${\cal K}_{\cre 2}$ and ${\cal K}_{\cg 2}$.
Now rule (c) is needed and emerging are the operator ${\cal K}_{{\cre m},{\cg n}}$
and, further, arbitrary red-green cycles
with {\it non-intersecting} thin blue shortcuts.
Thus all tree-operators are  single planar  cycles:

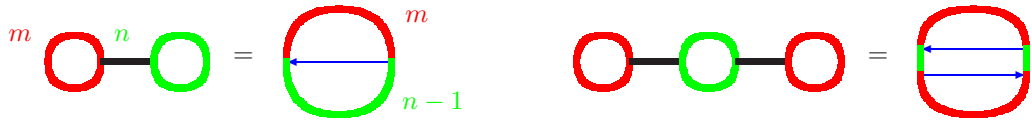
\begin{picture}(400,100)(-45,-50)

\put(120,0){\cb\vector(-1,0){40}}

\put(360,5){\cb\vector(-1,0){40}}
\put(320,-5){\cb\vector(1,0){40}}

\linethickness{0.8mm}
\put(0,0){\cre\qbezier(-10,0)(-10,10)(0,10)}
\put(0,0){\cre\qbezier(10,0)(10,10)(0,10)}
\put(0,0){\cre\qbezier(-10,0)(-10,-10)(0,-10)}
\put(0,0){\cre\qbezier(10,0)(10,-10)(0,-10)}
\put(0,0){\cre\put(-25,8){\mbox{$m$}}}

\put(10,0){\line(1,0){20}}

\put(40,0){
\put(0,0){\cg\qbezier(-10,0)(-10,10)(0,10)}
\put(0,0){\cg\qbezier(10,0)(10,10)(0,10)}
\put(0,0){\cg\qbezier(-10,0)(-10,-10)(0,-10)}
\put(0,0){\cg\qbezier(10,0)(10,-10)(0,-10)}
\put(0,0){\cg\put(-25,8){\mbox{$n$}}}
}

\put(60,0){\mbox{$=$}}

\put(100,0){
\put(25,15){\cre\mbox{$m$}}
\put(0,0){\cre\qbezier(-20,0)(-20,20)(0,20)}
\put(0,0){\cre\qbezier(20,0)(20,20)(0,20)}
\put(0,0){\cg\qbezier(-20,0)(-20,-20)(0,-20)}
\put(0,0){\cg\qbezier(20,0)(20,-20)(0,-20)}
\put(24,-18){\cg\mbox{$n-1$}}
}

\put(200,0){

\put(0,0){\cre\qbezier(-10,0)(-10,10)(0,10)}
\put(0,0){\cre\qbezier(10,0)(10,10)(0,10)}
\put(0,0){\cre\qbezier(-10,0)(-10,-10)(0,-10)}
\put(0,0){\cre\qbezier(10,0)(10,-10)(0,-10)}

\put(10,0){\line(1,0){20}}

\put(40,0){
\put(0,0){\cg\qbezier(-10,0)(-10,10)(0,10)}
\put(0,0){\cg\qbezier(10,0)(10,10)(0,10)}
\put(0,0){\cg\qbezier(-10,0)(-10,-10)(0,-10)}
\put(0,0){\cg\qbezier(10,0)(10,-10)(0,-10)}
}
\put(50,0){\line(1,0){20}}

\put(80,0){
\put(0,0){\cre\qbezier(-10,0)(-10,10)(0,10)}
\put(0,0){\cre\qbezier(10,0)(10,10)(0,10)}
\put(0,0){\cre\qbezier(-10,0)(-10,-10)(0,-10)}
\put(0,0){\cre\qbezier(10,0)(10,-10)(0,-10)}
}

\put(100,0){\mbox{$=$}}

\put(140,0){
\put(0,0){\cre\qbezier(-20,5)(-20,20)(0,20)}
\put(0,0){\cre\qbezier(20,5)(20,20)(0,20)}
\put(0,0){\cg\qbezier(-20,5)(-20,0)(-20,-5)}
\put(0,0){\cg\qbezier(20,5)(20,0)(20,-5)}
\put(0,0){\cre\qbezier(-20,-5)(-20,-20)(0,-20)}
\put(0,0){\cre\qbezier(20,-5)(20,-20)(0,-20)}
}

}

\end{picture}

The loop operators are
either the red-green cycles with the intersecting blue shortcuts or
several such red-green cycles with the shortcuts connected by thin blue lines.

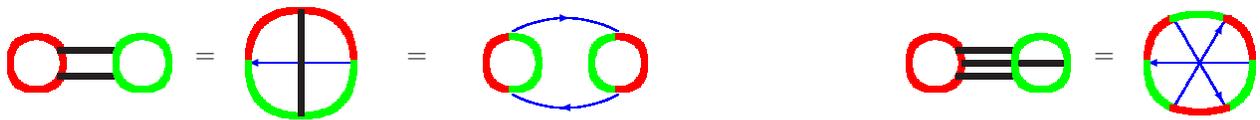
\begin{picture}(400,100)(0,-50)

\put(120,0){\cb\vector(-1,0){40}}

\put(460,0){\cb\vector(-1,0){40}}
\put(-20,0){\cb \qbezier(450,-17)(460,0)(470,17) \qbezier(450,17)(460,0)(470,-17)
\put(466.7,-11){\vector(1,-2){2}} \put(466.7,11){\vector(1,2){2}} }

\put(180,0){\cb
\qbezier(0,12)(20,22)(40,12) \put(20,17){\vector(1,0){2}}
\qbezier(0,-12)(20,-22)(40,-12) \put(20,-17){\vector(-1,0){2}}
}

\linethickness{0.8mm}
\put(0,0){\cre\qbezier(-10,0)(-10,10)(0,10)}
\put(0,0){\cre\qbezier(10,0)(10,10)(0,10)}
\put(0,0){\cre\qbezier(-10,0)(-10,-10)(0,-10)}
\put(0,0){\cre\qbezier(10,0)(10,-10)(0,-10)}

\put(8,5){\line(1,0){24}}
\put(8,-5){\line(1,0){24}}

\put(40,0){
\put(0,0){\cg\qbezier(-10,0)(-10,10)(0,10)}
\put(0,0){\cg\qbezier(10,0)(10,10)(0,10)}
\put(0,0){\cg\qbezier(-10,0)(-10,-10)(0,-10)}
\put(0,0){\cg\qbezier(10,0)(10,-10)(0,-10)}
}

\put(60,0){\mbox{$=$}}

\put(100,0){
\put(0,0){\cre\qbezier(-20,0)(-20,20)(0,20)}
\put(0,0){\cre\qbezier(20,0)(20,20)(0,20)}
\put(0,0){\cg\qbezier(-20,0)(-20,-20)(0,-20)}
\put(0,0){\cg\qbezier(20,0)(20,-20)(0,-20)}
\put(0,-20){\line(0,1){40}}
}

\put(140,0){\mbox{$=$}}

\put(180,0){
\put(0,0){\cre\qbezier(-10,0)(-10,10)(0,10)}
\put(0,0){\cg\qbezier(10,0)(10,10)(0,10)}
\put(0,0){\cre\qbezier(-10,0)(-10,-10)(0,-10)}
\put(0,0){\cg\qbezier(10,0)(10,-10)(0,-10)}


\put(40,0){
\put(0,0){\cg\qbezier(-10,0)(-10,10)(0,10)}
\put(0,0){\cre\qbezier(10,0)(10,10)(0,10)}
\put(0,0){\cg\qbezier(-10,0)(-10,-10)(0,-10)}
\put(0,0){\cre\qbezier(10,0)(10,-10)(0,-10)}
}

}

\put(340,0){
\put(0,0){\cre\qbezier(-10,0)(-10,10)(0,10)}
\put(0,0){\cre\qbezier(10,0)(10,10)(0,10)}
\put(0,0){\cre\qbezier(-10,0)(-10,-10)(0,-10)}
\put(0,0){\cre\qbezier(10,0)(10,-10)(0,-10)}

\put(8,5){\line(1,0){24}}
\put(8,-5){\line(1,0){24}}
\put(10,0){\line(1,0){40}}

\put(40,0){
\put(0,0){\cg\qbezier(-10,0)(-10,10)(0,10)}
\put(0,0){\cg\qbezier(10,0)(10,10)(0,10)}
\put(0,0){\cg\qbezier(-10,0)(-10,-10)(0,-10)}
\put(0,0){\cg\qbezier(10,0)(10,-10)(0,-10)}
}

\put(60,0){\mbox{$=$}}

\put(100,0){
\put(0,0){\cre\qbezier(-20,0)(-20,13)(-10,17)}
\put(0,0){\cg\qbezier(-10,17)(0,20)(10,17)}
\put(0,0){\cre\qbezier(20,0)(20,13)(10,17)}
\put(0,0){\cg\qbezier(20,0)(20,-13)(10,-17)}
\put(0,0){\cg\qbezier(-20,0)(-20,-13)(-10,-17)}
\put(0,0){\cre\qbezier(-10,-17)(0,-20)(10,-17)}
}
}

\end{picture}

This describes the set of operators in the extended action of the RG-completed
model (\ref{Zpertrg1}).
Clearly, they have an interpretation in terms of some quantum mechanics:
a one-dimensional QFT defined on a collection of circles (which, in turn,
can be thought of as boundaries of holes on a plane).
The circles are equipped with lengths of their segments
(thick red and green propagators),
and the field in this new effective 1d theory is responsible for the thin blue
lines connecting arbitrary points of arbitrary circles.
Since the relevant lengths are integer, this quantum mechanics should be
discrete and, perhaps, $p$-adic.
Following this line one can also approach the old problem of relating the
BZ forest formulas with the Bruhat-Tits trees, i.e. finding a $p$-adic interpretation
of the measures on the space of trees which the BZ theory associates
with arbitrary QFT with a chosen keystone operator.
This can bring us back to the old attempts of \cite{BT}
in the simplest string models, to their reformulation in terms of matrix
models and further generalizations to tensor models.

\subsection{Towards Ward identities }

The recursion relations (\ref{basicrecursion}), i.e. the equations of motion in the theory
(\ref{Zpertrg1}), are now the defining relations (equivalencies) between the contributions of decorated
tri-valent graphs.

The simplest recursion relation we already encountered in s.\ref{mmp},
it remains just the same:
\be
\mu \Big<{\cal K}_{\cre m}\Big> =
\sum_{m_1+m_2=m-1} = \Big< {\cal K}_{\cre m_1} {\cal K}_{\cre m_2}\Big>
\ee
or, pictorially,

\begin{picture}(300,100)(-150,-50)

\put(-90,0){\qbezier(0,0)(3,10)(6,20)\qbezier(0,0)(3,-10)(6,-20)}
\put(-48,0){\qbezier(0,0)(-3,10)(-6,20)\qbezier(0,0)(-3,-10)(-6,-20)}

\put(30,0){\qbezier(0,0)(3,10)(6,20)\qbezier(0,0)(3,-10)(6,-20)}
\put(75,0){\qbezier(0,0)(-3,10)(-6,20)\qbezier(0,0)(-3,-10)(-6,-20)}

\put(170,0){\qbezier(0,0)(5,15)(10,30)\qbezier(0,0)(5,-15)(10,-30)}
\put(235,0){\qbezier(0,0)(-5,15)(-10,30)\qbezier(0,0)(-5,-15)(-10,-30)}

\put(-105,-2){\mbox{$\mu\ \cdot $}}
\put(-40,-2){\mbox{$= \  \sum_{m_1+m_2'=m}$}}
\put(-80,20){\cre\mbox{$m$}}
\put(55,20){\cre\mbox{$m_1$}}
\put(55,-25){\cre\mbox{$m_2'$}}
\put(215,35){\cre\mbox{$m_1$}}
\put(215,-40){\cre\mbox{$m_2$}}

\put(82,-2){\mbox{$= \ \ \sum_{m_1+m_2=m-1}$}}

\linethickness{0.8mm}

\put(-70,0){\cre
\qbezier(-14,0)(-14,14)(0,14) \qbezier(14,0)(14,14)(0,14)
\qbezier(-14,0)(-14,-14)(0,-14)\qbezier(14,0)(14,-14)(0,-14)
 }

 \put(50,0){\cre
 \qbezier(-14,0)(-14,14)(0,14) \qbezier(14,0)(14,14)(0,14)
\qbezier(-14,0)(-14,-14)(0,-14)\qbezier(14,0)(14,-14)(0,-14)
 }
 \put(36,0){\line(1,0){28}}

\put(200,20){\cre
 \qbezier(-14,0)(-14,14)(0,14) \qbezier(14,0)(14,14)(0,14)
\qbezier(-14,0)(-14,-14)(0,-14)\qbezier(14,0)(14,-14)(0,-14)
 }

\put(200,-20){\cre
 \qbezier(-14,0)(-14,14)(0,14) \qbezier(14,0)(14,14)(0,14)
\qbezier(-14,0)(-14,-14)(0,-14)\qbezier(14,0)(14,-14)(0,-14)
 }

\end{picture}

\noindent
A similar relation holds in the green sector.

The first recursion which mixes the red and green colorings is
\be
\mu\cdot ({\cre m} + {\cg n})\cdot\Big<{\cal K}_{\cre m}{\cal K}_{\cg n}\Big>
= {\cre m}\cdot \!\!\! \sum_{m_1+m_2=m-1} \Big<{\cal K}_{\cre m_1}{\cal K}_{\cre m_2}{\cal K}_{\cg n}\Big>
+ {\cg n}\cdot \!\!\! \sum_{n_1+n_2=n-1} \Big<{\cal K}_{\cre m}{\cal K}_{\cg n_1}{\cal K}_{\cg n_2}\Big>
+ 2{\cre m}{\cg n}\cdot \Big<{\cal K}_{\cre m,\cg n} \Big>
\label{recumn}
\ee
Pictorially, with coefficients and summations omitted,

\begin{picture}(400,135)(-230,-100)

\put(-235,-2){\mbox{$\mu\ \cdot$}}

\put(-220,0){\qbezier(0,0)(3,10)(6,20)\qbezier(0,0)(3,-10)(6,-20)}
\put(-130,0){\qbezier(0,0)(-3,10)(-6,20)\qbezier(0,0)(-3,-10)(-6,-20)}
\put(-120,0){\mbox{$=$}}
\put(-100,0){\qbezier(0,0)(3,10)(6,20)\qbezier(0,0)(3,-10)(6,-20)}
\put(-10,0){\qbezier(0,0)(-3,10)(-6,20)\qbezier(0,0)(-3,-10)(-6,-20)}
\put(0,0){\mbox{$+$}}
\put(20,0){\qbezier(0,0)(3,10)(6,20)\qbezier(0,0)(3,-10)(6,-20)}
\put(110,0){\qbezier(0,0)(-3,10)(-6,20)\qbezier(0,0)(-3,-10)(-6,-20)}
\put(120,0){\mbox{$+$}}
\put(140,0){\qbezier(0,0)(3,10)(6,20)\qbezier(0,0)(3,-10)(6,-20)}
\put(240,0){\qbezier(0,0)(-3,10)(-6,20)\qbezier(0,0)(-3,-10)(-6,-20)}

\put(-140,-70){\qbezier(0,0)(3,10)(6,20)\qbezier(0,0)(3,-10)(6,-20)}
\put(-15,-70){\qbezier(0,0)(-3,10)(-6,20)\qbezier(0,0)(-3,-10)(-6,-20)}

\put(5,-70){\qbezier(0,0)(3,10)(6,20)\qbezier(0,0)(3,-10)(6,-20)}
\put(140,-70){\qbezier(0,0)(-3,10)(-6,20)\qbezier(0,0)(-3,-10)(-6,-20)}

\put(160,-70){\qbezier(0,0)(3,10)(6,20)\qbezier(0,0)(3,-10)(6,-20)}
\put(225,-70){\qbezier(0,0)(-3,10)(-6,20)\qbezier(0,0)(-3,-10)(-6,-20)}

\put(-60,-37){\mbox{$||$}} \put(60,-37){\mbox{$||$}}  \put(180,-37){\mbox{$||$}}
\put(171,0){\cb \put(0,-70){\vector(1,0){38}}   }

\linethickness{0.8mm}

\put(-200,0){\cre
\put(14,15){\mbox{$m$}}
\qbezier(-14,0)(-14,14)(0,14) \qbezier(14,0)(14,14)(0,14)
\qbezier(-14,0)(-14,-14)(0,-14)\qbezier(14,0)(14,-14)(0,-14)
}

\put(-160,0){\cg
\put(14,15){\mbox{$n$}}
\qbezier(-14,0)(-14,14)(0,14) \qbezier(14,0)(14,14)(0,14)
\qbezier(-14,0)(-14,-14)(0,-14)\qbezier(14,0)(14,-14)(0,-14)
}

\put(-80,0){\cre
\put( 14,15){ \mbox{$m$}}
\qbezier(-14,0)(-14,14)(0,14) \qbezier(14,0)(14,14)(0,14)
\qbezier(-14,0)(-14,-14)(0,-14)\qbezier(14,0)(14,-14)(0,-14)
}
\put(-94,0){\line(1,0){28}}

\put(-40,0){\cg
\put(14,15){\mbox{$n$}}
\qbezier(-14,0)(-14,14)(0,14) \qbezier(14,0)(14,14)(0,14)
\qbezier(-14,0)(-14,-14)(0,-14)\qbezier(14,0)(14,-14)(0,-14)
}

\put(40,0){\cre
\put( 14,15){ \mbox{$m$}}
\qbezier(-14,0)(-14,14)(0,14) \qbezier(14,0)(14,14)(0,14)
\qbezier(-14,0)(-14,-14)(0,-14)\qbezier(14,0)(14,-14)(0,-14)
}

\put(80,0){\cg
\put(14,15){\mbox{$n$}}
\qbezier(-14,0)(-14,14)(0,14) \qbezier(14,0)(14,14)(0,14)
\qbezier(-14,0)(-14,-14)(0,-14)\qbezier(14,0)(14,-14)(0,-14)
}
\put(66,0){\line(1,0){28}}

\put(160,0){\cre
\put( 14,15){ \mbox{$m$}}
\qbezier(-14,0)(-14,14)(0,14) \qbezier(14,0)(14,14)(0,14)
\qbezier(-14,0)(-14,-14)(0,-14)\qbezier(14,0)(14,-14)(0,-14)
}
\put(174,0){\line(1,0){22}}
\put(210,0){\cg
\put(14,15){\mbox{$n$}}
\qbezier(-14,0)(-14,14)(0,14) \qbezier(14,0)(14,14)(0,14)
\qbezier(-14,0)(-14,-14)(0,-14)\qbezier(14,0)(14,-14)(0,-14)
}

\put(-120,-70){\cre
\put( 7,12){ \mbox{$m_1$}}
\qbezier(-10,0)(-10,10)(0,10) \qbezier(10,0)(10,10)(0,10)
\qbezier(-10,0)(-10,-10)(0,-10)\qbezier(10,0)(10,-10)(0,-10)
}
\put(-80,-70){\cre
\put( 7,12){ \mbox{$m_2$}}
\qbezier(-10,0)(-10,10)(0,10) \qbezier(10,0)(10,10)(0,10)
\qbezier(-10,0)(-10,-10)(0,-10)\qbezier(10,0)(10,-10)(0,-10)
}
\put(-40,-70){\cg
\put(11,15){\mbox{$n$}}
\qbezier(-14,0)(-14,14)(0,14) \qbezier(14,0)(14,14)(0,14)
\qbezier(-14,0)(-14,-14)(0,-14)\qbezier(14,0)(14,-14)(0,-14)
}

\put(30,-70){\cre
\put( 11,15){ \mbox{$m$}}
\qbezier(-14,0)(-14,14)(0,14) \qbezier(14,0)(14,14)(0,14)
\qbezier(-14,0)(-14,-14)(0,-14)\qbezier(14,0)(14,-14)(0,-14)
}
\put(70,-70){\cg
\put(7,12){ \mbox{$n_1$}}
\qbezier(-10,0)(-10,10)(0,10) \qbezier(10,0)(10,10)(0,10)
\qbezier(-10,0)(-10,-10)(0,-10)\qbezier(10,0)(10,-10)(0,-10)
}
\put(110,-70){\cg
\put(7,12){\mbox{$n_2$}}
\qbezier(-10,0)(-10,10)(0,10) \qbezier(10,0)(10,10)(0,10)
\qbezier(-10,0)(-10,-10)(0,-10)\qbezier(10,0)(10,-10)(0,-10)
}

\put(190,-70){\cre
\put( 14,18){ \mbox{$m$}}
\qbezier(-20,0)(-20,20)(0,20) \qbezier(20,0)(20,20)(0,20)
}
\put(190,-70){\cg
\put( 14,-22){ \mbox{$n$}}
\qbezier(-20,0)(-20,-20)(0,-20)\qbezier(20,0)(20,-20)(0,-20)
}

\end{picture}

Each correlator at the both sides of such recursions is by itself a result
of calculations with the help of the Wick theorem.
The recursion describes insertion of just a single propagator:
it connect the Feynman diagram with "smaller" Feynman diagrams,
which contain less background fields (by two) and smaller power of
$\mu^{-1}$ (by one).
Instead, it converts the tree operators into the loop ones, i.e. increases
the number of thick colored circles.
In the above examples, they are disconnected, but in general thin blue lines
will appear between them, e.g.

\begin{picture}(400,140)(-270,-100)

\put(-40,0){
\put(-235,-2){\mbox{$\mu\ \cdot$}}

\put(-220,0){\qbezier(0,0)(3,10)(6,20)\qbezier(0,0)(3,-10)(6,-20)}
\put(-130,0){\qbezier(0,0)(-3,10)(-6,20)\qbezier(0,0)(-3,-10)(-6,-20)}
\put(-120,0){\mbox{$=$}}
\put(-100,0){\qbezier(0,0)(3,10)(6,20)\qbezier(0,0)(3,-10)(6,-20)}
\put(-10,0){\qbezier(0,0)(-3,10)(-6,20)\qbezier(0,0)(-3,-10)(-6,-20)}
\put(0,0){\mbox{$+$}}
\put(20,0){\qbezier(0,0)(3,10)(6,20)\qbezier(0,0)(3,-10)(6,-20)}
\put(110,0){\qbezier(0,0)(-3,10)(-6,20)\qbezier(0,0)(-3,-10)(-6,-20)}
\put(120,0){\mbox{$+$}}
\put(140,0){\qbezier(0,0)(3,10)(6,20)\qbezier(0,0)(3,-10)(6,-20)}
\put(230,0){\qbezier(0,0)(-3,10)(-6,20)\qbezier(0,0)(-3,-10)(-6,-20)}

\put(-160,-70){\qbezier(0,0)(3,10)(6,20)\qbezier(0,0)(3,-10)(6,-20)}
\put(-32,-70){\qbezier(0,0)(-3,10)(-6,20)\qbezier(0,0)(-3,-10)(-6,-20)}

\put(0,-70){\qbezier(0,0)(3,10)(6,20)\qbezier(0,0)(3,-10)(6,-20)}
\put(124,-70){\qbezier(0,0)(-3,10)(-6,20)\qbezier(0,0)(-3,-10)(-6,-20)}

\put(140,-70){\qbezier(0,0)(3,10)(6,20)\qbezier(0,0)(3,-10)(6,-20)}
\put(230,-70){\qbezier(0,0)(-3,10)(-6,20)\qbezier(0,0)(-3,-10)(-6,-20)}

\put(-60,-37){\mbox{$||$}} \put(60,-37){\mbox{$||$}}  \put(180,-37){\mbox{$||$}}
}

\put(0,-70){\cb
\put(134,0){\vector(1,0){21}}
\qbezier(106,0)(145,-30)(182,0)
\put(145,-15){\vector(-1,0){2}}
}

\linethickness{0.8mm}

\put(-240,0){
\put(0,0){\cre
\put( 10,15){ \mbox{$m$}}
\qbezier(-14,0)(-14,14)(0,14) \qbezier(14,0)(14,14)(0,14)
\qbezier(-14,0)(-14,-14)(0,-14)\qbezier(14,0)(14,-14)(0,-14)
}
\put(14,0){\line(1,0){22}}
\put(50,0){\cg
\put(-15,15){\mbox{$n$}}
\qbezier(-14,0)(-14,14)(0,14) \qbezier(14,0)(14,14)(0,14)
\qbezier(-14,0)(-14,-14)(0,-14)\qbezier(14,0)(14,-14)(0,-14)
}
}

\put(-120,0){
\put(0,0){\cre
\put( 10,15){ \mbox{$m$}}
\qbezier(-14,0)(-14,14)(0,14) \qbezier(14,0)(14,14)(0,14)
\qbezier(-14,0)(-14,-14)(0,-14)\qbezier(14,0)(14,-14)(0,-14)
}
\put(14,0){\line(1,0){22}}
\put(0,-14){\line(0,1){28}}
\put(50,0){\cg
\put(-15,15){\mbox{$n$}}
\qbezier(-14,0)(-14,14)(0,14) \qbezier(14,0)(14,14)(0,14)
\qbezier(-14,0)(-14,-14)(0,-14)\qbezier(14,0)(14,-14)(0,-14)
}
}

\put(0,0){
\put(0,0){\cre
\put( 10,15){ \mbox{$m$}}
\qbezier(-14,0)(-14,14)(0,14) \qbezier(14,0)(14,14)(0,14)
\qbezier(-14,0)(-14,-14)(0,-14)\qbezier(14,0)(14,-14)(0,-14)
}
\put(14,0){\line(1,0){22}}
\put(0,-14){\line(0,1){28}}
\put(50,0){\cg
\put(-15,15){\mbox{$n$}}
\qbezier(-14,0)(-14,14)(0,14) \qbezier(14,0)(14,14)(0,14)
\qbezier(-14,0)(-14,-14)(0,-14)\qbezier(14,0)(14,-14)(0,-14)
}
}

\put(120,0){
\put(0,0){\cre
\put(10,15){ \mbox{$m$}}
\qbezier(-14,0)(-14,14)(0,14) \qbezier(14,0)(14,14)(0,14)
\qbezier(-14,0)(-14,-14)(0,-14)\qbezier(14,0)(14,-14)(0,-14)
}
\put(13,-5){\line(1,0){24}}
\put(13,5){\line(1,0){24}}
\put(50,0){\cg
\put(-15,15){\mbox{$n$}}
\qbezier(-14,0)(-14,14)(0,14) \qbezier(14,0)(14,14)(0,14)
\qbezier(-14,0)(-14,-14)(0,-14)\qbezier(14,0)(14,-14)(0,-14)
}
}

\put(-150,-70){
\put(-30,0){\cre
\put(7,12){ \mbox{$m_1$}}
\qbezier(-10,0)(-10,10)(0,10) \qbezier(10,0)(10,10)(0,10)
\qbezier(-10,0)(-10,-10)(0,-10)\qbezier(10,0)(10,-10)(0,-10)
}
\put(4,0){\cre
\put(7,12){ \mbox{$m_2$}}
\qbezier(-10,0)(-10,10)(0,10) \qbezier(10,0)(10,10)(0,10)
\qbezier(-10,0)(-10,-10)(0,-10)\qbezier(10,0)(10,-10)(0,-10)
}
\put(14,0){\line(1,0){22}}
\put(50,0){\cg
\put(14,15){\mbox{$n$}}
\qbezier(-14,0)(-14,14)(0,14) \qbezier(14,0)(14,14)(0,14)
\qbezier(-14,0)(-14,-14)(0,-14)\qbezier(14,0)(14,-14)(0,-14)
}
}

\put(-20,-70){
\put(0,0){\cre
\put( 10,15){ \mbox{$m$}}
\qbezier(-14,0)(-14,14)(0,14) \qbezier(14,0)(14,14)(0,14)
\qbezier(-14,0)(-14,-14)(0,-14)\qbezier(14,0)(14,-14)(0,-14)
}
\put(14,0){\line(1,0){22}}
\put(46,0){\cg
\put(7,12){\mbox{$n_1$}}
\qbezier(-10,0)(-10,10)(0,10) \qbezier(10,0)(10,10)(0,10)
\qbezier(-10,0)(-10,-10)(0,-10)\qbezier(10,0)(10,-10)(0,-10)
}
\put(80,0){\cg
\put(7,12){\mbox{$n_2$}}
\qbezier(-10,0)(-10,10)(0,10) \qbezier(10,0)(10,10)(0,10)
\qbezier(-10,0)(-10,-10)(0,-10)\qbezier(10,0)(10,-10)(0,-10)
}
}

\put(120,-70){
\put(0,0){\cre
\put(10,15){ \mbox{$m$}}
\qbezier(-14,0)(-14,14)(0,14) \qbezier(14,0)(14,14)(0,14)
}
\put(50,0){\cre
\qbezier(-14,0)(-14,-14)(0,-14)\qbezier(14,0)(14,-14)(0,-14)
}
\put(50,0){\cg
\put(-15,15){\mbox{$n$}}
\qbezier(-14,0)(-14,14)(0,14) \qbezier(14,0)(14,14)(0,14)
}
\put(0,0){\cg
\qbezier(-14,0)(-14,-14)(0,-14)\qbezier(14,0)(14,-14)(0,-14)
}
}

\end{picture}

\noindent
(combinatorial coefficients can be easily restored).

\bigskip

These are the simplest examples of recursions in the simplest non-trivial tensor model
(\ref{Zpertrg1}).
The next problems to look at will be:

(1) To calculate particular averages as it was done in s.\ref{aveAr} and to check that they
satisfy these recursions. Already this is a rather tedious exercise.

(1a) As an important deviation, one can look at particular large-$N$ limits,
where the averages simplify.
What needs to be found there is an analogue of the factorization properties,
allowing one to simplify disconnected correlators and those with the intersecting blue lines.
This is also a line leading to a description in terms of the spectral curves
and the AMM/EO topological recursion.
The subsequent steps below can also be done separately for generic $N$ and in the limit,
and then lifted back to generic $N$ with the help of the genus expansion
(which will be not literally {\it genus} beyond matrix models).

(1b) Alternatively, one can put one of $N$'s equal to one and study the emerging
(not quite trivial) reduction to the complex matrix model, or, what is the same,
to the "red" quasi-tensor model in s.\ref{redmod}.
This is a simpler, still an exciting exercise, and this model has its own large-$N$
limits, AMM/EO-topological recursions, check operators etc, all being under-investigated (see, however, \cite{BD} for some models).
What facilitates this particular study are the known Virasoro constraints (\ref{Virpertr})
and their amusing indirect corollary (\ref{factRCM}), which drastically simplifies the
examination of the recursion when some circle are of unit length.

(2) To find the analogue of general formulas like (\ref{HZ1pH})
or (\ref{sum})
and check that the simple
recursions of this subsection are satisfied {\it functorially} in the parameters $m$, $n$
and promote them to particular generating functions.

(3) To find generating function(al) of another level, describing the entire set of
recursion relations, of which the ones mentioned in this subsection are just the simplest
examples.

(4) To proceed to a more complicated rainbow and then to the uncolored tensor models.

\subsection{Towards solvability of tensor models
\label{solvten}}

Actually, in our consideration of matrix models in s.\ref{mamorem} and s.\ref{redmod},
we discovered that the infinite Virasoro recursion
is not the maximal structure one can look for:
there are also {\it finite} and {\it  linear} relations (\ref{sum})
between Gaussian correlators.
The reason for their existence is presumably the interplay, a combination of infinite linear
Virasoro constraints and of the quadratic Hirota relations that reflect integrability.
The matrix model $\tau$-functions are peculiar objects in the intersection of these two worlds,
and now we recognized that, as was long expected, this indeed leads to a full {\it solvability}.

It is natural to look in the same direction in analysis of the tensor models.
In this paper, we discussed in some detail what the recursion means in this case.
What substitutes integrability for the tensor models is still a mystery.
However, we can attempt to bypass this problem, and look directly at finite
relations between correlators.
A part of the problem is that the relations like (\ref{sum})
are not homogeneous: the correlators are expressed through some more fundamental
objects, dimensions, i.e. the values characters at the topological locus \cite{RJ,DMMSS,MMM}.
However, it is not {\it a priori} clear what should play their role in the tensor case.
We postpone a detailed discussion on this subject in order to avoid mixing clear {\it facts}
reported in the present paper, with speculations and fantasies.
Here me provide just a very simple evidence that things can work.

The averages $<{\cal K}_{\cre \Lambda}>$  are already expressed through quantities
like $D_R({\cre N_r})$ and $D_R({\cg N_g}{\cb N_b})$.
A more accurate characteristic of emerging quantities
is that their {\it double} Fourier transforms factorize,
see eq.(\ref{suruN1N2}) and discussion after it.
The same is true for expressing $<{\cal K}_{\cg \Lambda}>$  through the
quantities from the class of $D_R({\cg N_g})$ and $D_R({\cre N_r}{\cb N_b})$,
and, actually, for their blue analogues $<{\cal K}_{\cb \Lambda}>$ through
 $D_R({\cb N_b})$ and $D_R({\cre N_r}{\cg N_g})$.
Thus, the question is about the {\it new}, essentially tensor model correlators,
which were not present in matrix models.
At the simple level, which we are at in the present paper,
we should look at least for expressions of the first non-trivial correlators
in the Aristotelian model,
$\Big<{\cal K}_{\cre m}{\cal K}_{\cg n}\Big>$ and $\Big<{\cal K}_{{\cre m},{\cg n}}\Big>$
through $\Big<{\cal K}_{\cre m}\Big>$ and $\Big<{\cal K}_{\cg n}\Big>$,
i.e. through the variables ${\alpha}$ and $\beta$.
Surprisingly or not, such expressions indeed exist: formulas from s.\ref{aveAr}
can be converted into
\be
\beta\cdot \Big< K_{{\cre 2},{\cg 2}}\Big>
= \beta^2\cdot \Big({\cre \alpha_r}\,{\cg \alpha_g} + {\cb \alpha_b}\Big)
= \Big<{\cal K}_{\cre 2}\Big>\Big<{\cal K}_{\cg 2}\Big>
+\Big<{\cal K}_{1}\Big>\Big<{\cal K}_{\cb 2}\Big>
\nn \\ \nn \\
\beta\cdot \Big< K_{{\cre 3},{\cg 2}}\Big>
= \beta^2\cdot\Big(({\cre \alpha_r}\!^2+\beta+1)\cdot {\cg \alpha_g}
+ 2\cdot({\cre\alpha_r}\,{\cb \alpha_b} + {\cg \alpha_g} )\Big)
= \Big<{\cal K}_{\cre 3}\Big>\Big<{\cal K}_{\cg 2}\Big>
+ 2\cdot\beta^2\cdot({\cre\alpha_r}\,{\cb \alpha_b} + {\cg \alpha_g} )
\nn \\ \nn \\
\beta\cdot \Big< K_{{\cre 4},{\cg 2}}\Big>
= \beta^2\cdot\Big({\cre\alpha_r}({\cre \alpha_r}\!^2+3\beta+5)\cdot {\cg \alpha_g}
+ (3{\cre\alpha_r}\!^2+2\beta+1)\,{\cb \alpha_b}
+ 7\cdot({\cre \alpha_r} {\cg \alpha_g}+{\cb \alpha_b})
\Big)
= \Big<{\cal K}_{\cre 4}\Big>\Big<{\cal K}_{\cg 2}\Big> +   \ldots
\nn \\ \nn \\
\ldots
\ee
and
\be
\beta\cdot\Big<{\cal K}_{\cre 2}{\cal K}_{\cg 2}\Big>
=  \beta^2\cdot\Big({\cre \alpha_r}{\cg \alpha_g} \cdot(\beta+4) +2\cdot  {\cb \alpha_b}\Big)
= \Big<{\cal K}_{\cre 2}\Big>\Big<{\cal K}_{\cg 2}\Big>\cdot\Big( 2\cdot 2+\beta\Big)
 + 2 \cdot \Big<{\cal K}_{1}\Big>\Big<{\cal K}_{\cb 2}\Big>
\nn \\ \nn \\
\beta\cdot\Big<{\cal K}_{\cre 3}{\cal K}_{\cg 2}\Big>
= \beta^2\cdot\Big(({\cre \alpha_r}\!^2+\beta+1)\cdot {\cg \alpha_g}\cdot(\beta+6)
+ 6\cdot ({\cre\alpha_r}{\cb \alpha_b}+{\cg \alpha_g})\Big)
= \Big<{\cal K}_{\cre 3}\Big>\Big<{\cal K}_{\cg 2}\Big>\cdot\Big(3\cdot 2+\beta\Big)
 + 6\beta^2\cdot ({\cre\alpha_r}{\cb \alpha_b}+{\cg \alpha_g})
\nn
\ee
\be
\beta\cdot\Big<{\cal K}_{\cre 4}{\cal K}_{\cg 2}\Big> = \beta^2\cdot \Big(
{\cre\alpha_r}({\cre \alpha_r}\!^2+3\beta+5)\cdot {\cg \alpha_g}\cdot (\beta+8)
+4\cdot (3{\cre\alpha_r}\!^2+2\beta+1)\cdot {\cb \alpha_b}
+28\cdot ({\cre\alpha_r}{\cg\alpha_g}+{\cb \alpha_b})\Big)
= \nn \\
= \Big<{\cal K}_{\cre 4}\Big>\Big<{\cal K}_{\cg 2}\Big>\cdot\Big( 4\cdot 2+\beta\Big)
+ \ldots
\nn \\ \nn \\
\ldots
\ee
Moreover, some combinations familiar from the table (\ref{corrZpertr}),
show up in these expressions.
The two next relations include $\Big<K_{3W}\Big>$ from (\ref{K3W}),
which, at the tensor model level, should probably be included into the set of dimension-like objects:
\be
\beta\cdot\Big<{\cal K}_{{\cre 3},{\cg 3}}\Big> = \beta^2\cdot \Big(
({\cre \alpha_r}\!^2+\beta+1)\cdot ({\cg \alpha_g}\!^2+\beta +1)
+ 4\cdot({\cre\alpha_r}\!^2+\beta+1) + ({\cb\alpha_b}\!^2+\beta+1)
+ 4\cdot ({\cre\alpha_r}{\cb\alpha_b}+{\cg\alpha_g})\cdot {\cg \alpha_g}
\Big)  +
\nn \\
+ \beta\cdot\Big<K_{3W}\Big>
=  \Big<{\cal K}_{\cre 3}\Big>\Big<{\cal K}_{\cg 3}\Big>
+  4\cdot\Big<{\cal K}_{1}\Big>\Big<{\cal K}_{\cre 3}\Big>
+ \Big<{\cal K}_{1}\Big>\Big<{\cal K}_{\cb 3}\Big>
+ \ldots + \Big<K_{1}\Big>\Big<K_{3W}\Big>
\nn
\ee

$$
\beta\cdot\Big<{\cal K}_{{\cre 4},{\cg 3}}\Big> = \beta^2\cdot \Big(
{\cre \alpha_r}({\cre \alpha_r}\!^2+3\beta+5)\cdot ({\cg \alpha_g}\!^2+\beta +1)
+ 6\cdot{\cre \alpha_r}({\cre\alpha_r}\!^2+3\beta+5)
+ 3\cdot{\cre \alpha_r}\cdot ({\cb\alpha_b}\!^2+\beta+1)
+ 2\cdot  (3{\cre \alpha_r}^2+2\beta+1)\,{\cb \alpha_b}\cdot {\cg\alpha_g}
+
$$
\vspace{-0.4cm}
\be
+14 ({\cre\alpha_r}{\cg\alpha_g}+{\cb\alpha_b})\cdot {\cg \alpha_g}
+12 ({\cg\alpha_g}{\cb\alpha_b}+{\cre\alpha_r})\Big)
+ 3\,{\cre\alpha_r}\beta\cdot\Big<K_{3W}\Big> \ =
\nn \\
\ = \  \Big<{\cal K}_{\cre 4}\Big>\Big<{\cal K}_{\cg 3}\Big>
+  6\cdot\Big<{\cal K}_{1}\Big>\Big<{\cal K}_{\cre 4}\Big>
+ 3\cdot\Big<{\cal K}_{{\cre 2}}\Big>\Big<{\cal K}_{\cb 3}\Big>
+ \ldots + 3\cdot\Big<K_{\cre 2}\Big>\Big<K_{3W}\Big>
\ee
and
$$
\beta\cdot \Big<{\cal K}_{\cre 3}{\cal K}_{\cg 3}\Big> = \beta^2 \cdot \Big(
 ({\cre \alpha_r}\!^2+ \beta+1)\cdot ({\cg \alpha_g}\!^2+\beta +1)\cdot(\beta+9)
+ 18\cdot({\cre\alpha_r}\!^2+\beta+1) + 3\cdot ({\cb\alpha_b}\!^2+\beta+1)
+ 18\cdot ({\cre\alpha_r}{\cb\alpha_b}+{\cg\alpha_g})\cdot {\cg \alpha_g}
\Big) +  \!\!\!\!\!\!\!\!\!\!\!\!\!\!\!\!\!\!\!
$$
\vspace{-0.4cm}
\be
+ 3\beta\cdot\Big<K_{3W}\Big>
\ = \ \Big<{\cal K}_{\cre 3}\Big>\Big<{\cal K}_{\cg 3}\Big>\cdot\Big( 3\cdot 3+\beta\Big)
 +  18\cdot\Big<{\cal K}_{1}\Big>\Big<{\cal K}_{\cre 3}\Big>
+ 3\cdot\Big<{\cal K}_{1}\Big>\Big<{\cal K}_{\cb 3}\Big>
+ \ldots + 3\cdot \Big<K_{1}\Big>\Big<K_{3W}\Big> \ \ \
\ee
We omitted $\mu$ factors to avoid overloading the formulas.

\bigskip

Is this exactly what we could dream about?

\bigskip

Not quite: there are four not quite expected new features.

\paragraph{First,} not only ${\cre \alpha_r} = {\cre N_r}+{\cg N_g}{\cb N_b}$
and ${\cg \alpha_g} = {\cg N_g} + {\cre N_r}{\cb N_b}$
appear in these expressions, but also ${\cb \alpha_b} = {\cb N_b}+{\cre N_r}{\cg N_g}$.
We remind that $\beta = {\cre N_r}{\cg N_g}{\cb N_b}$ is symmetric in the three colorings.
The four quantities ${\cre \alpha_r}$, ${\cg\alpha_g}$, ${\cb \alpha_b}$ and $\beta$
are, of course, not independent, but relation between them is irrational.
More than that, additional quantities like $\Big<{\cal K}_{3W}\Big>$ can need
to be added to the set of "dimensions".

\paragraph{Second,}  new peculiar combinations emerge involving the blue ${\cb\alpha_b}$,
which are not immediately seen in (\ref{corrZpertr}).
They are explicitly written in the intermediate formulas and are substituted by dots
at the r.h.s.
Presumably, they are made from correlators which mix the green and blue
operators and were not calculated in s.\ref{aveAr}.

\paragraph{Third,} grading is not fully respected.
Still, these formulas are quite different from the Virasoro-related
recursions like (\ref{recumn}) and have a potential to reach the capacity
of (\ref{sum}) after more examples are worked out and the structure
is fully revealed.

\paragraph{Fourth,} expressions look at best {\it quadratic} in correlators
from (\ref{corrZpertr}).
To see this better, one can interpret the additional $\beta$ factors
at the l.h.s. as $\,<{\cal K}_1>\,=\beta$.

\bigskip

Thus, the relations we are searching for seem to respect the symmetry
between the three colors (what is natural to expect from a fundamental property of the model),
but they can actually be non-linear.
Since in the matrix models these relations were linear and there they substituted the quadratic
Hirota equations, this can mean that the substitute of integrable structure in the tensor
case is going to be not quadratic,
but have a higher degree of non-linearity,
perhaps, in a spirit of the generalized Nambu structure.

All this opens a new exciting perspective for further development of the tensor models.

\section{Conclusion}

In this paper, we reviewed the general notion of RG-completeness and the
BZ-induced theory of the universal Virasoro-like constraints
(with the Virasoro algebra substituted by that of the rooted trees)
in application to matrix and tensor models.
A relatively detailed presentation was given of a simple Aristotelian
("red-green" or RGB) tensor model
(\ref{Zpertrg1}), with explicit examples of the Gaussian averages
and relations between them, some related to the Ward identities,
some to a still hidden integrability-like structure.
This illustrates a possibility of identifying the RG-completions of the tensor
models as potentially solvable, perhaps to the extent of solvability
of the simplest matrix models,
which we also raised in this paper to a qualitatively new level.
However, this solvability needs still to be studied
and much more remains to be done in the case of more
interesting rainbow models with tetrahedron-like vertices.
Fortunately, referring to Aristotle's celebrated thesis that
{\it ``... three completes the series of colours (as we
find three does in most other things), and the change into the rest
is imperceptible to sense ...''} \cite{Arist2},
we can hope that the study of RGB model advanced in the present paper
rightly captures the most important sides of the story.

\section*{Acknowledgements}

The work of H.I. was supported in part by JSPS KAKENHI Grant Number 15K05059. A.Mor. acknowledges the support of JSPS (\#  S16124) and hospitality of OCU. Our work was partly supported by RFBR grants 16-01-00291 (A.Mir.), 16-02-01021 (A.Mor.) and by joint grants
17-51-50051-YaF, 15-51-52031-NSC-a, 16-51-53034-GFEN, 16-51-45029-IND-a. Support from JSPS/RFBR
bilateral collaborations "Faces of matrix models in quantum field theory and statistical mechanics" is
gratefully appreciated.

\end{document}